\documentclass[12pt,a4paper,english,nofootinbib,,superscriptaddress]{revtex4-2}
\usepackage{lmodern}

\usepackage[T1]{fontenc}
\usepackage[latin9]{inputenc}
\usepackage{xcolor}
\usepackage{babel}

\usepackage{xcolor}
\usepackage{amsmath,mathtools}
\usepackage{amssymb,latexsym,amsmath,cancel} 
\usepackage{graphicx}
\usepackage[scr=rsfso,cal=zapfc,frak=euler,bb=ams]{mathalfa}
\usepackage{amssymb}
\usepackage{esint}
\usepackage{stmaryrd}
\usepackage[unicode=true, pdfusetitle,
 bookmarks=true,bookmarksnumbered=false,bookmarksopen=false,
 breaklinks=false,pdfborder={0 0 1},backref=false,colorlinks=false]
 {hyperref}
\def\b{\begin{equation}}
\def\e{\end{equation}}

\makeatletter
\@ifundefined{textcolor}{}
{%
 \definecolor{BLACK}{gray}{0}
 \definecolor{WHITE}{gray}{1}
 \definecolor{RED}{rgb}{1,0,0}
 \definecolor{GREEN}{rgb}{0,1,0}
 \definecolor{BLUE}{rgb}{0,0,1}
 \definecolor{CYAN}{cmyk}{1,0,0,0}
 \definecolor{MAGENTA}{cmyk}{0,1,0,0}
 \definecolor{YELLOW}{cmyk}{0,0,1,0}
 }

\usepackage{latexsym}\usepackage{bm}

\makeatletter
\providecommand*{\Dashv}{%
  \mathrel{%
    \mathpalette\@Dashv\vDash
  }%
}
\newcommand*{\@Dashv}[2]{%
  \reflectbox{$\m@th#1#2$}%
}

\makeatother

\makeatother

\begin{document} 

\title{Cosmological Weyl-Einsteinian-Cubic Gravity as a Gauge Theory of Gravity}

\author{Suat Dengiz}

\email{suat.dengiz@ostimteknik.edu.tr}

\affiliation{OSTIM Technical University, Department of Computer Engineering, 06374 Ankara, Turkey}

\date{\today}
\begin{abstract} 
We construct a Weyl-Einsteinian-Cubic Gravity (ECG) as a cubic gauge theory of gravity via abelian gauge and properly tuned compensating real scalar fields. The model is free from any dimensionful parameters. The bare ECG emerges as the lower energy limit of the Weyl-ECG in the local {\it non}-conformal-invariant vacua (i.e., broken phase) in the maximally symmetric spacetimes fixing the vacuum expectation value of the scalar field to the Planck mass scale. Here, the natural presence of (anti-) de Sitter backgrounds spontaneously breaks Weyl's local conformal symmetry akin to the Higgs mechanism, while it is radiatively broken at the renormalization scale at the one-loop level in flat vacua through the Coleman-Weinberg mechanism. The model allows anti-de Sitter and flat spaces but does not allow de Sitter to be vacuum spacetime solutions. The properties of the model deserve further exploration, specifically, those of nonperturbative (e.g., instantons and/or anti-instantons) contributions, for example, in the resurgence or tachyon condensation context requires detailed study.

% \tableofcontents{}\[\]

\end{abstract}

\maketitle

\section{Introduction}

The biggest obstacle to a UV-complete quantum field theory of general relativity is its inability to achieve unitarity and renormalizability simultaneously. Recall that, during the first and second corrections to Compton scattering due to graviton loops at the radiative scale in general relativity within the context of perturbative quantum field theory, the vertices turn out to be exclusively momenta-dependent, unlike quantum electrodynamics. Consequently, quartic infinities arise at the one-loop level, and sextic infinities emerge at the two-loop level at higher momenta in the UV limit. The renormalization of quantum general relativity unavoidably requires an infinite number of counter-terms in the compact form as $\triangle S\sim \sum_{n=0}^{\infty} \int d^4x\, (\kappa^2 R)^n R^2,$ where $\kappa$ is the dimensionful Newton's constant. As is known, the one-loop divergence is fixed through the topological Gauss-Bonnet term and a particular field redefinition. But the existing catastrophic infinities at the two-loop cannot be renormalized. Further, the theory gets out of control even at the one-loop level as one takes into account all the possible internal matter (scalars, spinors, photons, Yang-Mills) loops, too. Thus, general relativity is non-renormalizable, albeit being unitary \cite{tHooft:1974toh, Deser:1974cz}. It has been demonstrated in \cite{Stelle:1976gc} that general relativity integrated by specific quadratic curvature combinations turns out to be power-counting renormalizable despite losing unitarity this time. Having a complete quantum theory of gravity inevitably necessitates the remedy of its very fundamentals [e.g., (loop-level) propagator structure \emph{etc.}] by assuming either extra appropriate symmetries or higher-order modifications in curvature (with or without other viable lower/higher spin fields) \emph{etc}., such that those annexed parts would be effective at the sufficiently higher frequencies but suppressed in the lower frequencies. Since the renormalization of general relativity requires infinite numbers of higher-curvature counter terms at each high energy scale, and also since it was shown in \cite{Stelle:1976gc} that the quadratic curvature modification absorbs the existing divergences despite the cost of losing unitarity, the higher-order curvature gravity models (with or without extra lower spin fields) have occupied a crucial place in this research area. In this respect, as the standout and relevant alternative approaches in the last decade, see, e.g., \cite{IDG1, IDG2, IDG3} for how the propagator structure of general relativity gets modified by \emph{non-local} infinite derivative curvature interactions through particular form factors putting forward intriguing explanations at both the classical and quantum levels. Higher-order curvature terms also arise in string theory and holography, such as specific quadratic and cubic curvature corrections in the low energy limits of distinct types of string theories \cite{string1,string2} or quasi-topological higher-derivative curvature corrections in holography \cite{qttholograpy1,qttholograpy2,qttholograpy3,qttholograpy4,qttholograpy5,qttholograpy6, qttholograpy7}). As toy models, higher-curvature modifications have also been widely studied in lower dimensions for a long time. Among those, despite its so-called problems in the holographic perspective at the chiral points, the most successful one is the Topologically Massive Gravity (TMG) by Deser, Jackiw, and Templeton which contains the higher-derivative gravitational Chern-Simons topological term \cite{tmg1,tmg2}. 

As is naturally expected, any viable covariant higher curvature gravity model must have the following characteristics: ($a$) The linearized particle content around the constant curvature vacua are identical to those of ordinary general relativity, namely, massless transverse spin-$2$ along with the extra emerging propagating degrees of freedom such as massive spin-$2$ or spin-$0$ mode. ($b$) The existing relative couplings among the compensating higher-order curvature invariant terms are dimension-independent. ($c$) It must not be trivial or topological in $4D$ for the sake of a dynamical theory. For the last decade, despite its shortcomings such as existing ghost modes (instabilities), a particular cubic curvature gravity model entitled ``Einsteinian Cubic Gravity (ECG)'' has been proposed in \cite{ECG}. ECG is interesting since it transcends the so-called Gauss-Bonnet or cubic-Lovelock  modifications and fulfills all three prerequisites aforesaid and thus turns out to also be a nontrivial and non-topological cubic-curvature model, unlike the cubic Lovelock model \cite{ECG, Edelstein, BeltranJimenezGCG}. (See \cite{ECGext1,seealso1,seealso2,seealso3,seealso4,seealso5,seealso6,seealso7,seealso8} for the static, stationary, and charged modified black hole solutions of ECG and \cite {ECGext2,ECGext3} for its quasi-topological generalizations. For the cosmological early and late expansion epochs as well as a smooth universe bouncing between two de Sitter vacua in ECG, see \cite{cosmology1,cosmology2,cosmology3,cosmology4,BeltranJimenezGCG}. For the studies on quasinormal modes, gravitational shadow and lensing in ECG, see \cite{quasishadowlens1,quasishadowlens2,quasishadowlens3}. See \cite{Braneworlds} for the Braneworlds in ECG and \cite{Jensko:2023lmn} for a comprehensive thesis in general on geometric modifications of gravity.) But, as is known, since ECG does not belong to the Lovelock family, it contains Ostrogradski ghost-like modes about arbitrary vacua as anticipated and involves pathologies \cite{Woodard1,Woodard2,BeltranJimenezpertviable}. First of all, one should notice that those instabilities turn out to be catastrophic as one approaches the theory as a UV-complete model, whereas it is secure as to the perturbative perspective via cautious fine-tuning of couplings. In this case, one can study that either the ghost modes are infinitely massive and hence not propagating degrees of freedom such that it decouples in strongly coupled regimes about some particular spacetimes and thus resolves the existing instability \cite{BeltranJimenezpertviable}\footnote{In fact, it is expected to have similar pathologies as extra (massive) degrees of freedom  are considered \cite{vdvz1,vdvz2}. In such cases, e.g., with the notable Vainshtein radius, \cite{vdvz3} shows that nonlinear effects could repress the problematic degrees of freedom in the linearized model at the perceptible scales.} Of course, this or other alternative approaches can be addressed to render the ghost modes to proceed toward a UV-complete ECG that is already safe to be studied at the perturbative scale. Due to its notable exclusive attributes mentioned above, ECG surely deserves to be further studied, at least as an effective theory. Furthermore, as for the failures encountered in the full quantum ECG, it is indeed expected to have such problems as one considers ECG as a UV-complete theory since it involves general relativity in its core. That is, the main obstacle against a well-behaved quantum theory of gravity is the \emph{dimensionful} nature of Newton's constant (i.e., $[\kappa]=m^{-2}$) which obliges the assumption of infinite numbers of counter-terms at each energy scale for such a desired legitimate complete quantum gravity. Thus, it is natural to anticipate that such a UV-complete theory be free from dimensionful parameters as one moves toward the high-frequency domains, (e.g., Planck's length)\footnote{Note that it is consistent with the eminent ``Dirac large numbers hypothesis'' stating that the universal constants are not constant at all, rather they rely on the distinct energy scales of the Universe \cite{Dirac1,Dirac2} as well as the Hermann Weyl's proposal in \cite{Weyl1,Weyl2,Weyl3}.}. Actually, this is also the case in special relativity, that is,  the particles' masses do not retain their significance for sufficiently higher energies. Thus, all these suggest boosting Newton's constant to a viable field augmented with the local conformal symmetry for generic spacetimes. 

In addition to all those observations about ECG, recall the following attribute of gravity once more: in general relativity with the torsionless and metric compatibility restrictions, the metric becomes the sole dynamical quantity controlling the sort of geometry and affine structure of the substance manifolds. So, the degrees of freedom associated with the torsion and non-metricity are omitted, and thereby the spacetimes with the Levi-Civita connections are taken as the only physical solutions. However, a broader geometrical description of gravity seems to be essential to shed light on such a desired ultimate theory which will affirmatively upgrade the basic characteristics of general relativity even deeper into extremely high energy regimes (i.e., Planck scale and beyond) where its regular form breaks down \cite{modeg1,modeg2,modeg3,modeg4}. Having a broader geometrical description of the ordinary general relativity without assuming any additional modification will require fully or partially lifting the limitations (namely, torsion-free and metricity) on the connection which were initially set for the sake of metric compatibility. Due to the points mentioned so far and many other reasons in the experimentally confirmed standard model gauge theories, the partial relaxation of metricity restriction in the regular general relativity predominantly deserves more attention as it supplies a natural geometrical environment for the \emph{local scale-invariant} field theories. This upgrading of gravitational connection is dubbed ``Weyl's conformal gauge modification''. Here, the so-called global rigid scale invariance mandating conformal flatness for Lorentz invariance\footnote{For a given field theory involving the specific configuration of field operator $\widehat{\varphi}$ to be scale-invariant (i.e., conformal to a complex plane) under the transformations $x \rightarrow x^{'}= \lambda x$ only if  $\widehat{\varphi} \rightarrow \widehat{\varphi} (x{'}) = \lambda^h \widehat{\varphi} (x)$ where $\lambda$ is any constant and $h$ is the scaling dimension of field operator.} is boosted to a local scale invariance to achieve the Poincar\'e invariant theories in any curved spacetimes \cite{Weyl1,Weyl2,Weyl3,Weyl4,Weyl5,Weyl6,Weyl7,affine1}. Notice also that \emph{Weyl's gauging provides not only local conformal symmetry but also a gauge theory of gravity as in the Standard Model of particles rooted on the internal gauge group symmetries}\footnote{ In fact,  in a parallel analogy, for instance, with the standard model Higgs mechanism, it is affirmative to expect the local scale symmetry to be one of the fundamental symmetries of nature, at least at the UV regimes, because it does not tolerate any \emph{dimensionful} couplings (say, Newton's constant, Higgs mass) which can be generated only after the local scale invariance is broken through a reasonable mechanism.}. See, e.g., \cite{WeylSee7,WeylSee8} for how the local scale symmetry is coherently incorporated into the standard model and \cite{WeylSee81} for a recent study where it is shown that the metric-affine gravity provides a compatible particle physics. See \cite{WeylSee9,WeylSee10} for the one-loop beta functions in Weyl-invariant gravity models and \cite{WeylSee11} for how to quantize a Weyl-invariant gravity model coupled to a Stuckelberg photon through background field method. See \cite{GrumillerJackiw} where Jackiw and Grumiller showed that the Einstein-Weyl equations emerge as a consequence of (anti-)self-duality of $3+1$-dimensional Weyl tensor in the Kaluza-Klein reduction from four to three dimensions. See also \cite{WeylTMG1} where it has been demonstrated that the related asymptotic symmetry generating Lee-Wald charges of Weyl-invariant TMG \cite{WeylTMG2} are both closed and integrable wherein the associated charge algebra is composed of a class of supertranslation, two families of Witt algebras and a family of multiple charges. For a generic enhanced Gauss-Bonnet gravity theory in Weyl's geometry giving the vector-tensor Horndeski interaction in $4-$dimensions, see \cite{WeylSee1}. Finally,  see, e.g., \cite{WeylSee2} for the n-dimensional extension of quadratic curvature gravity in Weyl's geometry, and \cite{WeylSee3, WeylSee4, WeylSee5, Dengiz:2011ig, WeylSee6} for some other related works. 

As is mentioned above, observe that the Standard Model of particles is predominantly established as a gauge theory augmented with the internal compact Lie group symmetries \cite{Gaugegravity1, Gaugegravity2}. This has ended up with several experimentally revealed inventions, including the exceptional Higgs mechanism for the generation of the existing masses of the particles through a legitimate symmetry-breaking mechanism. Hence, almost all the fundamental interactions in universe seem to be underlined by the gauge field paradigm, whereas gravity stands out as a curious and notable exclusion. Corollary to this, a reasonable gauge theory of gravity seems to be essential. See particularly \cite{Gaugegravity3} (and references therein) for a recent and a comprehensive study on the physical reasons why the upgrading of the paradigm is required, especially why \emph{nonmetricity is essential to further shed light on high energy regimes beyond Standard Model}.

Combining all the above-mentioned crucial points and observations, we will formulate a $3+1$-dimensional Weyl-invariant ECG in Weyl's geometry-as a cubic gauge theory of gravity-with the help of appropriately tuned weights of compensating real scalar fields and Weyl's curvature tensors composed of both the regular Riemannian curvature tensors and abelian gauge fields in this paper. We will see that the model does not involve any dimensionful parameter and the regular ECG comes about in the local {\it non}-conformal-invariant vacua (i.e., broken phase) of the model in the maximally symmetric spacetimes. Here, Weyl's local conformal symmetry will be spontaneously broken akin to the Higgs mechanism in (anti)-de Sitter spaces and radiatively broken at one-loop level in flat spaces via the notable Coleman-Weinberg mechanism. We will also see that the anti-de Sitter and flat spaces are permitted but de Sitter is excluded to be vacuum spacetime solutions.

The layout of the paper is as follows: In Sec. \ref{sectwo}, we present the preliminaries necessary during calculations. In Sec. \ref{secthree}, we formulate the Weyl-Einsteinian-Cubic gravity and introduce its fundamental characteristics. Sec. \ref{secfour} is devoted to the derivation of the field equations for the dynamical fields for \emph{arbitrary} background spacetimes. In Sec. \ref{secfive}, we study the field equations in the maximally symmetric vacuum spacetimes and the corresponding symmetry-breaking mechanism. In Sec. \ref{secsix}, we conclude our results and examine the prospective road maps. Lastly, all the essential expressions required to analyze and understand the model are given in the Appendix. 

\section{Preliminaries}\label{sectwo}
Before progressing further, as is explicitly shown in the rest of paper, it is necessary to emphasize here that the Weyl-Einsteinian-Cubic gravity (WECG) provides not only the Poincar\'e invariant ECG in \emph{any arbitrary} curved spacetimes but also how the ordinary ECG emerges as the vacuum (lower-energy) limit of WECG as a result of a viable symmetry breaking mechanism (namely, either spontaneously in the (Anti)-de Sitter spacetimes like the Standard Model Higgs Mechanism or radiatively at one-loop level in flat backgrounds as in the Coleman-Weinberg mechanism for the quantum electrodynamics of a scalar field in four-dimensions). One should also note that since the Weyl's gauge symmetry is achieved with the help of extra gauge and scalar fields, extra tachyonic modes may emerge. But, as it is clearly explained in the conclusion, analogous phenomena (i.e., unstable tachyonic modes with \emph{imaginary masses and so vacuum expectation values}) also emerge in $D\bar{D}$-brane pairs in string field theory, ending up with the rolling down of the field to a stable vacuum ground state that are free from any observable tachyonic mode; or, in the resurgence theory which combines the perturbative and nonperturbative contributions such that the two-fold ambiguous imaginary part of the Borel resummation of the perturbation theory and the relevant part of instanton-anti-instanton amplitude completely cancel each others leading to a stable system. (For a related detailed discussion and also possible future direction, see the conclusion.) One shall also note that, we will study the foremost fundamentals of model (e.g., construction of the action, field equations, vacuum analysis, etc.) in this work, but of course, since the model suggests intriguing outcomes and all the analyses cannot be done in a single paper, further extra studies are apparently required to understand WECG in all essential aspects. 

As for the quantitative base of the model, let us first note that, as in the explicit reasons given below, it is a priori natural to expect a highly complicated action describing WECG since, as is given in (\ref{wecgaction2}), its explicit form contains various non-minimal interactions among higher-order curvatures, vector and scalar fields as well as their higher-order self-interactions. As we shall see, the explicit version of WECG includes not only the bare cubic curvature terms but also additional higher-order curvature terms up to second order, nonminimally coupled to gauge and scalar fields, as well as self-interacting gauge and scalar field terms up to sextic and quartic order, respectively. For this reason, before dwelling on the details of the model, let us now present some crucial preliminaries that supply convenience during calculations. Firstly, our metric signature convention is $(-,+,+,+)$ and we use $[\nabla_\mu,\nabla_\nu] V_\alpha=R_{\mu\nu\alpha\beta}V^\beta$.  Secondly, let us also define two classes of two-fold indices operations which will be very useful particularly during the derivations of the field equations for $(g^{\mu\nu}, A^\mu, \Phi)$: the first one is called as the ``two-fold symmetric-(anti)symmetric indices operations'' providing the following operations between indices 
\begin{equation}
X_{\llparenthesis \, ab} Y_{cd\,\rrparenthesis}=\frac{1}{2} \left[ X_{(ab)}Y_{(cd)}+X_{(cd)}Y_{(ab)} \right]\quad \rm{and} \quad X_{\{ \!\!\{ \, ab} Z_{cd\,\}\!\!\}ef\cdots}=\frac{1}{2} \left[ X_{(ab)}Z_{[cd]ef\cdots}-X_{(cd)}Z_{[ab]ef\cdots} \right].
\label{twofoldder1}
\end{equation}
The second class is dubbed the ``two-fold cyclic symmetric-(anti)symmetric indices operations'' described by the following compact form of two-fold cyclic operations
\begin{equation}
 X_{\lessdot \mu\nu\alpha  \gtrdot} =\frac{1}{2} \left[X_{(\mu\nu)\alpha}+X_{(\mu\alpha)\nu}\right] \quad \rm{and} \quad  X_{ \mu\nu\alpha \vDash}=\frac{1}{2} \left[X_{\mu(\nu\alpha)}-X_{\alpha(\mu\nu)}\right]. 
\label{twfldcylclc}
\end{equation}
At this point, one shall observe that the locations and order of indices are crucial in (\ref{twfldcylclc}). That is, as a simple sample, let $ X_{\mu\nu\alpha}=(\nabla_\mu A_\nu)A^\alpha$ and $X_{\alpha \mu\nu}=A^\alpha (\nabla_\mu A_\nu)$. Then, notice that $X_{ \mu\nu\alpha \vDash} \neq X_{\alpha \mu\nu \vDash}$ and $X_{\lessdot \mu\nu\alpha  \gtrdot} \neq  X_{\lessdot \alpha \mu\nu  \gtrdot}$ in both two-fold cyclic operations.

\section{Weyl-Einsteinian Cubic Gravity: Construction}\label{secthree}
\subsection{Cosmological Einsteinian Cubic Gravity: A Review}
Due to being the base model, let us shortly recall that the usual ECG is promoted to arbitrary cosmological scenarios (FLRW) with additional cubic terms in the {\it Cosmological Einsteinian Cubic Gravity} (Cosmological ECG) \cite{Edelstein}. While keeping the intriguing characteristics of usual ECG intact, the Cosmological ECG also provides a well-defined initial value problem and further a geometric mechanism driving an inflationary period in the radiation dominated region to a late time cosmology converging to $\Lambda CDM$. In doing so, they start with the plausible cubic combinations
\begin{equation}
\begin{array}{ll}
\mathcal{L}_1 = {{{R_a}^c}_b}^d {{{R_c}^e}_d}^f {{{R_e}^a}_f}^b ~, \qquad & \mathcal{L}_2 = {R_{ab}}^{cd} {R_{cd}}^{ef} {R_{ef}}^{ab} ~, \\ [0.43em]
\mathcal{L}_3 = R_{abcd} {R^{abc}}_e R^{de} ~, \qquad & \mathcal{L}_4 = R_{abcd} R^{abcd} R ~, \\ [0.43em]
\mathcal{L}_5 = R_{abcd} R^{ac} R^{bd} ~, \qquad & \mathcal{L}_6 = {R_a}^b {R_b}^c {R_c}^a ~, \\ [0.43em]
\mathcal{L}_7 = R_{ab} R^{ab} R ~, \qquad & \mathcal{L}_8 = R^3 ~, \\ [0.43em]
\mathcal{L}_9 = \nabla_a R_{bc} \nabla^a R^{bc} ~, \quad & \mathcal{L}_{10} = \nabla_a R\, \nabla^a R ~.
\label{cosmologecg}
\end{array}
\end{equation}
Referring \cite{Edelstein} for details, let us quote their results of that due to the topological effects and the requirement that it shall preserve all the promising characteristics of ECG, one ends up with the following particular cubic combinations
\begin{equation}
\mathcal{P} = 12 \mathcal{L}_1 + \mathcal{L}_2 -12 \mathcal{L}_5 + 8 \mathcal{L}_6 ~, \qquad \mathcal{C} = \mathcal{L}_3 - \frac14 \mathcal{L}_4 - 2 \mathcal{L}_5 + \frac12 \mathcal{L}_7 ~,
\label{cecgterms}
\end{equation}
which describes a unitary massless graviton around maximally symmetric vacua. Observe that the terms $\mathcal{L}_8=R^3, \mathcal{L}_9= \nabla_a R_{bc} \nabla^a R^{bc} $ and $\mathcal{L}_{10}= \nabla_a R\, \nabla^a R$ in \ref{cosmologecg} drop in the construction of the Cosmological ECG. Moreover, it has been shown in \cite{Edelstein} that, in addition to preserving all the promising characteristics of ECG, a particular combination of $\mathcal{P} $ and $\mathcal{C}$ also provides a cosmological scenario (in FLRW ansatz) wherein the set of Friedmann equations remains second order.

\subsection{Weyl-Invariant Extension of Einsteinian Cubic Gravity}\label{secthree}

Now that we have briefly presented the preliminaries for our construction, let us now proceed with our focal purpose of concretizing the Weyl-invariant extention of the Cosmological ECG which we abbreviate as WECG. First of all, note that we construct the Weyl's gauging of the \emph{particular} cubic model of the Cosmological ECG \cite{Edelstein, BeltranJimenezGCG} and thus take the specific cubic curvature terms in (\ref{cecgterms}). Accordingly, it will be primarily expected from the WECG to recover the main attributes of the ordinary Cosmological ECG in the subregion of its broken phase. At this point, one shall observe that the main projective of the paper is to build the Weyl-invariant extension of the above-mentioned {\it particular} cubic model, namely, the Cosmological ECG which describes $7$ terms and so $7$ generic couplings. But, of course, this is a very active research fields and hence there are other alternative directions to shed further lights on the local conformal invariant cubic curvature theories. Here, having a viable Weyl-invariant extension of the other cubic models based on the Weyl tensors and traceless Ricci tensor is particularly interesting because it has potential to provide less number of independent higher-derivative terms and so relatively shorter field equations: in this regard, e.g., see \cite{Modestosixderiv} which posits that there are $6$ independent terms in $3+1$ dimensions or \cite{Maldacenacubic1} (and more explicitly in (3.6) of \cite{Maldacenacubic2}) for a $6$ dimensional conformal quantum gravity describing certain cubic terms of Weyl tensors, Ricci tensors and particular divergence terms. At this point, see also \cite{Rachwal1, Rachwal2} for another interesting alternative conformal quantum gravity model constructed with the help of (nonlocal) higher derivative gravitational corrections wherein intriguing outcomes such as resolutions of catastrophic problems towards a complete theory are presented. Observe that the Weyl's gauge invariant cubic gravity that we shall construct and study, in general, is different in many aspects: in our construction, we shall take the basics (particularly, the local gauge field nature) of the $3+1$-dimensional Standard model of fundamental particles as one of guiding (reference) paths. This is mainly because of the following main reasons: firstly, as was mentioned in the introduction, the Standard model has been verified by countless number of experiments and almost all the fundamental interactions are based on the {\it gauge field} paradigm, while gravity stands out as a curious and notable exclusion. So, a reasonable {\it gauge theory of gravity} seems to be essential. (See, e.g., \cite{Gaugegravity3} [and references therein] for a recent intriguing study on the requirement of such a gauge theory of gravity.)\footnote{Observe that despite many appealing higher dimensional models, since the observed universe is $3+1$, we mainly focus on the construction of a complete well-behaved $3+1$ dimensional gauge gravity model.}. Thus, as in the Standard model gauge theory, the WECG that we dwell on is achieved with the help of extra {\it abelian gauge field} unlike those in \cite{Maldacenacubic1, Maldacenacubic2, Rachwal1, Rachwal2}. Hence, this construction supplies a gauge theory of gravity model analogous to Standard model gauge theories. This approach of local conformal-invariance also provides Poincare invariant theories for not only certain curved spacetimes having asymptotical conformal group but also {\it any} arbitrary spacetimes, whereas those approaches, e.g., in \cite{Maldacenacubic1, Maldacenacubic2, Rachwal1, Rachwal2} do not assume any extra gauge field and they require the spacetime solutions to be asymptotically conformal invariant\footnote{For example, note that the construction in \cite{Maldacenacubic1} requires asymptotically (A)dS spacetime solution together with the selection of some certain Neumann boundary conditions for the sake of viable limits.} Of course, since all those distinct approaches are notable ones and deserves to be studied in all possible aspects, we suggest all of them to get further possible insights towards a ultimate theory.

As for our main model of WECG, the relative coupling constants will be selected as those in the strongly coupled (Cosmological) extension of ECG upgraded here by the dimensionless free parameters $\alpha_i $ (with $i=1\cdots 8$) to get all the possible configurations of relative couplings among the higher order terms in the model. [Note that assuming $\alpha_i$'s will also become useful as one studies the unitary (ghost and tachyon-free) parameter regions of WECG]. Referring to \cite{WeylSee2,Dengiz:2011ig} for the details of Weyl's local conformal invariance via real scalar and abelian gauge fields, one can write the WECG gravity action with the help of properly tuned weight for the compensating scalar field as follows
\begingroup
\allowdisplaybreaks
\begin{align}
		S_{\mbox{WECG}}&=  \int \mbox{d}^4x \sqrt{-g} \,\Big\{\alpha_1 \Phi^2\widehat{R}+\Phi^{-2} \Big[12 \alpha_2 \widehat{R}_\mu{^\rho}{_\nu}{^\sigma} \widehat{R}_\rho{^\tau}{_\sigma}{^\eta}\widehat{R}_\tau{^\mu}{_\eta}{^\nu}+\alpha_3 \widehat{R}{_{\mu\nu}}{^{\rho\sigma}}\widehat{R}{_{\rho\sigma}}{^{\tau\eta}} \widehat{R}{_{\tau\eta}}{^{\mu\nu}}\nonumber\\
		&+2\alpha_4\widehat{R}\widehat{R}_{\mu\nu\rho\sigma}\widehat{R}^{\mu\nu\rho\sigma}-8 \alpha_5 \widehat{R}^{\mu\nu}\widehat{R}_\mu{^{\rho\sigma\tau}} \widehat{R}_{\nu\rho\sigma\tau}+4\alpha_6 \widehat{R}^{\mu\nu}\widehat{R}^{\rho\sigma} \widehat{R}_{\mu\rho\nu\sigma}-4\alpha_7\widehat{R}\widehat{R}_{\mu\nu} \widehat{R}^{\mu\nu}\nonumber\\
		&+8\alpha_8 \widehat{R}_\mu{^\nu}\widehat{R}_\nu{^\rho}\widehat{R}_\rho{^\mu}\Big]  \Big\}+S_{\Phi}+S_{A^\mu},
\label{wecgaction1}
\end{align}
\endgroup
which is invariant under the ensuing Weyl's local transformations in $3+1$-dimension
\begingroup
\allowdisplaybreaks
\begin{align}
& g_{\mu\nu} \rightarrow g^{'}_{\mu\nu}=e^{2 \lambda(x)} g_{\mu\nu}, \hskip 0.7 cm \Phi \rightarrow \Phi^{'} =e^{-\lambda(x)} \Phi\nonumber\\
&{\cal{D}}_\mu \Phi =\partial_\mu\Phi-A_\mu \Phi, \hskip 0.9 cm A_\mu \rightarrow A^{'}_\mu = A_\mu - \partial_\mu \lambda(x),
\label{weyltrnsf}
\end{align}
\endgroup
where $\lambda(x) $ is an arbitrary real function and ${\cal{D}}_\mu$ is the so-called gauge covariant derivative to achieve local gauge invariance. Also, note that the volume part transforms as $\sqrt{-g} \rightarrow (\sqrt{-g})^{'}=e^{4 \lambda(x)} \sqrt{-g}$.  Here, the Weyl's extensions of curvature terms in generic $n-$dimensions are
\begingroup
\allowdisplaybreaks
\begin{align}
\widehat{R}^\mu{_{\nu\rho\sigma}} (g_{\mu\nu},A_\mu) &=R^\mu{_{\nu\rho\sigma}}+\delta^\mu{_\nu}F_{\rho\sigma}+2\delta^\mu_{[\sigma} \nabla_{\rho]} A_\nu+2 g_{\nu[\rho}\nabla_{\sigma]} A^\mu+2 A_{[\sigma} \delta_{\rho]}^\mu A_\nu+2 g_{\nu[\sigma} A_{\rho]} A^\mu\nonumber\\
& +2 g_{\nu[\rho} \delta_{\sigma]}^\mu  A^2,\nonumber\\
\widehat{R}_{\nu\sigma} (g_{\mu\nu},A_\mu)&= \widehat{R}^\mu{_{\nu\mu\sigma}}=R_{\nu\sigma}+F_{\nu\sigma}-(n-2)\Big [\nabla_\sigma A_\nu - A_\nu A_\sigma +A^2  g_{\nu\sigma} \Big ]-g_{\nu\sigma}\nabla \cdot A,\nonumber\\
\widehat{R}(g_{\mu\nu},A_\mu)&=R-2(n-1)\nabla \cdot A-(n-1)(n-2) A^2,
\label{wcurvterms}
\end{align}
\endgroup   
where $A_{[\mu}B_{\nu]}= (A_\mu B_\nu-A_\nu B_\mu)/2$, $\nabla\cdot A\equiv \nabla_\mu A^\mu$ and $A^2=A_\mu A^\mu$. Observe also that due to the relaxation of metricity constraint on connection, the enhanced Riemann tensor in Weyl's geometry does not possess the regular symmetries of the Riemann tensor for now. Actually, this is not required at the current stage since the primary aim is to build the WECG first. The Riemann tensor would acquire its genuine symmetries in the subregion of vacua where the symmetry is broken. Notice also that the Weyl version of Riemann and Ricci curvature tensors are invariant under the Weyl's transformations but Ricci scalar is not and transforms as $(\widehat{R}[g,A])^{'} = e^{-2\zeta(x)}\widehat{R}[g,A]$. Moreover, the Weyl's local conformal invariant scalar and gauge field actions in generic $n$ dimensions respectively are 
\begingroup
\allowdisplaybreaks
\begin{align}
S_\Phi=\pi \int d^n x \sqrt{-g}\Big ({\cal{D}}_\mu \Phi {\cal{D}}^\mu\Phi +\nu \,\Phi^{\frac{2n}{n-2}}\Big ), \quad S_{A^\mu} =\varepsilon \int d^n x \sqrt{-g}\,\Phi^{\frac{2(n-4)}{n-2}} F_{\mu\nu}F^{\mu\nu},
\label{scalarwithpot}
\end{align}
\endgroup
where $\nu \ge 0$ is a dimensionless coupling constant ensuring the presence of a ground state, and $F_{\mu\nu}=\partial_\mu A_\nu-\partial_\nu A_\mu $ is the ordinary abelian field strength tensor. Of course, one can also set the parameters $\pi$ and $\epsilon$ to their so-called canonically normalized values, but we shall keep them as free for now for the sake of future works such as the (tree-level) unitarity analysis. Notice also that,  due to being conformally invariant in $3+1$ dimensions, the compensating scalar field factor in Maxwell action disappears here \cite{WeylSee2}. In fact, one could leave out Maxwell's kinetic term initially since such a term would also emerge on its own from the nature of the model. But, we will keep assuming it for the sake of generalization since the coefficient can always be tuned to get canonically normalized Maxwell's action. Observe that, for the sake of future studies (say, unitarity analysis, black hole solutions, quantum corrections, \emph{etc.}), all the dimensionless coefficients in (\ref{wecgaction1}) are taken to be arbitrary since extra restrictions on the relative couplings may be enforced by those prospective studies. At this point, one shall notice that it also deserves to generalize WECG in (\ref{wecgaction1}) (as well as its higher dimensional extensions) with the ensuing Weyl's extension of the dimensionally-extended Euler quadratic and cubic densities (namely, $\widehat{\chi}_4$ and $\widehat{\chi}_6$) associated with Weyl-invariant analogous of Gauss-Bonnet and Cubic Lovelock combinations \cite{Edelstein,BeltranJimenezGCG}
\begingroup
\allowdisplaybreaks
\begin{align}
\widehat{\chi}_4&=\widehat{R}^2-4 \widehat{R}_{ab}\widehat{R}^{ab}+\widehat{R}_{abcd}\widehat{R}^{abcd}\nonumber\\
\widehat{\chi}_6&=-8 \widehat{R}_a{^c}{_b}{^d} \widehat{R}_c{^e}{_d}{^f}\widehat{R}_e{^a}{_f}{^b}+4 \widehat{R}{_{ab}}{^{cd}}\widehat{R}{_{cd}}{^{ef}}\widehat{R}{_{ef}}{^{ab}}-24 \widehat{R}_{abcd}\widehat{R}{^{abc}}{_{e}}\widehat{R}^{de}\nonumber\\
&\hskip 0.6 cm +3 \widehat{R}_{abcd} \widehat{R}^{abcd}R+24 \widehat{R}_{abcd} \widehat{R}^{ac}\widehat{R}^{bd}+16 \widehat{R}_a{^{b}} \widehat{R}_b{^{c}}\widehat{R}_c{^{a}}-12\widehat{R}_{ab}\widehat{R}^{ab}\widehat{R}+\widehat{R}^3, 
\end{align}
\endgroup
because although the \emph{usual bare} $\chi_4$ and $\chi_6$ respectively are topological in $n=3+1$ and $5+1$ dimensions, whereas they are trivial in $n\le 2+1$ and $n\le 4+1$, respectively, their Weyl's enhancements come with additional particular geometric configurations of gauge fields in addition to $\chi_4$ and $\chi_6$ (that is, $\widehat{\chi}_4=\chi_4+f_1(A_\mu)$ and $\widehat{\chi}_6=\chi_6+f_2(A_\mu)$).

Observe that Weyl's local conformal invariance of the action in (\ref{wecgaction1}) in general comes up with $11$ free dimensionless parameters, two of which could be fixed according to their canonically normalized kinetic actions for the scalar and vector fields. But, we keep taking them as generic to get the most general parameter regions, and, in fact, unitarity \emph{etc}. will fix them to their viable values ultimately. In contrast to the seemingly misleading compact form of action in (\ref{wecgaction1}), even though the existing scalar and vector fields are real and abelian, respectively, it actually describes a very complicated model as is expressed by the dynamical degrees of freedom ($g_{\mu\nu}, A_\mu, \Phi $), which explicitly represent various non-minimal interactions between higher-order curvature, vector, and scalar fields as well as their higher-order self-interactions. More precisely, substituting the explicit expressions of Weyl's extended curvature terms (\ref{wcurvterms}) into the WECG action (\ref{wecgaction1}) and subsequently after some straightforward computations, one can show that (\ref{wecgaction1}) explicitly turns out to be   
\begingroup
\allowdisplaybreaks
\begin{align}
S_{\mbox{WECG}}&= \int \mbox{d}^4 x \sqrt{-g} \,\Big\{\alpha_1 \Phi^2 (R-6\nabla \cdot A-6 A^2 )+\Phi^{-2}\Big [\mathfrak{L}_{\mbox{ECG}}+\Xi_{(1)} R^2_{\mu\nu\alpha\beta}\nonumber\\
&+\Xi_{(2)\mu}{^\lambda} R^{\mu\nu\alpha\beta}R_{\alpha\beta\nu\lambda }+\Xi_{(3)\mu}{^\lambda} R^{\mu\nu\alpha\beta}R_{\nu\alpha\beta\lambda}+\Xi_{(4)} R^{\mu\nu\alpha\beta}R_{\nu\alpha\beta\mu}+\Xi_{(5)}{^{\beta\alpha}} R^{\mu\nu} R_{\mu\alpha\nu\beta}\nonumber\\
&+\Xi_{(6)}R^2_{\mu\nu}+\Xi_{(7)\mu\nu} R^{\mu\alpha}R_\alpha{^\nu}+\Xi_{(8)\mu\nu} R R^{\mu\nu}+\Xi_{(9)} R^2+\Xi_{(10)\mu\alpha\nu\beta} R^{\mu\alpha\nu\beta}+\Xi_{(11)\mu\nu} R^{\mu\nu}\nonumber\\
&+\Xi_{(12)} R+\Xi_{(13)}F^2_{\mu\nu}+\Xi_{(14)\nu}{^\alpha} F^{\mu\nu}F_{\mu\alpha}+\Xi_{(15)\mu\nu} F^{\mu\nu}+\Xi_{(16)\mu\nu} (\nabla^\mu A^\alpha)(\nabla_\alpha A^\nu)\nonumber\\
&+\Xi_{(17)}(\nabla^\mu A^\nu)(\nabla_\nu A_\mu)+\Xi_{(18)}A_\mu A_\nu (\nabla^\mu A^\nu)+\Xi_{(19)} (\nabla \cdot A)^3+\Xi_{(20)} (\nabla \cdot A)^2 A^2\nonumber\\
&+\Xi_{(21)} (\nabla \cdot A) A^4+\Xi_{(22)} A^6 \Big] \Big \}+S_{\Phi}+S_{A^\mu},
\label{wecgaction2}
\end{align}
\endgroup
where $\mathfrak{L}_{\mbox{ECG}}$ is the Lagrangian density of the strongly-coupled ECG \cite{BeltranJimenezGCG} upgraded by the dimensionless couplings $\alpha_i$ which involves specific cubic curvature combinations only and is described as below
\begingroup
\allowdisplaybreaks
\begin{align}
\mathfrak{L}_{\mbox{ECG}}&=12 \alpha_2 R_\mu{^\rho}{_\nu}{^\sigma} R_\rho{^\tau}{_\sigma}{^\eta}R_\tau{^\mu}{_\eta}{^\nu}+\alpha_3 R{_{\mu\nu}}{^{\rho\sigma}}R{_{\rho\sigma}}{^{\tau\eta}} R{_{\tau\eta}}{^{\mu\nu}}+2\alpha_4 R R_{\mu\nu\rho\sigma}R^{\mu\nu\rho\sigma}  \nonumber\\
&-8 \alpha_5 R^{\mu\nu}R_\mu{^{\rho\sigma\tau}} R_{\nu\rho\sigma\tau}+4\alpha_6 R^{\mu\nu} R^{\rho\sigma} R_{\mu\rho\nu\sigma}-4\alpha_7 R R_{\mu\nu} R^{\mu\nu}+8\alpha_8 R_\mu{^\nu} R_\nu{^\rho} R_\rho{^\mu},
\label{ordecgact} 
\end{align}
\endgroup
and the terms $\Xi_{(i)}=\Xi_{(i)}(\zeta_j,A_\mu) $ (where $i=1 \cdots 22$) which in general are composed of particular higher order gauge fields are  
\begingroup
\allowdisplaybreaks
\begin{align}
&\Xi_{(1)}=-\zeta_1 (\nabla \cdot A)+\zeta_2 A^2, \quad \Xi_{(2)\mu}{^\lambda}=\zeta_3 F_\mu{^\lambda}+\zeta_4 (\nabla^\lambda A_\mu), \quad \Xi_{(3)\mu}{^\lambda}=72 \alpha_2 (\nabla^\lambda A_\mu)-\zeta_5 A_\mu A^\lambda \nonumber\\
&\Xi_{(4)}=36\alpha_2 A^2, \quad \Xi_{(5)}{^{\beta\alpha}}=-\zeta_6 [\nabla^\beta A^\alpha- A^\beta A^\alpha], \quad\Xi_{(6)}=-[\zeta_7 (\nabla \cdot A)+\zeta_8 A^2],\nonumber\\
&\Xi^{(7)}_{\mu\nu}=\zeta_9 [\nabla_\mu A_\nu-A_\mu A_\nu], \quad \Xi^{(8)}_{\mu\nu}=-\zeta_{10} [\nabla_\mu A_\nu-A_\mu A_\nu], \quad \Xi_{(9)}=8\alpha_7 (\nabla \cdot A)-\zeta_{11} A^2\nonumber\\
&\Xi^{(10)}_{\mu\alpha\nu\beta}=18 \alpha_2 F_{\mu\alpha} F_{\nu\beta}+\zeta_{12}F_{\mu\nu}F_{\alpha\beta}+\zeta_{13} F_{\mu\nu} (\nabla_\beta A_\alpha)+\zeta_{14} (\nabla_\mu A_\nu)(\nabla_\beta A_\alpha)+\zeta_{15} A_\alpha A_\nu (\nabla_\mu A_\beta) \nonumber\\
&\Xi^{(11)}_{\mu\nu}=\zeta_{16} F_{\mu\alpha}F^\alpha{_{\nu}}+\zeta_{17} F_{\mu\alpha}(\nabla_\nu A^\alpha)-\zeta_{18} F_{\mu\alpha}(\nabla^\alpha A_\nu)+\zeta_{19} (\nabla_\mu A_\alpha)(\nabla^\alpha A_\nu) \nonumber\\
&+\zeta_{20} (\nabla_\mu A_\nu)(\nabla \cdot A)-\zeta_{20}A_\mu A_\nu (\nabla \cdot A)+\zeta_{21} F_{\mu\alpha}A_\nu A^\alpha+\zeta_{22}A_\mu A_\alpha (\nabla_\nu A^\alpha)-\zeta_{23}(\nabla_\mu A_\nu)A^2 \nonumber\\
&-\zeta_{24}A_\mu A_\nu A^2, \quad \Xi_{(12)}=\zeta_{25} F^2_{\mu\nu}+\zeta_{26} (\nabla_\mu A_\nu)^2-\zeta_{27}A_\mu A_\nu (\nabla^\mu A^\nu)+\zeta_{28} (\nabla \cdot A)^2\nonumber\\
&+\zeta_{29}(\nabla \cdot A) A^2+\zeta_{30}A^4, \quad \Xi_{(13)}=\zeta_{31} (\nabla \cdot A)+\zeta_{32}A^2, \quad \Xi_{(14)\nu}{^\alpha}=-\zeta_{33} \nabla_\nu A^\alpha+\zeta_{34} A_\nu A^\alpha,\nonumber\\
& \Xi^{(15)}_{\mu\nu}=\zeta_{35} A_\mu A_\alpha (\nabla_\nu A^\alpha), \quad \Xi^{(16)}_{\mu\nu}=\zeta_{36} \nabla_\mu A_\nu-\zeta_{37} A_\mu A_\nu, \quad \Xi_{(17)}=\zeta_{38}\nabla \cdot A+\zeta_{39}A^2,\nonumber\\
& \Xi_{(18)}=2 \zeta_{24}[\nabla \cdot A+A^2], \quad \Xi_{(19)}=-\zeta_{40}, \quad \Xi_{(20)}=-\zeta_{41}, \quad\Xi_{(21)}=-\zeta_{42}, \quad\Xi_{(22)}=-\zeta_{43}.
\label{xiterms}
\end{align}
\endgroup
Note that the dimensionless coefficients $\zeta_j$ (with $j=1 \cdots 43$) in (\ref{xiterms}) are composed of unique combinations of the relative coupling constants $\alpha_i$ and explicitly presented in (\ref{AppenA}). These various specific combinations of the couplings will turn out to be particularly important as one searches for the ghost-free parameter regimes in the WECG since it supplies several possible configurations of the couplings. Note also that the WECG comes with not only the bare cubic curvature terms in regular ECG but also all the possible curvature terms up to the second order nonminimally coupled to vector and scalar fields together with the self-interacting scalar and vector fields up to quartic and sextic powers. Hence, it is a very rich dynamical gravity model supplying a wide geometrical plateau of higher-order interacting lower and higher spin fields. 

\section{Field Equations}\label{secfour}
As expected from any new approach, the WECG also requires successive studies in various aspects at the classical and quantum levels to ultimately arrive at a firm model in all the essential facets.  With this viewpoint, in this section, we first concentrate on the imperative study of finding and analyzing the field equations for the propagating degrees of freedom described in the WECG due to the ensuing reasons: observe that, by freezing the scalar field to its vacuum expectation value $\Phi=\left \langle\Phi_{vac} \right\rangle=\sqrt{{\cal M}}$ and simply picking $A_\mu=0$, the WECG recovers the bare strongly coupled ECG \cite{BeltranJimenezGCG} in the vacuum with an effective Planck's mass $M_{Pl} \sim \left \langle\Phi_{vac} \right\rangle$ and the proper tuning of dimensionless constants. Intriguingly, this rough observation suggests analyzing first and foremost whether this conventional lower-energy limit of WECG comes about as its vacuum solution or not. Observe that, as in the $3+1$-dimensional conformally-coupled scalar field in \cite{DeserSymBreak}, one could have initially assumed a symmetry-breaking term (e.g. mass of the scalar field) and proceeded accordingly. In lieu, we will progress to search for whether there occurs any such spontaneous symmetry-breaking mechanism akin to that of the standard model of particles and conduct further investigations of WECG accordingly. For this purpose, the field equations for the existing dynamical degrees of freedom  $(g_{\mu\nu}, A_\mu, \Phi)$ must be computed in the first place. At this point, since the action of WECG in (\ref{wecgaction1}) or (\ref{wecgaction2}) consists of cubic orders of Weyl's curvature terms that contain both the usual pure Riemannian curvature tensors and vector fields and that also nonminimally couple to the scalar field with specific weight in the action, it is natural and inevitable to have very long and highly complex field equations in their {\it generic} forms as they will give {\it all possible} higher-order interactions of curvatures, vector, and scalar fields. Later, as we study them in the maximally-symmetric backgrounds, they will reduce to {\it simpler} expressions. As is pointed out, due to their rather complex nature, it will be appropriate to work on the generic field equations separately. Hence, let us proceed accordingly.

\subsection{Metric Field Equations}

As we shall see, since the field equation for the \emph{generic} background becomes highly intricate, we relegate all the necessary expressions in the field equations to the appendix and only present the compact forms of the field equations in the body of the paper. As for the derivation of the first field equation, by making use of the fundamental symmetries of the Riemannian tensors together with the ensuing convention we follow and the so-called identities of variation of curvature tensors
\begingroup
\allowdisplaybreaks
\begin{align}
[\nabla_\mu,\nabla_\nu] V_\alpha&=R_{\mu\nu\alpha\beta}V^\beta, \quad \delta \Gamma^\alpha_{\mu\nu}=\frac{1}{2} g^{\alpha\sigma} (\nabla_\mu \delta g_{\sigma\nu}+\nabla_\nu \delta g_{\sigma\mu}-\nabla_\sigma \delta g_{\mu\nu}), \nonumber \\
\delta R^\mu{_{\nu\sigma\rho}}&=\nabla_\sigma \delta \Gamma^\mu_{\nu\rho}-\nabla_\rho \delta \Gamma^\mu_{\nu\sigma}, \qquad \delta R_{\mu\nu}=\nabla_\alpha \delta \Gamma^\alpha_{\mu\nu}-\nabla_\mu \delta \Gamma^\alpha_{\alpha\nu},
\end{align}
\endgroup
during the variation of the action with respect to $g^{\mu \nu}$, one can show that, posterior to a long process of calculations, the metric field equation up to a boundary term can be written in the compact form according to the certain expressions of scalar field terms as follows 
\begingroup
\allowdisplaybreaks
\begin{align}
&\mathscr{O}^{(1)}_{\mu\nu} \square \Phi^{-2} + \mathscr{O}_{(2)} \nabla_\mu \nabla_\nu \Phi^{-2}+ \mathscr{O}^\alpha_{(3)\mu} \nabla_\nu \nabla_\alpha \Phi^{-2}+\mathscr{O}^{\alpha\beta}_{(4)\mu\nu} \nabla_\alpha \nabla_\beta \Phi^{-2} + \mathscr{O}^{(5)}_{\mu} \nabla_\nu \Phi^{-2}\nonumber\\
&+\mathscr{O}^\alpha_{(6)\mu\nu} \nabla_\alpha \Phi^{-2}+\mathscr{O}^{(7)}_{\mu\nu} \Phi^{-2}+g_{\mu\nu}[ \mathscr{O}_{(8)}-(1/2) \mathfrak{L}_{\mbox{ECG}}]+\mathscr{O}_{(9)\mu\nu}=0.
\label{meticfeterms}
\end{align}
\endgroup
Here, $\mathscr{O}_{i}=\mathscr{O}_{i}(R, A, \Phi)$ where $i=1\cdots 9$ in general. As sequentially presented below, each $\mathscr{O}_{i}$ term further contains the other kind of terms compactly denoted as ``$\Theta_{[i]j}A=\Theta^{[i]j}A$'' where the first indices ``$i$'' in the square bracket stands for specific $\mathscr{O}_{i}$-term in (\ref{meticfeterms}) it belongs to, and the second index ``$j$'' signifies the number of the sub-term for same $\mathscr{O}_{i}$. Also, the pure specific curvature combinations in $\mathscr{O}_{i}$-terms are compactly denoted by $\widehat{\Omega}_{(i)}$ and presented in (\ref{omegatrms}), explicitly. To be more precise, let us first notice that the term $\mathscr{O}^{(1)}_{\mu\nu}$ reads as 
\begingroup
\allowdisplaybreaks 
\begin{align}
\mathscr{O}^{(1)}_{\mu\nu}&=\widehat{\Omega}^{(1)}_{\mu\nu}+(\Theta^{\alpha}_{[1]1} A^\beta) R_{\mu\alpha\nu\beta}-(\Theta_{[1]2}A)R_{\mu\nu}+(\Theta^{[1]3}_{\mu}A_\nu) R+\frac{1}{2}[(\Theta^{[1]4}_\alpha A_\mu)(\nabla_\nu A^\alpha)\nonumber\\
&+(\Theta^{[1]5}_\alpha A_\mu)(\nabla^\alpha A_\nu)+ (\Theta^{[1]6}_\mu A_\nu)(\nabla \cdot A)+(\Theta^{[1]7}_\mu A_\alpha) A_\nu A^\alpha-(\Theta^{[1]8}_\mu A_\nu)A^2],
\label{o1partexplicit}
\end{align}
\endgroup
where $\widehat{\Omega}^{(1)}_{\mu\nu}$ describing a particular combination of pure higher-order curvatures is given in (\ref{firstexpression}). Here, all the related coefficient functions $\Theta_{\alpha}^{[1]i} A_\beta $ (with $i=1\cdots 8$) containing certain combinations of gauge fields with distinct couplings are provided in (\ref{AppenB}) explicitly. 

Secondly, the compact term $\mathscr{O}_{(2)}$ becomes as follows
\begin{equation}
\begin{aligned}
\mathscr{O}_{(2)}&=\widehat{\Omega}_{(2)}-2 (\Theta^{[1]3}_{\alpha}A_\beta) R^{\alpha\beta}-2(\Theta^{[2]1} A) R-(\Theta^{[2]2}_{\mu}A_\nu) \nabla^\mu A^\nu-(\Theta^{[2]3} A) \nabla \cdot A-\zeta_{30}A^4.
\label{o2partexplicit}
\end{aligned}  
\end{equation}
Here, the expression for $\widehat{\Omega}_{(2)}$, which signifies a certain configuration of quadratic curvatures, can be found in (\ref{secondexpression}). The associated functions $\Theta^{[1]3}_{\alpha}A_\beta $ and $\Theta_{\alpha}^{[2]i} A_\beta $ (where $i=1\cdots 3$) are presented in (\ref{AppenB}) and (\ref{AppenC}), respectively. 

Next, the term $ \mathscr{O}^\alpha_{(3)\mu}$  turns out to be as below  
\begingroup
\allowdisplaybreaks
\begin{align}
\mathscr{O}^{\alpha}_{(3)\mu}&=\widehat{\Omega}^\alpha_{(3)\mu}-2 (\Theta^{(\sigma}_{[1]1} A^{\beta)})R^\alpha{_{\sigma\mu\beta}}+(\Theta^{[3]1}_{\llparenthesis \mu} A_\sigma) (R^{\alpha \sigma\rrparenthesis})+2 (\Theta_{[1]2}A) R^\alpha{_\mu}-2 (\Theta^{[1]3}_{(\mu} A^{\alpha)}) R \nonumber\\
&-(\Theta^{[1]6}_{(\mu} A^{\alpha)}) \nabla \cdot A+(\Theta^{[3]2}_{(\mu} A^{\alpha)}) A^2-(\Theta^{[3]3}_{\mu} A_{\beta})\nabla^\beta A^\alpha-(\Theta^{[3]4}_{\mu} A_{\beta})\nabla^\alpha A^\beta\nonumber\\
&-\frac{1}{2}(\Theta^{[1]7}_{\mu} A_{\beta})A^\alpha A^\beta-\frac{1}{2}(\Theta_{[1]7}^{\alpha} A^{\beta})A_\mu A_\beta,
\label{o3partexplicit}
\end{align}
\endgroup
where the bare curvature term $\widehat{\Omega}_{(3)\mu}^\alpha$ can be found in (\ref{thirdexpression}), and the relevant coefficient functions $\Theta^{[1]i}_{\alpha} A_{\beta}$ (with $i=1, 2, 3, 6, 7$) and $\Theta_{\alpha}^{[3]j} A_\beta$ (for $j=1\cdots 4$) are presented in (\ref{AppenB}) and (\ref{AppenD}), respectively. Notice that the two-fold symmetric-symmetric index operation $``\llparenthesis \cdots \rrparenthesis"$ is already described in (\ref{twofoldder1}). 

As for the compact term $\mathscr{O}^{\alpha\beta}_{(4)\mu\nu}$ in (\ref{meticfeterms}), one arrives at 
\begingroup
\allowdisplaybreaks
\begin{align}
\mathscr{O}^{\alpha\beta}_{(4)\mu\nu}&=\widehat{\Omega}^{\alpha\beta}_{(4)\mu\nu}+4 (\Theta^{[4]1}A)R{_\mu}{^{\alpha\beta}}{_\nu}+(\Theta^{[4]2}_{\{\!\!\{\mu}A_\sigma) (R^{\beta\sigma\}\!\!\}\alpha}{_\nu})-2 (\Theta^{[1]1}_{(\mu}A^{\beta)}) R^\alpha{_\nu}\nonumber\\
&+2(\Theta^{[1]1}_{\llparenthesis \mu}A_{\nu}) (R^{\alpha\beta\rrparenthesis}) +(\Theta^{[4]3\alpha}_\mu A^{[\beta})(A_{\nu]})-(\Theta_{[4]3\mu}^\beta A^{[\alpha})(A_{\nu]})+(\Theta^{[4]4}_\mu A^{\beta}) \nabla_\nu A^\alpha\nonumber\\
&+(\Theta^{[4]5}_\mu A_{\nu}) \nabla^\alpha A^\beta+(\Theta_{[4]6}^\alpha A_{\mu}) \nabla^\beta A_\nu.
\label{o4partexplicit}
\end{align}
\endgroup
Here, the corresponding term $\widehat{\Omega}^{\alpha\beta}_{(4)\mu\nu}$ is given in (\ref{fourthexpression}), whereas the coefficient functional $\Theta^{[1]1}_{\alpha}A_{\beta}$ and $\Theta_{\alpha}^{[4]i} A_\beta $ (running as $i=1\cdots 6$) are independently provided in (\ref{AppenB}) and (\ref{AppenE}). Similarly, the two-fold symmetric-anti symmetric index operation $``\{\!\!\{ \cdots \}\!\!\}"$ is defined in (\ref{twofoldder1}). 

Afterwards, the compact term $\mathscr{O}^{(5)}_{\mu}$ are defined expressly as follows
\begingroup
\allowdisplaybreaks
\begin{align}
\mathscr{O}^{(5)}_{\mu}&=\widehat{\Omega}^{(5)}_{\mu}+[\frac{\zeta_6}{4} \nabla^\beta F^{\sigma\alpha}+\frac{1}{2}\kappa_{1({Q_5})} A^\beta F^{\sigma\alpha}+\Theta_{[5]1}^{\beta\sigma}A^\alpha+\Theta_{[5]2}^{\sigma\alpha}A^\beta ] R_{\sigma\alpha\mu\beta}\nonumber\\
&+(\Theta_{[5]3}^{\alpha}A^\beta) \nabla_\mu R_{\alpha\beta}+(\Theta_{[5]4}^{(\alpha}A^{\beta)})\nabla_\alpha R_{\beta\mu}+(\Theta_{[5]5} A^\alpha) R_{\alpha\mu}+(\Theta_{[5]6} A) \nabla_\mu R\nonumber\\
&+(\Theta^{[5]7}_{\mu\alpha} A_\beta) R^{\alpha\beta}+(\Theta^{[5]8}_{(\mu} A_{\alpha)}) \nabla^\alpha R+(\Theta^{[5]9} A_\mu) R+(\Theta^{[5]10}_{\mu} A_\beta) \square A^\beta\nonumber\\
&-\frac{1}{2}(\Theta^{[5]11} A) \square A_\mu+(\Theta_{[5]12}^{\alpha} A^\beta)\nabla_\mu\nabla_\alpha A_\beta+(\Theta_{[5]13}^{\alpha}A^\beta) \nabla_\alpha \nabla_\beta A_\mu\nonumber\\
&+(\Theta^{[5]14}_{\mu} A_\alpha)\nabla^\alpha \nabla \cdot A-(\Theta^{[5]15} A) \nabla_\mu \nabla \cdot A+(\Theta_{[5]16} A^\alpha-\Theta_{[5]17} A^\alpha ) \nabla_\mu A_\alpha\nonumber\\
&+(\Theta_{[5]18} A^\alpha+\Theta_{[5]19} A^\alpha) \nabla_\alpha A_\mu-\frac{1}{2}(\Theta^{[5]20}_{\alpha\beta}A_\mu) \nabla^\alpha A^\beta+(\Theta_{[5]21} A)  A_\mu \nabla \cdot A,
\label{o5partexplicit}
\end{align}
\endgroup
wherein the related term $\widehat{\Omega}^{(5)}_{\mu}$ comprising specific curvature interactions only is presented in (\ref{fifthexpression}). Here, the explicit forms of coupling $\kappa_{1({Q_5})}$ and coefficient functional $\Theta_{\alpha}^{[5]i} A_\beta $ (with $i=1\cdots 21)$ are respectively exhibited in (\ref{kapQ5}) and (\ref{AppenF}). 

On the other side, the compact term $\mathscr{O}^\alpha_{(6)\mu\nu}$ turns out to be as below
\begingroup
\allowdisplaybreaks
\begin{align}
\mathscr{O}_{(6)\mu\nu}^\alpha&=\widehat{\Omega}_{\mu\nu}^{(6)\alpha}+(\Theta_{[6]3} A^\beta) R^\alpha{_{\mu\beta\nu}}-4 (\Theta_{[6]4} A) \nabla^{[\alpha}R_{\mu]\nu}+(\Theta^\sigma_{[6]7\mu} A^\beta) R^\alpha{_{\nu\sigma\beta}}+(\Theta^\sigma_{[6]9\mu} A_\beta)R_{\sigma\nu}{^{\alpha\beta}}\nonumber\\
&+(\Theta^{[6]10}_{(\mu} A_{\beta)})(2 \nabla_{[\nu}R^{\alpha]\beta}-\nabla^\alpha R^\beta{_\nu})+\frac{1}{2}(\Theta_{[4]2}^{(\sigma} A^{\beta)}) \nabla_\sigma R^\alpha{_{\mu\beta\nu}}+\frac{1}{2}(\Theta_{[6]12}^{\alpha\beta} A^{\sigma}) R_{\sigma\mu\beta\nu}+\nonumber\\
&(\Theta^{[6]13}_{(\mu} A_{\beta)}) \nabla^\beta R^\alpha{_\nu}+(\Theta_{[6]10}^{(\alpha} A^{\beta)})\nabla_\mu R_{\beta\nu}-(\Theta_{[6]13}^{(\alpha} A^{\beta)}) \nabla_\beta R_{\mu\nu}+4(\Theta_{[1]1}^{(\beta} A^{\sigma)}) \nabla_{[\mu}R^{\alpha]}{_{\sigma\beta\nu}}\nonumber\\
&+(\Theta^{[6]15}A_{\mu})R^\alpha{_\nu}-\frac{1}{2} (\Theta^{[6]19}_{\mu\beta} A_\nu) R^{\alpha\beta}-\frac{1}{2}(\Theta^{[5]3}_{(\mu} A_{\nu)})\nabla^\alpha R+\frac{1}{2}(\Theta^{[6]25\beta}_{\mu} A^\alpha) R_{\beta\nu}+\nonumber\\
&\frac{1}{2} (\Theta^{[5]3}_{(\mu} A^{\alpha)}) \nabla_\nu R-(\Theta_{[6]15} A^\alpha) R_{\mu\nu}+ (\Theta^{[6]26}_{\mu\nu} A^\alpha) R+\frac{1}{2} (\Theta^{[6]27}_\mu A^\alpha) \nabla_\nu \nabla \cdot A+(\Theta^{[6]29}_\mu A^\alpha) \square A_\nu \nonumber\\
&+(\Theta_{[6]29}^\beta A_\mu)\nabla_\beta \nabla_\nu A^\alpha+(\Theta^{[6]28}_{(\mu} A_{\nu)}) \square A^\alpha-2 (\Theta^{[4]3}_{\Dashv \mu\nu} A^{\alpha\vDash}) \nabla \cdot A+(\Theta^{[6]30}_\mu A_\beta) \nabla^\beta \nabla^\alpha A_\nu \nonumber\\
&+\frac{1}{2} (\Theta_{[6]31}^\alpha A^\beta) \nabla_\beta \nabla_\mu A_\nu+ (\Theta^{[6]32}_\mu A^\nu) \nabla^\alpha \nabla \cdot A+ (\Theta^{[6]33}_{\mu\nu} A^\beta) \nabla_\beta A^\alpha+(\Theta_\mu^{[6]34\alpha} A^\beta) \nabla_\beta A_\nu \nonumber\\
&-(\Theta^{[6]35}_\mu A^\beta)\nabla^\alpha F_{\nu\beta}+\frac{1}{2}  (\Theta^\alpha_{[6]36} A^\beta) F_{\nu\beta}+ (\Theta^{[6]37}_\beta A_\mu) \nabla_\nu \nabla^\alpha A^\beta+ (\Theta^{[6]38} A_\mu) (\nabla_\nu A^\alpha-\nonumber\\
&2 \nabla^\alpha A_\nu)-\frac{1}{2} (\Theta^\alpha_{[6]35} A^\beta) \nabla_\mu F_{\beta\nu}+\frac{1}{2} (\Theta^{[5]11} A) \nabla_\mu F^\alpha{_\nu}+\frac{1}{2} A_{(\mu}(\Theta^{[6]39}_{\nu)} A_\beta) \nabla^\alpha A^\beta-\nonumber\\
&\frac{1}{2} (\Theta_{[6]40} A^\alpha)\nonumber \nabla_\mu A_\nu-\frac{1}{2} (\Theta^{[6]41}_\mu A_\beta) \nabla_\nu F^{\alpha\beta}-\frac{1}{2} (\Theta^\alpha_{[6]6} A^\beta) \nabla_\mu \nabla_\nu A_\beta-\frac{1}{2} (\Theta^{[6]42\beta}_\mu A^\alpha) \nabla_\nu A_\beta,\nonumber\\
\label{o6partexplicit}
\end{align}
\endgroup
wherein the corresponding pure curvature part $\widehat{\Omega}_{\mu\nu}^{(6)\alpha}$ is explicitly provided in (\ref{sixthexpression}). Here, the distinct relevant coefficients functional $\Theta_{[1]1}^{\beta} A^{\sigma}$, $\Theta^{[4]i}_{\alpha} A_{\beta}$ (for $i=2, 3$), $\Theta^{[5]j}_{\alpha} A_{\beta} $ (with $j=3, 11$) and $\Theta_{\alpha}^{[6]k} A_\beta $ (where $k=1\cdots 42$) are given in (\ref{AppenB}), (\ref{AppenE}), (\ref{AppenF}) and (\ref{AppenG}), successively. Observe also that the two-fold cyclic symmetric-(anti)symmetric indices operation $"\Dashv \cdots \vDash"$ is described in (\ref{twfldcylclc}). 

The succeeding term $\mathscr{O}^{(7)}_{\mu\nu}$ explicitly reads as 
\begingroup
\allowdisplaybreaks
\begin{align}
\mathscr{O}^{(7)}_{\mu\nu}&=\widehat{\Omega}^{(7)}_{\mu\nu}-(\Theta^{[7]1}_\mu A_\nu)R^2_{\alpha\beta\sigma\lambda}+(\Theta_{[1]1}^\alpha A^\beta) \square R_{\mu\alpha\nu\beta}+(\Theta^{[7]2}_{\mu\alpha\nu\beta} A_\lambda) \nabla_\sigma R^{\alpha\sigma\beta\lambda}+(\Theta^{[7]3\alpha}_\mu A^\sigma) \nabla^\beta R_{\alpha\beta\nu\sigma}\nonumber\\
&-(\Theta^{[7]4\alpha}_\mu A^\sigma) \nabla^\beta R_{\nu\beta\alpha\sigma}+(\Theta_{[7]5}^{\alpha\beta} A^\sigma) \nabla_\alpha R_{\mu\beta\nu\sigma}+(\Theta_{[7]6}^{\alpha\sigma} A^\beta) \nabla_\mu R_{\alpha\nu\beta\sigma}-\frac{1}{2}(\Theta_{[6]2} A) A^\sigma \nabla^\alpha R_{\mu\alpha\nu\sigma}\nonumber\\
&-(\Theta_{[6]4} A) \square R_{\mu\nu}-(\Theta^{[6]10}_{(\mu} A^{\alpha)}) \square R_{\nu\alpha}-\frac{2\kappa_{5(7)}}{\zeta_6}\Xi^{(\alpha\beta)}_{(5)}\nabla_\alpha \nabla_\mu R_{\nu\beta}-(\Theta^{[7]7\alpha\sigma\beta}_\mu A^\lambda) \nabla_\sigma R_{\nu\lambda\alpha\beta}\nonumber\\
&+ (\Theta_{[7]8}^\alpha A^\beta) \nabla_\alpha\nabla_\beta R_{\mu\nu}+\frac{\kappa_{6(7)}}{\kappa_{3(5)}} (\Theta_{[5]3}^\alpha A^\beta) \nabla_\mu \nabla_\nu R_{\alpha\beta}-\frac{1}{4} (\Theta^{[5]3}_\mu A_\nu) \square R+(\Theta_{[7]9} A) \nabla_\mu \nabla_\nu R\nonumber\\
&-\frac{2\kappa_{9(7)}}{\zeta_6} \Xi_{(\mu}^{(5)\alpha)}\nabla_\alpha \nabla_\nu R-\frac{\zeta_5}{72 \alpha_2} (\Theta^{[7]7}_{\mu\alpha\nu\beta} A_\sigma) \nabla^\alpha R^{\beta\sigma}+(\Theta_{[7]10}A^\alpha) \nabla_\alpha R_{\mu\nu}+(\Theta^{[7]11}_{\mu\alpha} A_\beta) \nabla^\alpha R^\beta{_\nu}\nonumber\\
&+(\Theta^{[7]12}_{\mu\alpha} A_\beta)\nabla_\nu R^{\alpha\beta}+(\Theta^{[7]13} A^\alpha)\nabla_\mu R_{\alpha\nu}+(\Theta_{[7]14\mu}^\alpha A_\nu)\nabla_\alpha R+(\Theta_{[7]15} A_\mu) \nabla_\nu R\nonumber\\
&+(\Theta_{[7]16}^\alpha A^\beta)R_{\mu\alpha\nu\beta}+(\Theta^{[7]17\alpha\beta}_\mu A^\sigma)R_{\nu\alpha\beta\sigma}-36\alpha_2 A_\mu A_\nu R^{\sigma\lambda\alpha\beta}R_{\alpha\lambda\beta\sigma}+(\Theta^{[7]18}_{\mu\nu\alpha} A_\beta)R^{\alpha\beta\nonumber}\nonumber\\
&+ (\Theta_{[7]19}^\beta A_\mu) R_{\beta\nu}+(\Theta_{[7]20} A) R_{\mu\nu}+(\Theta^{[7]21}_\mu A_\nu) R+(\Theta^{[7]22}_\mu A^\alpha) \nabla_\nu \nabla_\alpha \nabla \cdot A+(\Theta^{[7]23} A_\mu)\square A_\nu\nonumber\\
&+(\Theta^{[7]24}_{\mu\alpha} A_\beta) \nabla^\alpha \nabla^\beta A_\nu+(\Theta_{[7]25}^\alpha A^\beta)\nabla^\alpha \nabla^\beta \nabla_\mu A_\nu +(\Theta_{[7]26}^\alpha A^\beta)\nabla_\mu \nabla_\nu \nabla_\alpha A_\beta\nonumber\\
&+(\Theta^{[7]27}_{\mu\nu} A_\alpha) \nabla^\alpha \nabla \cdot A+(\Theta^{[7]28}_\mu A^\alpha)\square \nabla_\alpha A_\nu+(\Theta^{[7]29}_{\mu\nu} A_\alpha)\square A^\alpha+(\Theta^{[7]30}_\mu A_\nu)\square \nabla \cdot A\nonumber\\
&+(\Theta^{[7]31}_{\mu\alpha} A_\beta)\nabla_\nu \nabla^\alpha A^\beta+(\Theta^{[7]32} A_\mu) \nabla_\nu \nabla \cdot A+(\Theta^{[7]33} A) \nabla_\mu \nabla_\nu \nabla \cdot A+(\Theta^{[7]34}_\mu A_\nu) (\nabla_\alpha A_\beta)^2\nonumber\\
&+(\Theta^{[7]35} A) \nabla_\mu A_\nu+(\Theta_{[7]36} A^\beta) \nabla_\beta \nabla_\mu A_\nu+(\Theta_{[7]37} A^\beta) \nabla_\mu \nabla_\nu A_\beta+(\Theta^{[7]38}_\mu A^\alpha) \nabla_\nu A_\alpha\nonumber\\
&+(\Theta^{[7]39}_\mu A^\alpha) \nabla_\alpha A_\nu+(\Theta^{[7]40} A) A_\mu A_\nu+(\Theta^{[7]41}_\mu A_\nu)(\nabla \cdot A)^2,
\label{o7partexplicit}
\end{align}
\endgroup
where the relevant bare curvature term $\widehat{\Omega}^{(7)}_{\mu\nu}$ and function $\Xi^{\alpha\beta}_{(5)}$ can be found in (\ref{seventhexpression}) and (\ref{xiterms}), manifestly. The constants $\zeta_i$'s, $\kappa_{3(5)}$ and $\kappa_{j(7)}$ (with $j=5, 6, 9$) are presented in (\ref{AppenA}), (\ref{kapQ234}) and (\ref{kapQ7}), respectively. Also, the different coefficient functional $\Theta_{[1]1}^\alpha A^\beta$, $\Theta_{[5]3}^{\alpha} A^{\beta} $, $\Theta^{\alpha}_{[6]k} A^\beta $ (for $k=2, 4, 10$) and $\Theta^{\alpha}_{[7]l} A^\beta $ (where $l=1\cdots 41$) are given in (\ref{AppenB}), (\ref{AppenF}), (\ref{AppenG}) and (\ref{AppenH}) in succession.  

As to the expression $\mathscr{O}_{(8)}$ associated with $ g_{\mu\nu}$ portion, observe that it independently contains all the related terms of scalar fields (namely, $\nabla^\alpha \nabla^\sigma \Phi^{-2}, \square \Phi^{-2}, \nabla_\alpha \Phi^{-2}$ and $\Phi^{-2} $ which are used to write the metric field equation in the compact form as in (\ref{meticfeterms})) together at the same time. Hence, these scalar parts of $\mathscr{O}_{(8)}$ could alternatively also be considered in the corresponding preceding parts of it instead of using them in the $ g_{\mu\nu}$-part and accordingly recast the compact form of the field equation further. But, it is just a matter of preference, so we keep up using the form given in (\ref{meticfeterms}). Anyway, let us note that the compact term $\mathscr{O}_{(8)}$ explicitly becomes
\begingroup
\allowdisplaybreaks
\begin{align}
\mathscr{O}_{(8)}&=\alpha_1 (\Theta^{[8]1}A)+ (\Theta^{[8]2}_\alpha A_\sigma) \cdot \nabla^\alpha \nabla^\sigma \Phi^{-2}+(\Theta^{[8]3} A) \cdot \square \Phi^{-2}+\nabla_\alpha \Phi^{-2} \cdot [\widehat{\Omega}^\alpha_{(g)(8)}+\nonumber\\
& (\Theta_{[8]4}^{\sigma\lambda} A^\beta) R^\alpha{_{\sigma\lambda\beta}}-(\Theta^{[5]3}_\beta A_\sigma)\nabla^\alpha R^{\beta\sigma}-(\Theta^{[5]4}_{(\beta} A_{\sigma)})\nabla^\beta R^{\alpha\sigma}-(\Theta^{[5]6} A)\nabla^\alpha R-(\Theta_{[5]8}^{(\alpha} A_{\beta)}) \nabla^\beta R \nonumber\\
&-(\Theta^{[5]5} A_\beta) R^{\alpha\beta}-(\Theta_{[5]7}^{\alpha\beta} A^\sigma) R_{\beta\sigma}-(\Theta_{[5]9} A^\alpha)R+\frac{1}{2} (\Theta_{[5]11} A)\square A^\alpha-(\Theta_{[5]10}^\alpha A^\beta)\square A_\beta \nonumber\\
&-(\Theta_{[5]14}^\alpha A^\beta)\nabla_\beta \nabla \cdot A+(\Theta_{[5]15} A) \nabla^\alpha \nabla \cdot A+(\Theta^{[8]5}_\sigma A_\beta)\nabla^\sigma \nabla^\beta A^\alpha+(\Theta^{[8]6}_\beta A_\sigma)\nabla^\alpha \nabla^\beta A^\sigma \nonumber\\
&-(\Theta^{[5]19} A_\beta)\nabla^\beta A^\alpha+(\Theta^{[5]17} A_\beta)\nabla^\alpha A^\beta-(\Theta_{[5]21} A)A^\alpha \nabla \cdot A+\frac{1}{2}(\Theta^{[5]20}_{\beta\sigma} A^\alpha) \nabla^\beta A^\sigma-\nonumber\\
&(\Theta^{[5]16} A_\sigma) \nabla^\alpha A^\sigma-(\Theta^{[5]18} A_\sigma) \nabla^\sigma A^\alpha ]+\Phi^{-2} \cdot [\widehat{\Omega}^{(g)}_{(8)}-(\Theta^{[8]7}_{\alpha\beta} A_\lambda)\nabla_\sigma R^{\sigma\alpha\beta\lambda}+(\Theta_{[8]8}^{\alpha\beta\sigma} A^\lambda)R_{\alpha\beta\sigma\lambda} \nonumber\\
&-\frac{1}{2}(\Theta_{[5]3}^\alpha A^\beta)\square R_{\alpha\beta}+(\Theta_{[8]9}^\alpha A^\beta)\nabla_\alpha \nabla_\beta R-\frac{1}{2}(\Theta_{[5]6} A)\square R+(\Theta_{[8]10}^{\alpha\beta} A^\sigma)\nabla_\sigma R_{\alpha\beta}+\nonumber\\
&(\Theta_{[8]11} A^\alpha)\nabla_\alpha R+(\Theta_{[8]12}^\alpha A^\beta) R_{\alpha\beta}+(\Theta_{[8]13} A) R+(\Theta_{[8]14}^\alpha A^\beta) \square \nabla_\alpha A_\beta+\nonumber\\
&(\Theta_{[8]15}^\alpha A^\beta)\nabla_\alpha \nabla_\beta \nabla \cdot A+(\Theta_{[8]16} A^\alpha)\square A_\alpha+(\Theta_{[8]17} A^\alpha)\nabla_\alpha \nabla \cdot A-(\Theta_{[8]18} A) \square \nabla \cdot A+\nonumber\\
&(\Theta_{[8]19}^{\alpha\beta} A^\sigma) \nabla_\alpha \nabla_\beta A_\sigma-\frac{\zeta_{20}}{2} (\nabla \cdot A)^3+(\Theta_{[8]20} A) (\nabla_\alpha A_\beta)^2-(\Theta_{[8]21} A)(\nabla_\alpha A_\beta)(\nabla^\beta A^\alpha) \nonumber\\
&-(\Theta_{[8]22}^\alpha A_\sigma)(\nabla_\alpha A_\beta)(\nabla^\beta A^\sigma)-\frac{\zeta_{24}}{2} A^2 (\nabla \cdot A)^2-2 (\Theta^{[8]23}_\alpha A_\beta)A^\alpha A^\beta ].
\label{o8partexplicit}
\end{align}
\endgroup
Here, the associated curvature terms $\widehat{\Omega}^\alpha_{(g)(8)}$ and $\widehat{\Omega}^{(g)}_{(8)}$ are exhibited in (\ref{eighthexpression}), explicitly. The specific constants $\zeta_{i}$'s are given in (\ref{xiterms}), while the corresponding coefficient functions $\Theta^{[5]j}_{\alpha} A_{\beta} $ (for $j=3 \cdots 21$ with $j\neq 12, 13$) and $\Theta_{\alpha}^{[8]k} A_\beta $ (with $k=1\cdots 23$) are given in (\ref{AppenF}) and (\ref{AppenI}), respectively. 

Finally, the last compact term $\mathscr{O}_{(9)\mu\nu}$ in (\ref{meticfeterms}) is expressed as follows
\begingroup
\allowdisplaybreaks
\begin{align}
\mathscr{O}_{(9)\mu\nu}&=\alpha_1 \{\Phi^2[R_{\mu\nu}-6 \nabla_\mu A_\nu +(\frac{\pi}{\alpha_1}-6) A_\mu A_\nu]-\nabla_\mu \nabla_\nu \Phi^2 \}-2\epsilon F_{\mu\alpha}F^\alpha{_\nu}\nonumber\\
&+\pi  \partial_\mu \Phi [\partial_\nu \Phi-2\Phi A_\nu].
\end{align}
\endgroup

\subsection{Field Equations for Weyl's Vector and Scalar Fields}
Now that we have found the most general metric field equations in all details, we can proceed to study the field equations associated with Weyl's real vector and fields in this part. Since they have relatively simpler forms than those of metric forms, we gather the gauge and scalar field equations here in the same section. 

Firstly, due to the highly complex higher-order nature of the WECG as was underlined, we shall similarly collect separately the out-coming terms in the gauge field equations mainly according to $ \nabla_\mu \Phi^{-2}$ and  $\Phi^{-2}$ even though they are more simpler in contrast to the metric field equations. To be more precise, by varying the action in (\ref{wecgaction2}) with respect to $A^\mu$, one gets the gauge field equation in compact form as follows
\begingroup
\allowdisplaybreaks
\begin{align}
&(\mathscr{X}_{[A]1}+\mathscr{X}_{[A]2}) \nabla_\mu \Phi^{-2}+(\mathscr{X}^{[A]3}_{\mu\nu}+\mathscr{X}^{[A]4}_{\mu\nu}) \nabla^\nu \Phi+(\mathscr{X}^{[A]5}_\mu+\mathscr{X}^{[A]6}_\mu+\mathscr{X}^{[A]7}_\mu+\mathscr{X}^{[A]8}_\mu+\mathscr{X}^{[A]9}_\mu \nonumber\\
&+\mathscr{X}^{[A]10}_\mu+\mathscr{X}^{[A]11}_\mu ) \Phi^{-2}+2 [\Phi(6 \alpha_1 {\cal D}_\mu \Phi-\pi {\cal D}_\mu \Phi^2)+2\epsilon \nabla^\nu F_{\mu\nu}]=0.
\label{vectfeqcmpc}
\end{align}
\endgroup
Here, the explicit forms of coefficient functional $\mathscr{X}_{[A]i}$ (where $i=1 \cdots 11$) are given in (\ref{functionalsofvectorfields}).

As for the scalar field equation, it is actually straightforward to show that the variation of WECG action in (\ref{wecgaction2}) with respect to $\Phi$ turns out to be 
\begingroup
\allowdisplaybreaks
\begin{align}
&\alpha_1 \Phi (R-6\nabla \cdot A-6 A^2 )-\Phi^{-3} [\mathfrak{L}_{\mbox{ECG}}+\Xi_{(1)} R^2_{\mu\nu\alpha\beta}+\Xi_{(2)\mu}{^\lambda} R^{\mu\nu\alpha\beta}R_{\alpha\beta\nu\lambda }+\Xi_{(3)\mu}{^\lambda} R^{\mu\nu\alpha\beta}R_{\nu\alpha\beta\lambda}\nonumber\\
&+\Xi_{(4)} R^{\mu\nu\alpha\beta}R_{\nu\alpha\beta\mu}+\Xi_{(5)}{^{\beta\alpha}} R^{\mu\nu} R_{\mu\alpha\nu\beta}+\Xi_{(6)}R^2_{\mu\nu}+\Xi_{(7)\mu\nu} R^{\mu\alpha}R_\alpha{^\nu}+\Xi_{(8)\mu\nu} R R^{\mu\nu}+\Xi_{(9)} R^2\nonumber\\
&+\Xi_{(10)\mu\alpha\nu\beta} R^{\mu\alpha\nu\beta}+\Xi_{(11)\mu\nu} R^{\mu\nu}+\Xi_{(12)} R+\Xi_{(13)}F^2_{\mu\nu}+\Xi_{(14)\nu}{^\alpha} F^{\mu\nu}F_{\mu\alpha}+\Xi_{(15)\mu\nu} F^{\mu\nu}\nonumber\\
&+\Xi_{(16)\mu\nu} (\nabla^\mu A^\alpha)(\nabla_\alpha A^\nu)+\Xi_{(17)}(\nabla^\mu A^\nu)(\nabla_\nu A_\mu)+\Xi_{(18)}A_\mu A_\nu (\nabla^\mu A^\nu)+\Xi_{(19)} (\nabla \cdot A)^3\nonumber\\
&+\Xi_{(20)} (\nabla \cdot A)^2 A^2+\Xi_{(21)} (\nabla \cdot A) A^4+\Xi_{(22)} A^6]-\pi [\partial^\mu {\cal D}_\mu \Phi+ A^\mu {\cal D}_\mu \Phi-2\nu \Phi^3]=0,
\label{sclrfildeqn}
\end{align}
\endgroup
where $\mathfrak{L}_{\mbox{ECG}}$ stands for the Lagrangian density of the ordinary ECG and is given in (\ref{ordecgact}) containing particular cubic curvature terms only. Also, the all the relevant coefficient terms $\Xi_{m}$'s are explicitly provided in (\ref{xiterms}). 

\section{Spontaneous Symmetry Breaking In the Maximally Symmetric Vacua}\label{secfive}

After obtaining the field equations for the existing dynamical degrees of freedom  in generic backgrounds, let us now turn our attention to the particular occasion that the background vacuum solution is the maximally symmetric spacetime (namely, (anti-) de Sitter or flat spaces) with the curvature tensors: 
\begin{equation}
R_{\mu\nu\alpha\beta}= \frac{\Lambda}{3} (g_{\mu\alpha} g_{\nu\beta}-g_{\mu\beta} g_{\nu\alpha}), \hskip 0.7 cm R_{\mu\nu}=\Lambda g_{\mu\nu}, \hskip 0.7 cm  R=4 \Lambda.
 \label{curvconstspc}
 \end{equation}
Notice that here we are \emph{not} searching for the exact solutions to the model which requires much more detailed separate studies, but instead we would like to determine whether the symmetry is broken in the classical vacua or not. In this part, we will see that the general field equations reduce to relatively \emph{simpler} expressions around constant curvature backgrounds (or at least when the vacuum solutions are those of maximally symmetric spacetimes) although they apparently are highly intricate for \emph{generic} backgrounds due to various higher-order non minimal interactions among the existing fields ($g_{\mu\nu}, A_\mu, \Phi$). In doing so, we follow \cite{Dengiz:2011ig}. To see that clearly, let us also pick $F_{\mu\nu}=0$ with the specific legitimate ansatz $A_\mu=0$ to avoid having any certain direction that would break the Lorentz-invariance of vacua and also freeze the scalar field to its vacuum expectation value as $ \Phi= \left<\Phi_{vac} \right>=\sqrt{{\cal M}}$ in addition to (\ref{curvconstspc}). With these proper setups, the vector field equation (\ref{vectfeqcmpc}) will be fulfilled immediately, while the metric and scalar field equations in (\ref{meticfeterms}) and (\ref{sclrfildeqn}) yield the \emph{identical} vacuum field equation incorporating background cosmological constant and vacuum expectation value of the scalar field\textendash which is also approval of the calculations\textendash as below   
\begin{equation}
\Psi \Lambda^3-4\alpha_1 {\cal M}^2 \Lambda-2\pi \nu {\cal M}^3=0,
\label{vaceqn1}
\end{equation}
where 
\begin{equation}
\Psi= \frac{32}{3} (\alpha_2+\frac{\alpha_3}{6}+2\alpha_4 -2 \alpha_5+\frac{3\alpha_6}{2}-6\alpha_7+3\alpha_8),
\end{equation}
with the imposed condition on the relative couplings as follows
\begin{equation}
\alpha_2-\frac{\alpha_3}{2}-\frac{13 \alpha_5}{8}-9\alpha_6+27 \alpha_{8}=0,
\label{condtion1}
\end{equation}
which fixes one of the $11$ free parameters in the generic action. Recall that $\nu>0$ for a viable ground state. As for the solutions of (\ref{vaceqn1}), note that the process is bilateral. That is, one may proceed by presupposing that the vacuum expectation value of the scalar field ${\cal M}$ is known and then find the cosmological constant $\Lambda$ or vice versa. For the first case, by setting coefficient of the kinetic term of the scalar field to its canonical value (i.e., $\pi=-1/2$) and also using Cardano's method for the depressed cubic equation, one gets the well-defined first generic solution for the background cosmological constant as
 \begin{equation}
\Lambda_{0}=-\frac{1}{3\Psi} \Big({\cal C}_{\Lambda}+\frac{\Delta_{0(\Lambda)}}{{\cal C}_{\Lambda}} \Big),
\end{equation}
where $\Delta_{0(\Lambda)}=12\alpha_1 \Psi {\cal M}^2$ and the term ${\cal C}_{\Lambda}$ is described as follows
\begingroup
\allowdisplaybreaks
\begin{align}
{\cal C}_{\Lambda}={\cal M}\Big[\frac{3\sqrt{3}}{2} \Psi^2 \Big(3\sqrt{3}\nu - \sqrt{27\nu^2-\frac{256 \alpha^3_1}{\Psi}}\Big)\Big]^{1/3},
\end{align}
\endgroup
which enforces the following reality condition on the relative coupling constants  
\begin{equation}
\Psi \ge \frac{256\alpha^3_1}{27\nu^2},
\label{condition2}
\end{equation}
on the free relative coupling constants to supply a real solution. Note also that since $\alpha_1 >0$, the negative values for $\Psi $ are not allowed which also imposes $\nu\neq 0$. Moreover, to have $AdS\,\,\, (\Lambda<0)$ and $dS \,\,\,(\Lambda>0)$ as background vacuum solutions, one must respectively satisfy 
\begin{equation}
{\cal C}^2_{\Lambda}>-\Delta_{0(\Lambda)} \quad \mbox{and} \quad {\cal C}^2_{\Lambda} <-\Delta_{0(\Lambda)}. 
\label{condition3}
\end{equation}
Notice that by combining the conditions in (\ref{condition2}) and (\ref{condition3}), one concludes that anti-de Sitter space is allowed but de Sitter space is excluded to be a vacuum background solution. Thus, WECG recovers the regular ECG (and so general relativity in its core) with positive and real Newton's constant in the vacuum possessing a ground state ($\nu>0$). As to the remaining two solutions (which are imaginary) for the sake of completeness, one can easily calculate them via the so-called primitive cube root of unity in general form as follows
 \begin{equation}
\Lambda_r=-\frac{1}{3\Psi} \Big(\widehat{\xi}^r \cdot {\cal C}_{\Lambda}+\frac{\Delta_{0(\Lambda)}}{\widehat{\xi}^r \cdot {\cal C}_{\Lambda}} \Big),
\label{allsollambda}
\end{equation}
where $\widehat{\xi}=(-1\pm i\sqrt{3})/2$, and $r=0, 1, 2$ for each root. 

As for the second case, let us now suppose that the cosmological constant is known and one wishes to find the vacuum expectation value of scalar field: by following the similar steps in the previous case, one gets the first well-defined generic real vacuum expectation value of the scalar field for the canonical value (i.e., $\pi=-1/2$) from (\ref{vaceqn1}) as below
\begin{equation}
{\cal M}_0= \frac{1}{3\nu} \Big(\sqrt{\Delta_{0({\cal M})}}+{\cal C}_{{\cal M}}+\frac{\Delta_{0({\cal M})}}{{\cal C}_{{\cal M}}}\Big),
\end{equation}
where $\Delta_{0({\cal M})}=16\alpha^2_1 \Lambda^2$ and the term ${\cal C}_{{\cal M}}$ turns out to be
\begingroup
\allowdisplaybreaks
\begin{align}
{\cal C}_{{\cal M}}=\mbox{sign} (\Lambda) \lvert \Lambda \rvert \Big[64 \alpha^3_1-\frac{3\sqrt{3}}{2} \nu \Psi \Big(3\sqrt{3}\nu - \sqrt{27\nu^2-\frac{256 \alpha^3_1}{\Psi}}\Big)\Big]^{1/3},
\end{align}
\endgroup
with the same reality condition given in (\ref{condition2}) as expected and also the positivity of the vacuum expectation value further implies ${\cal C}_{{\cal M}} \ge 0$. Here, $\mbox{sign} (\Lambda)=\Lambda/\lvert \Lambda \rvert$. Similarly, the other two imaginary roots can be computed from the following general equation as
\begin{equation}
{\cal M}_r= -\frac{1}{6\pi\nu} \Big(\sqrt{\Delta_{0({\cal M})}}+\widehat{\xi}^r \cdot{\cal C}_{{\cal M}}+\frac{\Delta_{0({\cal M})}}{\widehat{\xi}^r \cdot {\cal C}_{{\cal M}}}\Big),
\end{equation}
where $\widehat{\xi}$ and $r$ are given in (\ref{allsollambda}). Observe that the two roots for both cases are imaginary so we will not study them in detail since they will provide non-hermitian complex operators associated with non-physical observable. (See conclusion for the relevant discussions and possible future directions.)

Lastly, let us now dwell on the situation when the vacuum solution is the flat spacetime. Observe that the vacuum field equation in (\ref{vaceqn1}) turns out to be trivial (${\cal M}=0$) for the flat spacetime ($\Lambda=0$) and thereby the local conformal symmetry is not broken about the flat vacua. As was mentioned previously, one could have initially assumed a hard symmetry-breaking dimensionful term (say, designating a hard mass term to the scalar field as in \cite{DeserSymBreak}). Rather, we wish to keep on proceeding analogously to the standard model and hence examine whether the quantum corrections at the loop-level assign a symmetry-breaking dimensionful nonzero vacuum expectation value to the scalar field or not. This suggests probing the so-called Coleman and Weinberg computations for $\emph{massless}$ $\Phi^4$-theory in $4D$ flat spacetime which gives the effective potential due to the radiative corrections at one-loop level \cite{ColeWein}. This is exactly our case. Then, after a long course of renormalization and regularization process, Coleman and Weinberg arrived at the effective potential for the $\Phi^4$ theory in $4D$ flat spacetime due to the one-loop quantum corrections explicitly as follows
\begin{equation}
V^{eff}_{one-loop}=\hat{\nu}(\hat{\mu})  \varphi^4_c+\frac{\hat{\nu}(\hat{\mu})^2 \varphi^4_c}{256 \pi^2} \Big(\log\frac{\varphi^2_c}{\hat{\mu}^2}-\frac{25}{6}\Big).
\end{equation}
Here, $\varphi_c=\left< \Phi_{vac} \right>$ is the classical or vacuum expectation value of $\Phi$ and $\hat{\mu}$ is the renormalized mass scale. Due to the present log singularity of $V^{eff}_{one-loop}$, $\hat{\nu}(\mu)$ must be defined at the point where the renormalization scale is different than zero. It is apparent that, unlike the tree-level, one-loop radiative contributions recast the effective potential (and so the couplings) such that the minimum of the potential is moved to a point where $\varphi_c\neq 0$. This shifting of the minimum designates a nonzero \emph{dimensionful} vacuum expectation value to the scalar field that spontaneously breaks the existing local conformal symmetry. As in the ordinary Coleman-Weinberg mechanism, here the perturbation approach will collapse for the higher values of $\nu(\hat{\mu})$. However, this will likely be resolved once the vector fields are also considered, which is also the case in the usual Coleman-Weinberg mechanism. Nevertheless, the symmetry is broken by the dimensionful vacuum expectation value of the scalar field emerging due to the radiative corrections at the one-loop level, which is within the scope of the current paper. Here, as one of the future directions, notice that one needs to perform the Renormalization Group flow to determine the critical points and hence explicit viable regions for the relative coupling constants at the one-loop level in WECG.

Observe that the existence of the AdS solution solely comes from the unitarity and consistency conditions. Fundamentally, we are not seeking an exact solution in this work. Definitely, finding exact solutions for the WECG would be an interesting problem, but  it is beyond our scope and needs a much more detailed separate study. Let us also note that there might be solutions that do not satisfy the unitary conditions. As mentioned, given its breadth, this could be the subject of another study. But, as a side comment and suggestion in finding exact solutions, since the field equations of WECG are quite complicated and difficult, one shall benefit from quite effective techniques. In the literature, there are various methods to obtain exact solutions. For example, since the field equations of nonlocal gravity theory are quite complex and difficult to find solutions as in our current case, using the methods in \cite{Li:2015bqa} for the exact solutions in nonlocal gravity or the so-called technique of Kerr-Schild form as in \cite{Kilicarslan1,Kilicarslan2}, where the exact pp-wave and impulsive wave solutions of nonlocal infinite derivative gravity in \cite{IDG3} are found, will significantly simplify the field equations. Such a selection may simplify finding a solution significantly, but for our current purposes, it will not change the results obtained.

%(S!!!! that is, whether the conformal symmetry here is broken by the spontaneous appearance of vacuum or not, or further-in the other way around-whether the spontaneous breaking of the local conformal symmetry genuinely generates the vacuum spacetimes fixed to certain energy scales or not)

\section{Conclusions and Discussions}\label{secsix}

By upgrading the bare curvature terms to those composed of abelian gauge fields along with appropriately tuned weights of real scalar fields, we have constructed a Weyl-gauging of strongly coupled Einsteinian-Cubic gravity in Weyl's geometry in $3+1$ dimensions, called Weyl-Einsteinian-Cubic gravity (WECG). This provides a cubic gauge theory of gravity. The model is invariant under the so-called Weyl's local conformal transformations and hence does not involve any dimensionful parameter. Due to the nature of construction, the WECG describes $g_{\mu\nu}, A_\mu$ and $\Phi$ as the fundamental dynamical fields. We have also obtained the most general field equations associated with the existing propagating fields about any {\it generic} spacetime background. Accordingly, we have demonstrated that the ordinary ECG emerges as the lower energy limit of WECG in its local {\it non}-conformal-invariant vacua about (anti-) de Sitter and flat spacetimes where the vacuum expectation value is fixed to the Planck mass scale. Additionally, we have gotten the \emph{same} cubic vacuum field equations in the maximally symmetric backgrounds (namely, (anti-) de Sitter and flat spaces) from the \emph{generic} field equations for the metric and real scalar field where that of the gauge field is automatically fulfilled as expected. Ensuingly, we have explicitly obtained the desired real cosmological constant and vacuum expectation value of the scalar field and also determined the reality conditions on the values of existing relative couplings in WECG for the (anti-) de Sitter vacuum spacetime. Here, the natural existence of (anti)-de Sitter space spontaneously breaks Weyl's local conformal symmetry which fixes all the dimensionful parameters. Moreover, we have observed that anti-de Sitter space is allowed to be a vacuum background solution of WECG, while de Sitter space is ruled out. As for the flat spaces, the local conformal invariance of the vacuum persists to be unbroken. Here, we have demonstrated that the essential dimensionful parameter arises at the renormalization scale at the one-loop level, and thus the symmetry is broken radiatively. 

Finally, as a new model suggesting interesting outcomes, the properties of WECG deserve further exploration. In addition to the desired real one, the phenomena of having two extra partially imaginary vacuum expectation values (or cosmological constants) particularly deserve further study. These additional imaginary parts can be excluded by imposing hermiticity, as they would not provide hermitian operators and thus no viable physical observables in the quantum context. However, note that analogous phenomena also emerge in certain cases in quantum mechanics, quantum field theory, and string theory, where the perturbative and non-perturbative (instantons and/or anti-instantons) contributions generally cancel out the existing imaginary ambiguities. Recall that a vacuum energy density with an imaginary part is related to physical instability, indicating the existence of another true vacuum. However, there are cases where the imaginary term is formed, but somehow due to the general structure of the theory, they cancel out. Thus, stability is achieved despite the existence of imaginary contributions. To exemplify, let us remember the following two widely studied illustrative situations:
\begin{enumerate}
\item The first example is the so-called ``Resurgence Theory'' incorporating perturbative and nonperturbative analysis coherently, where there occurs cancellation between the emerging imaginary ambiguous parts. More precisely, here the two-fold ambiguous imaginary part of the  Borel resummation of perturbation theory and the corresponding part of the instanton-anti-instanton amplitude completely cancel out each other \cite{Zinn1, Zinn2, Zinn3, Dorigoni, DunneUnsal1,DunneUnsal2,DunneUnsal3,DunneUnsal4}. (Also, see, e.g., \cite{Aniceto} for an intriguing study on the resurgence in string theory.) To capture the idea, let us observe two simple explanatory samples in quantum mechanics: firstly, recall that, if one takes an inverted double-well, there is a real tachyonic instability. Here, one can find a bounce solution with a negative mode of the fluctuation operator related to the instability in vacuum energy. This is related to the fact that potential is unbounded from below. Now, consider a standard double-well potential, for which the potential is bounded from below and has no instability. In this case, instanton gives the level splitting if one takes a normal double-well and performs the ground state energy calculation. Accordingly, if one finds the instanton-anti-instanton contribution, it has a real and an imaginary ambiguous part. But, it is known that this system is stable. This imaginary part cancels out the ambiguity of the Borel resummation of perturbation theory. So, it does not cause any harm in the big picture and does not cause any instability \cite{Zinn1, Zinn2, Zinn3, Dorigoni, DunneUnsal1,DunneUnsal2,DunneUnsal3,DunneUnsal4}. 
\item As a second example, one can take the so-called \emph{tachyon condensation} in string field theory \cite{ASen1, ASen2}: remember that analogous phenomena (namely, unstable tachyonic modes with imaginary masses and so vacuum expectation values) also emerge in the $D\bar{D}$-brane pairs in string theory \cite{DDbranetachy1, DDbranetachy2, DDbranetachy3, DDbranetachy4, DDbranetachy5}. It has been shown in \cite{ASen1, ASen2} that the energy transmitted by tachyonic modes through the tachyon condensation compensates for the tensions in the $D$-brane pairs and hence supplies the zero point energy and thus achieving spacetime supersymmetry. Due to the instability based on being imaginary, the course of tachyon condensation triggered by viable quantum distortions provides the rolling down of the field to a stable ground state devoid of any observable tachyonic mode. (See \cite{ApprvofSen1, ApprvofSen2, ApprvofSen3} as the confirmative studies of tachyon condensation, and also \cite{Lee:2001cs} for an informative study and \cite{Erler:2019xof} for a recent interesting study.) 
\end{enumerate} 

Having those analogous phenomena suggests that nonperturbative (e.g., instantons and/or anti-instantons) effects in the WECG particularly deserve to be studied in detail in a viable approach, for example, in the resurgence or tachyon condensation framework to shed light on the emerging two extra partially imaginary vacuum expectation values (or cosmological constants).

Furthermore, note that the section \ref{secfive} is {\it not} devoted to find the exact solutions of the model, but rather the existing symmetry-breaking mechanism and unitarity analysis have been analyzed to some level here to solely determine how the local Weyl's conformal symmetry is broken and accordingly which constant curvature vacua are allowed or not, as in, e.g., \cite{Dengiz:2011ig}. But, of course, as is well-known from literature, determining explicit unitary (ghost and tachyon-free) parameter regions and also the emerging symmetry-breaking mechanism require separate elaborated studies as in, e.g., \cite{WeylSee2, Tanhayi:2011aa}. Also, finding exact solutions and spacetime singularities of the model seems to also be essential. Since the field equations of WECG are quite complex and cumbersome, one can utilize the effective techniques used in nonlocal gravity theories as in \cite{Li:2015bqa} or the Kerr-Schild form as in, e. g., \cite{Kilicarslan1,Kilicarslan2}. As was mentioned in the beginning, as another possible future direction, it will also be very interesting to study the plausible links between the current cosmological WECG and foremost alternative higher-order (non-local) conformal (quantum) gravity (in different dimensions) rooted in the Weyl tensors and traceless Ricci tensor as in \cite{Maldacenacubic1, Maldacenacubic2, Rachwal1, Rachwal2}. Finally, as is conducted in \cite{Rachwal1, Rachwal2}, it will also be interesting to analyze the model at the loop-level and figure out the modifications to the propagator structures. Since all the characteristics of a model cannot be throughly studied just in a single paper, we suggest them as the potentially interesting future projects about the model.

\section*{Acknowledgments}
This article is dedicated to Roman JACKIW's second death anniversary who was one of the world-leading scientists of the century and my former postdoc advisor at MIT/CTP.  We thank Daniel Grumiller, Mithat Unsal, Daniel Kabat, Anupam Mazumdar, Ercan Kilicarslan, Heeseung Zoe and Paul Brown for useful discussions and suggestions. We would also like to thank an anonymous referee for her/his constructive remarks, critics and suggestions on the paper.

%%%%%%%%%%%%%%%%%%%%%%%%%%%%%%%%%%%%%%%%%%%%%%%%%%%%%%%%%%%%%%%%%%%%%%%%%%%%%%%%
%%%%%%%%%%%%%%%%%%%%%%%%%%%%%%%%%%%%%%%%%%%%%%%%%%%%%%%%%%%%%%%%%%%%%%%%%%%%%%%%
%% Appendix

\appendix \label{secseven}

\section{Coefficients $\zeta_i$ in $\Xi_i$ Terms}\label{AppenA}

Note that it is indeed natural to end up with {\it highly-complicated} field equations for {\it generic backgrounds} here since it discloses all the possible lineage ties of the higher-order curvature, gauge and scalar fields. That is to say, since the WECG action integral in (\ref{wecgaction1}), which consists of the non-minimal couplings of cubic order Weyl curvature terms involving ordinary Riemannian tensors and gauge fields with the scalar fields, reveals all the possible fundamental higher-order interactions between curvature tensors, gauge fields, and scalar fields up to cubic, quartic and sextic order, respectively, having very complex field equations in generic case is an expected outcome. (Of course, they will reduce to relatively simpler expressions as one considers them in the particular constant curvature backgrounds thereafter.) Hence, we have collected all the relevant crucial complicated parts and coefficients of the field equations here in the appendices to avoid making the bulk of the paper messy. However, all are coherently and appropriately referred to and directed at the related points throughout the paper. In this perspective, let us notice here that the coefficients $\zeta_i$ (where $j=1 \cdots 43$) terms in (\ref{xiterms}) which are composed of certain combinations of relative couplings of $\alpha_j$ of the WECG action explicitly read 
\begingroup
\allowdisplaybreaks
\begin{align}
& \zeta_1=4 (3\alpha_4-2\alpha_5); \quad \zeta_2=2 [3 (\alpha_3+2\alpha_4)-8\alpha_5]; \quad  \zeta_3=4 [3(\alpha_2 +\alpha_3)-4\alpha_5]; \quad \zeta_4=4(3\alpha_3-\nonumber\\
&4\alpha_5); \quad  \zeta_5=8 [3(3\alpha_2-\alpha_3)+4\alpha_5];\quad  \zeta_6=8 [9\alpha_2-2(2\alpha_5-\alpha_6)]; \quad \zeta_7=8 [\alpha_6-3\alpha_7+3 \alpha_8];\nonumber\\
&\zeta_8=4 [9\alpha_2-8 \alpha_5+3(\alpha_6-2\alpha_7+4 \alpha_8)]; \quad  \zeta_9=8[2 (2 \alpha_5-3\alpha_8)+\alpha_6]; \quad \zeta_{10}=8[2(\alpha_4-\alpha_7)+\nonumber\\
&\alpha_6]; \quad \zeta_{11}=4[2(\alpha_4-2\alpha_7)+\alpha_6]; \quad \zeta_{12}=4\{3[2(\alpha_2+\alpha_3)+3\alpha_6]-8\alpha_5\}; \quad \zeta_{13}=8[3(\alpha_2+\alpha_3)\nonumber\\
&-4(\alpha_5-\alpha_6)]; \quad  \zeta_{14}=8 [3(\alpha_2+\alpha_3)-2(4\alpha_5-\alpha_6)]; \quad  \zeta_{15}=16 [3\alpha_3-2(4\alpha_5-\alpha_6)]; \quad \zeta_{16}=\nonumber\\
&8 [3(7\alpha_2+2\alpha_6)+4(\alpha_5+2\alpha_8)];\quad  \zeta_{17}=8[3(3\alpha_2+\alpha_6)+2(\alpha_5+\alpha_8)]; \quad \zeta_{18}=8 [3(4\alpha_2+\alpha_6)+\nonumber\\
&2(\alpha_5+\alpha_8)]; \quad \zeta_{19}=-8[3(7\alpha_2-\alpha_3)+4(2\alpha_5+\alpha_6-3\alpha_8)]; \quad \zeta_{20}=24 \{3\alpha_2+2[\alpha_6+2(\alpha_4-\nonumber\\
&\alpha_5-\alpha_7+\alpha_8)\}; \quad \zeta_{21}=-8 [12(\alpha_2+ \alpha_5-\alpha_8)-3\alpha_3+4 \alpha_6]; \quad \zeta_{22}=16[3(6\alpha_2-\alpha_3)+4(2\alpha_5\nonumber\\
&+\alpha_6-3\alpha_8)]; \quad  \zeta_{23}=8\{3(3\alpha_2-2\alpha_3)-4[3(\alpha_4-\alpha_7+2\alpha_8)-7\alpha_5+\alpha_6]\}; \quad \zeta_{24}=8\{3[\alpha_3+\nonumber\\
&4(\alpha_2+\alpha_4-\alpha_7+\alpha_8)]-4 (5\alpha_5-2\alpha_6)\}; \quad \zeta_{25}=4 [2(\alpha_4-\alpha_5)+\alpha_6-3\alpha_7]; \quad \zeta_{26}=16 [3\alpha_2+\nonumber\\
&\alpha_4-\alpha_5+\alpha_6-\alpha_7]; \quad \zeta_{27}=8 [9\alpha_2+4(\alpha_4-\alpha_5+\alpha_6-\alpha_7)]; \quad \zeta_{28}=4[2(\alpha_4-10\alpha_7)+3(\alpha_6+\nonumber\\
&2\alpha_8)]; \quad \zeta_{29}=8 \{9\alpha_2+2[5\alpha_4-2(2(\alpha_5-\alpha_6)+7\alpha_7-3\alpha_8)]\}; \quad \zeta_{30}=12 \{\alpha_3+2[3(\alpha_2+\alpha_4-\nonumber\\
&2\alpha_7)-2(2(\alpha_5-\alpha_8)-\alpha_6)]\}; \quad \zeta_{31}=-4[7(3\alpha_2-4\alpha_5-2\alpha_8)+6(4\alpha_4+2\alpha_6-5\alpha_7)]; \quad \zeta_{32}=\nonumber\\
&-4 \{3(\alpha_2+3\alpha_6-10\alpha_7)+4[2(3\alpha_4-4\alpha_5)-7\alpha_8]\}; \quad \zeta_{33}=8 (3\alpha_2+\alpha_6-14\alpha_8); \quad \,\, \zeta_{34}=\nonumber\\
&-8 [3(5\alpha_2+\alpha_6)+14\alpha_8], \quad \zeta_{35}=-48\alpha_2; \quad \zeta_{36}=32[2(3\alpha_2-\alpha_8)+\alpha_6]; \quad \zeta_{37}=96 [2(2\alpha_2-\nonumber\\
&\alpha_8)+\alpha_6]; \quad \zeta_{38}=-8 \{3(7\alpha_2+\alpha_3)+4[3(\alpha_4-\alpha_7+\alpha_8)-5\alpha_5+2\alpha_6]\}; \quad \zeta_{39}=8\{3(3\alpha_2-\alpha_3)\nonumber\\
&-2[\alpha_6+6(\alpha_4-2(\alpha_5-\alpha_8)-\alpha_7)]\}; \quad \zeta_{40}=8 [3(\alpha_2+2\alpha_4-8\alpha_7)-4\alpha_5+5(\alpha_6+2\alpha_8)];\nonumber\\
&\zeta_{41}=8 [21(\alpha_2-4\alpha_7+2\alpha_8)+3\alpha_3+8(3\alpha_4-4\alpha_5)+23\alpha_6 ]; \quad \zeta_{42}=16\{7[3(\alpha_2+\alpha_4)+2\alpha_6]+\nonumber\\
&3[\alpha_3-2(8\alpha_7-5\alpha_8)]-26\alpha_5 \}; \quad \zeta_{43}=24\{\alpha_3+2[3(\alpha_4-2\alpha_7)+2 (\alpha_6-2(\alpha_5-\alpha_8))]\}.
\end{align}
\endgroup

\section{The Pure Curvature Terms $\widehat{\Omega}_{(i)}$'s }\label{omegatrms}
 In this part, we explicitly present the expressions for the pure curvature terms associated with the metric field equations which are compactly used in the bulk of the paper. For this purpose, let us first notice that the mere curvature part $\widehat{\Omega}^{(1)}_{\mu\nu}$ of the coefficient functional $\mathscr{O}^{(1)}_{\mu\nu}$ in (\ref{o1partexplicit}) reads
\begingroup
\allowdisplaybreaks
\begin{align}
\widehat{\Omega}^{(1)}_{\mu\nu}=-4\alpha_5 R_{\mu\alpha\beta\rho}R_\nu{^{\alpha\beta\rho}}+4\alpha_6 R^{\alpha\beta}R_{\mu\alpha\nu\beta}-4\alpha_7 R R_{\mu\nu}+12 \alpha_8 R_{\mu\alpha}R^\alpha{_\nu}.
\label{firstexpression}
\end{align}
Secondly, the specific curvature term $\widehat{\Omega}_{(2)}$ associated with the coefficient $\mathscr{O}_{(2)}$ in (\ref{o2partexplicit}) manifestly turns out to be as follows
\begin{align}
\widehat{\Omega}_{(2)}=-2\alpha_4 R^2_{\alpha\beta\rho\sigma}+4\alpha_7 R^2_{\alpha\beta}.
\label{secondexpression}
\end{align}
As for the relevant curvature part $\widehat{\Omega}_{(3)\mu}^\alpha$ that of the functionals $\mathscr{O}^{\alpha}_{(3)\mu}$ in (\ref{o3partexplicit}), one gets
\begin{align}
\widehat{\Omega}_{(3)\mu}^\alpha =8 (\alpha_5 R^{\alpha\beta\rho\sigma}R_{\mu\beta\rho\sigma}-\alpha_6 R^{\beta\rho} R^\alpha{_{\beta\mu\rho}}+\alpha_7 R R^\alpha{_\mu}-3 \alpha_8 R^{\alpha\beta} R_{\beta\mu}).
\label{thirdexpression}
\end{align}
Next, the particular bare higher-order curvature term $\widehat{\Omega}^{\alpha\beta}_{(4)\mu\nu}$ associated with the $\mathscr{O}^{\alpha\beta}_{(4)\mu\nu}$ in (\ref{o4partexplicit}) becomes as below
\begingroup
\allowdisplaybreaks
\begin{align}
\widehat{\Omega}^{\alpha\beta}_{(4)\mu\nu}&=36\alpha_2 (R_{\mu\rho\nu\sigma} R^{\rho\alpha\sigma\beta}+R_{\mu\rho\sigma}{^\alpha}R^{\rho\beta\sigma}{_\nu})+6\alpha_3 R_\mu{^{\alpha\rho\sigma}}R_{\rho\sigma\nu}{^\beta}+8\alpha_4 R R_\mu{^\alpha}{_\nu}{^\beta}\nonumber\\
&+16\alpha_5(R_{\mu\sigma}R^{\sigma\alpha\beta}{_\nu}-R^{\alpha\sigma}R_{\mu\sigma\nu}{^\beta})+4\alpha_6 (R_{\mu\nu}R^{\alpha\beta}-R_\mu{^\alpha}R_\nu{^\beta}).
\label{fourthexpression}
\end{align}
\endgroup
Subsequently, the relevant curvature part $\widehat{\Omega}^{(5)}_{\mu}$ belonging to the compact expression $\mathscr{O}^{(5)}_{\mu}$ in (\ref{o5partexplicit}) is
\begingroup
\allowdisplaybreaks
\begin{align}
\widehat{\Omega}^{(5)}_{\mu}&=8\{ -[\alpha_4 \nabla_\mu R_{\alpha\beta\rho\sigma}-\alpha_5 \nabla_ \alpha R_{\mu\beta\rho\sigma}] R^{\alpha\beta\rho\sigma}+(2\alpha_5-\alpha_6) R_{\mu\alpha\beta\rho}\nabla^\beta R^{\alpha\rho}\nonumber\\
&-[(\alpha_6-2\alpha_7) \nabla_\mu R_{\alpha\beta}-(\alpha_6-3\alpha_8)\nabla_\alpha R_{\mu\beta}] R^{\alpha\beta}+(\alpha_7-\frac{3}{2}\alpha_8)R_{\mu\alpha} \nabla^\alpha R+\frac{\alpha_7}{2} R \nabla_\mu R\}.
\label{fifthexpression}
\end{align}
\endgroup
Later on, the associated pure higher-derivative curvature term $\widehat{\Omega}^{(6)\sigma}_{\mu\nu}$ of the coefficient functional $\mathscr{O}_{(6)\mu\nu}^\alpha$ in (\ref{o6partexplicit}) 
\begingroup
\allowdisplaybreaks
\begin{align} 
\widehat{\Omega}^{(6)\sigma}_{\mu\nu}&=[(36\alpha_2+8\alpha_6)\nabla_\nu R^{\alpha\beta}-36\alpha_2 \nabla^\beta R^\alpha{_\nu}-16\alpha_5 \nabla^\alpha R^\beta{_\nu}] R_{\mu\alpha\beta}{^\sigma}+[(72\alpha_2+8\alpha_6)\nabla^\sigma R^{\alpha\beta}\nonumber\\
&-(72\alpha_2+16\alpha_5) \nabla^\beta R^{\alpha\sigma}] R_{\mu\alpha\nu\beta}+[72\alpha_2 \nabla^\rho R_\mu{^{\beta\alpha\sigma}}+12\alpha_3 \nabla^\alpha R_\mu{^{\sigma\rho\beta}}+8\alpha_5 (\nabla_\mu R^{\sigma\alpha\rho\beta}-\nonumber\\
&2\nabla^\sigma R_\mu{^{\alpha\rho\beta}})]R_{\nu\alpha\rho\beta}+8 [\alpha_5 \nabla_\mu R_{\nu\alpha\beta\rho}-9\alpha_2\nabla_\rho R_{\mu\beta\nu\alpha}] R^{\sigma\alpha\beta\rho}+8(-3\alpha_3+2\alpha_5) (\nabla_\alpha R_{\mu\beta})R_\nu{^{\sigma\alpha\beta}}\nonumber\\
&+8(2\alpha_4-\alpha_5)(\nabla_\rho R) R_\mu{^\rho}{_\nu}{^\sigma}+8 (2\alpha_4-\alpha_7) [\nabla^\sigma R_{\mu\nu}-\nabla_\mu R_\nu{^\sigma}]R+8[(2\alpha_5-3\alpha_8)(\nabla_\mu R^{\sigma\alpha}\nonumber\\
&-2 \nabla^\sigma R^\alpha{_\mu})+(2\alpha_5-\alpha_6)\nabla^\alpha R^\sigma{_\mu}] R_{\nu\alpha}+8 [-2\alpha_5 \nabla_\alpha R_{\mu\beta\nu}{^\sigma}+\alpha_6 (\nabla^\sigma R_{\mu\alpha\nu\beta}+\nabla_\mu R_{\nu\beta\alpha}{^\sigma})]R^{\alpha\beta}\nonumber\\
&+8 [(2\alpha_5-3\alpha_8) \nabla_\mu R_{\nu\alpha}-(2\alpha_5-\alpha_6) \nabla_\alpha R_{\mu\nu}] R^{\sigma\alpha}+4 (\alpha_6-2\alpha_7)[R_{\mu\nu} \nabla^\sigma R-R_\mu{^\sigma}\nabla_\nu R].
\label{sixthexpression}
\end{align}
\endgroup
Afterward, the ensuing curvature part $\widehat{\Omega}^{(7)}_{\mu\nu}$ included in the compact function $\mathscr{O}^{(7)}_{\mu\nu}$ in (\ref{o7partexplicit}) explicitly reads as follows
\begingroup
\allowdisplaybreaks
\begin{align}
\widehat{\Omega}^{(7)}_{\mu\nu}&=\{8 \alpha_5 \square R_\mu{^{\alpha\beta\rho}}+8(-9\alpha_2+2\alpha_5-\alpha_6) \nabla^\rho \nabla_\mu R^{\alpha\beta}+8(3\alpha_3-2\alpha_5)\nabla^\alpha\nabla^\beta R^\rho{_\nu}+\nonumber\\
&4 [9 \alpha_2 R^{\rho\beta\alpha\sigma}+4\alpha_5 R^{\rho\alpha\beta\sigma}] R_{\mu\sigma}+6 (6\alpha_2+\alpha_3) R^{\alpha\sigma} R_{\mu\sigma}{^{\rho\beta}}-72 \alpha_2 R_\mu{^\beta}{_{\sigma\lambda}}R^{\sigma\rho\alpha\lambda}+4[(9\alpha_2\nonumber\\
&-4\alpha_5+2\alpha_6) R_\mu{^{\rho\sigma\alpha}}+4\alpha_5 R_\mu{^{\alpha\sigma\rho}}-2\alpha_6 R_\mu{^{\sigma\alpha\rho}}]R^\beta{_\sigma}+4 [9\alpha_2  R_\sigma{^{\alpha\lambda\beta}}+(9\alpha_2-3 \alpha_3-\nonumber\\
&4\alpha_5) R_\sigma{^{\beta\lambda\alpha}}] R_{\mu\lambda}{^{\rho\sigma}}-[2(3\alpha_3+4 \alpha_5) R_\mu{^{\lambda\sigma\alpha}}-3 \alpha_3 R_\mu{^{\alpha\lambda\sigma}}]R_{\lambda\sigma}{^{\rho\beta}}+ 4\alpha_4 R R_\mu{^{\alpha\rho\beta}}+8(2\alpha_5\nonumber\\
&-3\alpha_8)R_\mu{^\beta}R^{\alpha\rho}\}R_{\alpha\nu\beta\rho}+8 \{(2\alpha_5-\alpha_6) \nabla_\mu R_{\nu\rho\alpha\beta}-(9\alpha_2+2\alpha_5) \nabla_\rho R_{\mu\alpha\nu\beta}+(9\alpha_2+\nonumber\\
&\alpha_6) \nabla_\alpha R_{\mu\rho\nu\beta}\} \nabla^\alpha R^{\beta\rho}-4\{(9\alpha_2+4\alpha_6)\nabla_\mu R^{\alpha\beta}-(18\alpha_2+4\alpha_5+2\alpha_6-6\alpha_8)\nabla^\alpha R^\beta{_\mu}\}\nonumber\\
& \cdot \nabla_\nu R_{\alpha\beta}+\{4(9\alpha_2+\alpha_6)\square R^{\alpha\beta}+2(-9\alpha_2+4\alpha_4-2\alpha_5)\nabla^\alpha \nabla^\beta R+4(9\alpha_2+4\alpha_5+\nonumber\\
&2\alpha_6) R_{\rho\sigma}R^{\alpha\rho\beta\sigma}-4(9\alpha_2+4\alpha_5-2\alpha_6)R^{\alpha\sigma}R_\sigma{^\beta}+8 (\alpha_4-\alpha_7) R R^{\alpha\beta}\}R_{\mu\alpha\nu\beta}+\nonumber\\
&4  \{(-9\alpha_2-3\alpha_3+4\alpha_5-\alpha_6)\nabla_\beta R_{\mu\alpha}+(3\alpha_3-8 \alpha_5+6\alpha_8) \nabla_\alpha R_{\mu\beta} \} \nabla^\alpha R^\beta{_\nu}+\nonumber\\
&2\{18\alpha_2 \nabla^\sigma R_\mu{^{\beta\rho\alpha}}-3\alpha_3 \nabla^\alpha R_\mu{^{\rho\sigma\beta}}+4 \alpha_5 \nabla^\rho R_\mu{^{\alpha\sigma\beta}} \} \nabla_\rho R_{\alpha\nu\sigma\beta}+4 \{9 \alpha_2 \nabla_\alpha \nabla_\rho R_{\mu\beta\nu\sigma}-\nonumber\\
&\alpha_4 \nabla_\mu  \nabla_\nu R_{\alpha\beta\rho \sigma}+2 \alpha_5 \nabla_\alpha \nabla_\mu  R_{\nu\beta\rho\sigma}\} R^{\alpha\beta\rho\sigma}-4\{\alpha_4 \nabla_\mu R^{\alpha\beta\rho\sigma}-2 \alpha_5 \nabla^\alpha R_\mu{^{\beta\rho\sigma}} \}\nabla_\nu R_{\alpha\beta\rho\sigma}\nonumber\\
&+2 \{(\alpha_6-2\alpha_7)\square R+\alpha_4 R^2_{\alpha\beta\rho\sigma}-2\alpha_7 R^2_{\alpha\beta} \} R_{\mu\nu}+4 \{(-4\alpha_4+2\alpha_5+2\alpha_7-3 \alpha_8) \nabla_\mu R_{\nu\alpha}\nonumber\\
&+(4\alpha_4-2\alpha_5+\alpha_6-2 \alpha_7) \nabla_\alpha R_{\mu\nu} \} \nabla^\alpha R+4 \{(-4\alpha_5+6\alpha_8)\square R_{\mu\alpha}+(2\alpha_5-\alpha_6+\nonumber\\
&2\alpha_7-3\alpha_8)\nabla_\mu \nabla_\alpha R+(4\alpha_5-\alpha_6)R_{\mu\sigma}R^\sigma{_\alpha}-2\alpha_4 R R_{\mu\alpha} \}R^\alpha{_\nu}+4  \{(2\alpha_4-\alpha_7)\square R_{\mu\nu}\nonumber\\
&-(\alpha_4-\alpha_7)\nabla_\mu \nabla_\nu R \}R+4 \{\alpha_6 \square R_{\mu\alpha\nu\beta}-2 (\alpha_6-\alpha_7)\nabla_\mu \nabla_\nu R_{\alpha\beta}+(4 \alpha_5+2 \alpha_6-\nonumber\\
&6\alpha_8) \nabla_\alpha \nabla_\mu R_{\beta\nu}-(4 \alpha_5-\alpha_6) \nabla_\alpha \nabla_\beta R_{\mu\nu} \} R^{\alpha\beta}-(\alpha_6-4\alpha_7) (\nabla_\mu R)(\nabla_\nu R).
\label{seventhexpression}
\end{align}
\endgroup
Lastly, the relevant pure curvature parts $\widehat{\Omega}^\sigma_{(g)(8)}$ and $\widehat{\Omega}^{(g)}_{(8)}$ associated with the term $\mathscr{O}_{(8)}$ in (\ref{o8partexplicit}) respectively turn out to be as follows
\begingroup
\allowdisplaybreaks
\begin{align} 
\widehat{\Omega}^\sigma_{(g)(8)}&=8 [\alpha_4 \nabla^\sigma R_{\alpha\beta\rho\lambda}-\alpha_5 \nabla_\alpha R^\sigma{_{\beta\rho\lambda}}] R^{\alpha\beta\rho\lambda}-8(2\alpha_5-\alpha_6) (\nabla^\alpha R^{\beta\rho})R^\sigma{_{\rho\alpha\beta}}+\nonumber\\
&8[(\alpha_6-2\alpha_7)\nabla^\sigma R_{\alpha\beta}-(\alpha_6-3\alpha_8)\nabla_\alpha R_\beta{^\sigma}] R^{\alpha\beta}-4(2\alpha_7-3\alpha_8)(\nabla_\alpha R) R^{\alpha\sigma}\nonumber\\
&-4\alpha_7 (\nabla^\sigma R)R, \nonumber\\
\widehat{\Omega}^{(g)}_{(8)}&=-2\alpha_7 R \square R+4\alpha_4 (\nabla_\alpha R_{\beta\rho\sigma\lambda})^2-8(\alpha_5-\alpha_6+\alpha_7) (\nabla_\alpha R_{\beta\rho})^2-(4 \alpha_7-3\alpha_8) (\nabla_\alpha R)^2\nonumber\\
&-4 \alpha_5 (\nabla_\alpha R_{\beta\rho\sigma\lambda})(\nabla^\beta R^{\alpha\rho\sigma\lambda})+4 (2\alpha_5-2\alpha_6+3\alpha_8) (\nabla_\alpha R_{\beta\rho})(\nabla^\rho R^{\alpha\beta})+4 \{\alpha_4 \square R_{\alpha\beta\rho\sigma}\nonumber\\
&-(4\alpha_5-\alpha_6)\nabla_\alpha \nabla_\rho R_{\beta\sigma}-4\alpha_5 [R_{\alpha\lambda}R^\lambda{_{\beta\rho\sigma}}+R^{\lambda\theta}{_{\rho\sigma}} R_{\alpha\lambda\theta\beta}+2 R^\lambda{_{\beta\theta\sigma}}R_{\alpha\lambda}{^\theta}{_\rho}]\nonumber\\
&+4(\alpha_6-3\alpha_8) R_{\alpha\rho}R_{\beta\sigma}  \} R^{\alpha\beta\rho\sigma}+2  \{2(\alpha_6-2\alpha_7)\square R_{\alpha\beta}-(\alpha_6+2\alpha_7-6\alpha_8)\nabla_\alpha \nabla_\beta R\nonumber\\
&-2(\alpha_6-3\alpha_8) R_{\alpha\rho}R^\rho{_\beta} \} R^{\alpha\beta}. 
\label{eighthexpression}
\end{align}
\endgroup

\section{Coefficient Functionals $\Theta_{k}^{[i]j} A $ of $\mathscr{O}_{(i)}$-Parts:} 
\subsection{Coefficient Functions of $\mathscr{O}_{(1)}$-Part:}\label{AppenB}
As for the coefficient functionals associated with the $\mathscr{O}_{(i)}$-parts in the compact representation of the metric field equation in (\ref{meticfeterms}), let us first observe that the functionals $\Theta_{k}^{[1]j} A_l $ (where $j=1\cdots 8$) those of the term $\mathscr{O}^{(1)}_{\mu\nu}$ in (\ref{o1partexplicit}) respectively turn out to be 
\begingroup
\allowdisplaybreaks
\begin{align}
\Theta^{\alpha}_{[1]1} A^\beta &= \Xi_{(5)}^{\alpha\beta}/2, \nonumber \\
 \Theta_{[1]2}A&=-\Xi_{6}, \nonumber \\
 \Theta^{[1]3}_{\mu}A_\nu&=\Xi^{(8)}_{\mu\nu}/2, \nonumber\\
 \Theta^{[1]4}_\alpha A_\mu&=\kappa_{1(1)} \nabla_\alpha A_\mu+\kappa_{2(1)}\nabla_\mu A_\alpha,\nonumber\\
\Theta^{[1]5}_\alpha A_\mu&= \kappa_{3(1)} \nabla_\alpha A_\mu, \nonumber\\
\Theta^{[1]6}_\mu A_\nu &=(-\zeta_{20}/\zeta_6)\, \Xi^{(5)}_{\mu\nu}, \nonumber\\
\Theta^{[1]7}_\mu A_\alpha&= \kappa_{4({1})} \nabla_\mu A_\alpha-\zeta_{21}\, \nabla_\alpha A_\mu, \nonumber\\
\Theta^{[1]8}_\mu A_\nu &= \zeta_{23} \nabla_\mu A_\nu +\zeta_{24}\,A_\mu A_\nu.
\label{appb1}
\end{align}
\endgroup
Here, the expressions for $\Xi_{(5)}^{\alpha\beta}, \Xi_{6}$, and $\Xi^{(8)}_{\mu\nu}$ are explicitly presented in (\ref{xiterms}). The certain constant terms $\kappa_{m(1)}$ (where $m=1\cdots4$) and $\zeta_n$'s are provided in (\ref{kapQ1}) and (\ref{AppenA}), respectively.

\subsection{Coefficient Functions of $\mathscr{O}_{(2)}$-Part:} \label{AppenC}
Ensuingly, the coefficient functionals $\Theta_{k}^{[2]j} A_l $ (with $j=1\cdots 3$) associated with the term $\mathscr{O}_{(2)}$ in (\ref{o2partexplicit}) manifestly become
\begingroup
\allowdisplaybreaks
\begin{align}
\Theta^{[2]1} A&=\Xi_{(9)}, \nonumber\\
\Theta^{[2]2}_{\mu}A_\nu&=\kappa_{5(1)} \nabla_\mu A_\nu-2 \zeta_{25} \nabla_\nu A_\mu-\zeta_{27} A_\mu A_\nu, \nonumber\\
 \Theta^{[2]3} A&=\zeta_{28} \nabla \cdot A+\zeta_{29}A^2,  
\label{appb2}
\end{align}
\endgroup
where the constants $\Xi_{(9)}$ and $\kappa_{5(1)}$ are respectively described in (\ref{xiterms}) and (\ref{kapQ1}) while the certain values of terms $\zeta_n$'s can be attained in (\ref{AppenA}).

\subsection{Coefficient Functions of $\mathscr{O}^{\alpha}_{(3)\mu}$-Part:}\label{AppenD}
Afterwards, the explicit expressions of functions $\Theta_{k}^{[3]j} A_l $ (where $j=1\cdots 4$) which appertain to the term $\mathscr{O}^{\alpha}_{(3)\mu}$ in (\ref{o3partexplicit}) read
\begingroup
\allowdisplaybreaks
\begin{align}
\Theta^{[3]1}_{\mu} A_\sigma &=-2 \Xi^{(7)}_{\mu\sigma}, \nonumber\\
 \Theta^{[3]2}_{\mu} A^\alpha &=(-\zeta_{23}/\zeta_6)\, \Xi^{(5)\alpha}_\mu+\kappa_{1(3)} A_\mu A^\alpha, \nonumber\\
\Theta^{[3]3}_{\mu} A_\beta&=(\kappa_{1(1)}/2) \nabla_\mu A_\beta+\kappa_{3(1)} \nabla_\beta A_\mu,\nonumber\\
\Theta^{[3]4}_{\mu} A_\beta &=\kappa_{2(1)} \nabla_\mu A_\beta+(\kappa_{1(1)}/2) \nabla_\beta A_\mu, 
\label{appb3}
\end{align}
\endgroup
wherein the explicit form of terms $\Xi^{(5)\alpha}_\mu$ and $\Xi^{(7)}_{\mu\sigma}$ are provided in (\ref{xiterms}). Also, the coupling constants $\kappa_{m(1)}$ (where $m=1 \cdots 3$) and $\kappa_{1(3)}$ are consecutively given in (\ref{kapQ1}) and (\ref{kapQ234}).

\subsection{Coefficient Functions of $\mathscr{O}^{\alpha\beta}_{(4)\mu\nu}$-Part:}\label{AppenE}
On the other side, the expression of coefficient functions $\Theta_{k}^{[4]j} A_l $ (for $j=1\cdots 6)$ relevant to the term $\mathscr{O}^{\alpha\beta}_{(4)\mu\nu}$ in (\ref{o4partexplicit}) successively are
\begingroup
\allowdisplaybreaks
\begin{align}
 \Theta^{[4]1}A &=\zeta_1 \nabla \cdot A-\kappa_{1(4)}A^2, \nonumber\\
\Theta^{[4]2}_\mu A_\sigma &=(-2\zeta_5/\zeta_6)\, \Xi^{(5)}_{\mu\sigma}, \nonumber \\
\Theta^{[4]3\alpha}_\mu A^\beta &=-\zeta_{15}A_\mu \nabla^\alpha A^\beta,\nonumber\\
\Theta^{[4]4}_\mu A^\beta&=4(9\alpha_2+4\alpha_5+3\alpha_6) \nabla_\mu A^\beta-(72\alpha_2+2 \zeta_{12}-\zeta_{13}) \nabla^\beta A_\mu, \nonumber\\
\Theta^{[4]5}_\mu A_\nu&=\zeta_{14} \nabla_\mu A_\nu, \nonumber\\
\Theta_{[4]6}^\alpha A_\mu&=2(15\alpha_2+8\alpha_5+6\alpha_6) \nabla^\alpha A_\mu.
\label{appb4} 
\end{align}
\endgroup
Here, the explicit form of terms $\Xi^{(5)}_{\mu\sigma}$ and $\kappa_{1(4)}$ are presented in (\ref{xiterms}) and (\ref{kapQ234}), respectively. As for the specific constant terms $\zeta_n$'s, they can be found in (\ref{AppenA}).

\subsection{Coefficient Functions of $\mathscr{O}_{(5)\nu}$-Part:}\label{AppenF} 
Subsequently, the coefficient functions $\Theta_{k}^{[5]j} A_l $ (where $j=1\cdots 21)$ associated with the compact term $\mathscr{O}^{(5)}_{\mu}$ in (\ref{o5partexplicit}) are described expressly as follows 
\begingroup
\allowdisplaybreaks
\begin{align}
\Theta_{[5]1}^{\beta\sigma}A^\alpha&=[\kappa_{2(1)} \nabla^\beta A^\sigma-\kappa_{2(5)} \nabla^\sigma A^\beta] A^\alpha, \nonumber\\
\Theta_{[5]2}^{\sigma\alpha}A^\beta &=\zeta_6 A_\lambda R^{\sigma(\alpha\beta)\lambda}-(\zeta_9/2) A^\sigma R^{\alpha\beta},\nonumber\\
 \Theta_{[5]3}^{\alpha}A^\beta&=(-\kappa_{3(5)}/\zeta_6)\,\Xi^{\alpha\beta}_{(5)},\nonumber\\
\Theta_{[5]4}^{\alpha}A^\beta&=(\kappa_{4(5)}/\zeta_6)\, \Xi^{\alpha\beta}_{(5)},\nonumber\\
\Theta_{[5]5} A^\alpha &=(-\zeta_9/2)\,\square A^\alpha+\kappa_{5(5)}\nabla^\alpha \nabla \cdot A+[4\zeta_8 \nabla^\alpha A^\beta+\zeta_9 \nabla^\beta A^\alpha]A_\beta+[\kappa_{6(5)} \nabla \cdot A+(\zeta_{23}/2)A^2]A^\alpha \nonumber\\
&-(\zeta_9/2)\,A_\beta R^{\beta\alpha}+(\zeta_{10}/2)\,A^\alpha R, \nonumber\\
\Theta_{[5]6} A&=\kappa_{7(5)} \nabla \cdot A+\kappa_{8(5)}A^2,\nonumber\\
\Theta^{[5]7}_{\mu\alpha} A_\beta &=\kappa_{9(5)}\nabla_\mu \nabla_\alpha A_\beta-(\zeta_9/2)\nabla_\alpha \nabla_\beta A_\mu+\zeta_9 A_\mu \nabla_\alpha A_\beta+[\kappa_{10(5)}\nabla_\mu A_\beta+\kappa_{11(5)} \nabla_\beta A_\mu \nonumber\\
&+(\zeta_{21}/2)A_\mu A_\beta]A_\alpha, \nonumber\\
\Theta^{[5]8}_{\mu} A_{\alpha}&=(-\kappa_{12(5)}/\zeta_6)\,\Xi^{(5)}_{\mu\alpha}, \nonumber\\
\Theta^{[5]9} A_\mu &=(\zeta_{10}/2)[\square A_\mu-2 A_\mu \nabla \cdot A]+\kappa_{13(5)} \nabla_\mu \nabla \cdot A+[8\zeta_{11}\nabla_\mu A_\alpha-\zeta_{10}\nabla_\alpha A_\mu]A^\alpha, \nonumber\\
\Theta^{[5]10}_{\mu} A_{\beta}&=- \Theta^{[3]4}_{\mu} A_{\beta}-(\kappa_{4(1)}/2)A_\mu A_\beta, \nonumber\\
\Theta_{[5]11} A&=\zeta_{20} \nabla \cdot A-\zeta_{23}A^2, \nonumber\\
\Theta_{[5]12}^{\alpha} A^\beta &=\kappa_{14(5)}\nabla^\alpha A^\beta+\kappa_{15(5)}\nabla^\beta A^\alpha+\kappa_{16(5)} A^\alpha A^\beta,\nonumber\\
\Theta_{[5]13}^{\alpha} A^\beta &=\kappa_{17(5)}\nabla^\alpha A^\beta+\kappa_{18(5)}\nabla^\beta A^\alpha+(\zeta_{21}/2) A^\alpha A^\beta, \nonumber\\
\Theta^{[5]14}_{\mu} A_\alpha&=\kappa_{19(5)}\nabla_\mu A_\alpha+\kappa_{20(5)}\nabla_\alpha A_\mu+\kappa_{21(5)} A_\mu A_\alpha, \nonumber\\
\Theta_{[5]15} A&=\kappa_{22(5)}\nabla \cdot A+\kappa_{23(5)} A^2, \nonumber\\
\Theta_{[5]16} A^\alpha &=[\kappa_{24(5)}\nabla^\alpha A^\beta+\kappa_{16(5)}\nabla^\beta A^\alpha]A_\beta, \nonumber\\
\Theta_{[5]17} A^\alpha&=[\kappa_{25(5)}\nabla \cdot A+8\zeta_{30}A^2]A^\alpha, \nonumber\\
\Theta_{[5]18} A^\alpha&=[\kappa_{26(5)}\nabla^\alpha A^\beta+\zeta_{21}\nabla^\beta A^\alpha]A_\beta, \nonumber\\
\Theta_{[5]19} A^\alpha&=[\kappa_{21(5)}\nabla \cdot A+\zeta_{24}A^2]A^\alpha,\nonumber\\
\Theta^{[5]20}_{\alpha\beta} A_\mu&=[\Theta^{[1]7}_\alpha A_\beta-4\zeta_{24}A_\alpha A_\beta]A_\mu, \nonumber\\
\Theta_{[5]21} A&=\zeta_{20}\nabla \cdot A+\zeta_{24} A^2, \nonumber\\
\Theta_{[5]22} A^\alpha&=\square A^\alpha+\nabla^\alpha \nabla \cdot A-2 A_\beta [\nabla^\beta A^\alpha-R^{\beta\alpha}],
\label{appb5}
\end{align}
\endgroup
where the constant coefficients $\kappa_{2(1)}, \kappa_{4(1)}$ and $\kappa_{m(5)}$ (for $m=2 \cdots 26$) are independently described in (\ref{kapQ1}) and (\ref{kapQ5}). Again, the explicit form of particular constants $\zeta_n$'s can be found in (\ref{AppenA}). Also, observe that the last term is not used in (\ref{o5partexplicit}) explicitly. However, since it emerges during the calculations for this part, we define it here but use it in the other parts, for example in the (\ref{AppenI}).

\subsection{Coefficient Functions of $\mathscr{O}_{(6)\mu\nu}^\alpha$-Part:}\label{AppenG}
Later on, as to those of the term $\mathscr{O}_{(6)\mu\nu}^\alpha$ in (\ref{o6partexplicit}), the relevant coefficient functionals $\Theta_{k}^{[6]j} A_l $ (with $j=1\cdots 42$) are given in succession as below
\begingroup
\allowdisplaybreaks
\begin{align}
 \Theta_{[6]1}^\sigma A^\beta &=-\nabla^\sigma A^\beta+(R^{\sigma\beta}/2), \nonumber\\
\Theta_{[6]2} A&=\Theta_{[5]11} A-\zeta_{10}R,\nonumber\\
\Theta_{[6]3} A^\beta&=(\zeta_5/2)\square A^\beta -\kappa_{9(6)} \nabla^\beta \nabla \cdot A+A_\sigma [16 \kappa_{1(4)}\nabla^\beta A^\sigma+\zeta_5 \Theta_{[6]1}^\sigma A^\beta]-A^\beta \Theta_{[6]2} A, \nonumber\\
\Theta_{[6]4} A&=\kappa_{10(6)} \nabla \cdot A-\kappa_{11(6)}A^2, \nonumber\\
\Theta^{[6]5}_\mu A^\beta&=(\zeta_5/2)\nabla_\mu A^\beta-\zeta_4 \nabla^\beta A_\mu, \nonumber\\
\Theta^{[6]6}_\mu A^\beta&=\zeta_{17}\nabla_\mu A^\beta-\kappa_{4(6)}\nabla^\beta A_\mu+\zeta_{22}A_\mu A^\beta, \nonumber\\
\Theta^\sigma_{[6]7\mu} A^\beta&=\nabla^\sigma (\Theta^{[6]5}_\mu A^\beta)-2\zeta_5 (\nabla^\sigma A_{(\mu})A^{\beta)}-A^\sigma [\Theta^{[6]6}_\mu A^\beta+\zeta_9 R_\mu{^\beta}]+18\alpha_2 A^\lambda R_{\mu\lambda}{^{\sigma\beta}}, \nonumber\\
\Theta^{[6]8}_\mu A_\beta&=\kappa_{7(6)}\nabla_\mu A_\beta-\zeta_5 \nabla_\beta A_\mu, \nonumber \\
\Theta^\sigma_{[6]9\mu}A_\beta&=-(1/2)\nabla^\sigma (\Theta^{[6]8}_\mu A_\beta)-(\zeta_6/2)[\nabla_\mu \nabla_\beta A^\sigma-2 A_\beta (\nabla_\mu A^\sigma)+A^\lambda R_\mu{^\sigma}{_{\beta\lambda}}-2 A_{(\beta}R^{\sigma)}{_\mu}]\nonumber\\
&-2\zeta_5 (\nabla^\sigma A_{(\mu})A_{\beta)}+\kappa_{12(6)} A^\sigma (\nabla_\mu A_\beta), \nonumber\\
\Theta^{[6]10}_\mu A_\beta&=2\kappa_{5(6)}\nabla_\mu A_\beta+\kappa_{6(6)}A_\mu A_\beta, \nonumber\\
\Theta_{[6]11}^\alpha A^\sigma&=\kappa_{13(6)} \nabla^\alpha A^\sigma-\kappa_{14(6)} \nabla^\sigma A^\alpha+2\zeta_6 R^{\alpha\sigma}, \nonumber\\
\Theta_{[6]12}^{\alpha\beta} A^{\sigma}&=-4\zeta_5 [\nabla^{\Dashv \alpha}\nabla^\beta A^{\sigma\vDash}+(1/2) A^\alpha \nabla^\sigma A^\beta ]+A^\beta (\Theta_{[6]11}^\alpha A^\sigma), \nonumber\\
\Theta^{[6]13}_{\mu} A_{\beta}&=(-\kappa_{7(6)}/\zeta_6)\, \Xi^{(5)}_{\mu\beta}, \nonumber\\
\Theta^{[6]14}_\mu A^\beta&=4 \alpha_8 \nabla_\mu A^\beta+\zeta_6 \Theta_{[6]1}^\beta A_\mu, \nonumber\\
\Theta^{[6]15}A_{\mu}&=(\zeta_6/2)[\square A_\mu-2 A_\mu \nabla \cdot A]+(\kappa_{3(6)}/2) \nabla_\mu \nabla \cdot A+A_\beta (\Theta^{[6]14}_\mu A^\beta), \nonumber\\
\Theta^{[6]16}_\beta A_\nu&=\kappa_{15(6)}\nabla_\beta A_\nu +\zeta_9 \nabla_\nu A_\beta, \nonumber\\
\Theta^{[6]17}_\beta A_\nu&=2\zeta_6 \nabla_\beta A_\nu+\zeta_9 \nabla_\nu A_\beta-(\zeta_{15}/2) A_\nu A_\beta, \nonumber\\
\Theta^{[6]18}_\mu A_\nu&=-2\kappa_{16(6)} \nabla_\mu A_\nu, \nonumber\\
\Theta^{[6]19}_{\mu\beta} A_\nu&=\nabla_\mu (\Theta^{[6]16}_\beta A_\nu)-2  (\Theta^{[6]17}_\beta A_\nu)A_\mu+ (\Theta^{[6]18}_\mu A_\nu)A_\beta, \nonumber\\
\Theta_{[6]20}^\beta A^\alpha&=\kappa_{17(6)} \nabla^\beta A^\alpha+\zeta_9 \nabla^\alpha A^\beta, \nonumber\\
\Theta_{[6]21}^\beta A_\mu&=\kappa_{18(6)} \nabla^\beta A_\mu, \nonumber\\
\Theta^{[6]22}_\mu A^\beta&=\zeta_9 \nabla_\mu A^\beta-\zeta_6 \nabla^\beta A_\mu, \nonumber\\
\Theta_{[6]23}^\alpha A^\beta&= \Theta_{[6]17}^\alpha A^\beta+\Theta_{[6]22}^\beta A^\alpha, \nonumber\\
\Theta^{[6]24}_\mu A^\alpha&=\kappa_{19(6)} \nabla_\mu A^\alpha-\kappa_{29(6)} \nabla^\alpha A_\mu, \nonumber\\
\Theta^{[6]25\beta}_{\mu} A^\alpha&=\nabla_\mu (\Theta_{[6]20}^\beta A^\alpha)+\nabla^\alpha (\Theta_{[6]21}^\beta A_\mu)+2 A^\alpha (\Theta^{[6]22}_\mu A^\beta)-2A_\mu (\Theta_{[6]23}^\alpha A^\beta)+2A^\beta (\Theta^{[6]24}_\mu A^\alpha), \nonumber\\
\Theta^{[6]26}_{\mu\nu} A^\alpha&=(\zeta_{10}/2)[\nabla_\mu F_\nu{^\alpha}-8(\nabla_{\Dashv \mu} A_\nu) A^{\alpha\vDash}], \nonumber\\
\Theta^{[6]27}_\mu A^\alpha&=\kappa_{20(6)} \nabla_\mu A^\alpha+\kappa_{21(6)} \nabla^\alpha A_\mu+\kappa_{22(6)} A_\mu A^\alpha, \nonumber\\
\Theta_{[6]28}^\alpha A_\mu&=(-\zeta_{14}/\zeta_6)\, \Xi^\alpha_{(5)\mu}+24\alpha_2 A^\alpha A_\mu, \nonumber\\
\Theta^{[6]29}_\mu A^\alpha&=-(72\alpha_2+2 \zeta_{12}-\zeta_{13})F_\mu{^\alpha}-\Theta_{[6]28}^\alpha A_\mu, \nonumber\\
\Theta^{[6]30}_\mu A_\beta&=\kappa_{23(6)} \nabla_\mu A_\beta+(72\alpha_2+2\zeta_{12}-\zeta_{13}-\zeta_{14}) \nabla_\beta A_\mu+(\zeta_{15}/2) A_\mu A_\beta, \nonumber\\
\Theta_{[6]31}^\alpha A^\beta&=\kappa_{24(6)} \nabla^\alpha A^\beta+\kappa_{25(6)} \nabla^\beta A^\alpha-2\zeta_{15} A^\alpha A^\beta, \nonumber\\
\Theta^{[6]32}_\mu A^\nu&=\kappa_{26(6)} \nabla_\mu A_\nu-(\kappa_{22(6)}/2) A_\mu A_\nu, \nonumber\\
\Theta^{[6]33}_{\mu\nu} A^\beta&=(\zeta_{15}/2) [A_\mu F_\nu{^\beta}-2 A^\beta \nabla_{(\mu}A_{\nu)}], \nonumber\\
\Theta_\mu^{[6]34\alpha} A^\beta&=2\zeta_{15} [(\nabla_{\lessdot \mu}A^\alpha) A^{\beta \gtrdot}]+\Theta_\mu^{[4]3\alpha} A^\beta, \nonumber\\
\Theta^{[6]35}_\mu A^\beta&=\kappa_{1(6)}\nabla_\mu A^\beta-\kappa_{2(6)}\nabla^\beta A_\mu-\zeta_{21}A_\mu A^\beta, \nonumber\\
 \Theta^\alpha_{[6]36} A^\beta&=\nabla_\mu [\kappa_{27(6)}\nabla^\alpha A^\beta-\kappa_{2(6)}\nabla^\beta A^\alpha]-2 \zeta_{18} \nabla^\alpha \nabla^\beta A_\mu -2 \zeta_{21} [\nabla_\mu A^{(\alpha}-2 \nabla^\alpha A_{(\mu}] A^{\beta)}, \nonumber\\
\Theta^{[6]37}_\beta A_\mu&=(1/2) [\zeta_{19} \nabla_\beta A_\mu+\zeta_{22}A_\mu A_\beta], \nonumber\\
\Theta^{[6]38} A_\mu&=(\Theta^{[5]21}A) A_\mu-(\zeta_{22}/2) A^\beta \nabla_\mu A_\beta, \nonumber\\
\Theta^{[6]39}_{\mu} A_\nu&=\zeta_{22} \nabla_\mu A_\nu-4 \zeta_{24} A_\mu A_\nu, \nonumber\\
\Theta_{[6]40} A^\alpha&=[\kappa_{28(6)}\nabla^\alpha A^\beta-\zeta_{21} \nabla^\beta A^\alpha]A_\beta-2 (\Theta^{[5]21} A) A^\alpha, \nonumber\\
\Theta^{[6]41}_\mu A_\beta&=\zeta_{17} \nabla_\mu A_\beta-\zeta_{18} \nabla_\beta A_\mu+\zeta_{21} A_\mu A_\beta,  \nonumber\\
\Theta^{[6]42\beta}_\mu A^\alpha&=\zeta_{19} \nabla_\mu \nabla^\beta A^\alpha+\zeta_{21} A_\mu F^{\alpha\beta}-4 \zeta_{23} A^\beta \nabla_{(\mu}A^{\alpha)}+(\Theta^{[6]39}_{(\mu}A^{\beta)})A^\alpha. 
\label{appb6}
\end{align}
\endgroup
Here, the two-fold cyclic symmetric-(anti)symmetric index operations "$\Dashv \cdots \vDash$" and "$\lessdot \cdots \gtrdot$" are described in (\ref{twfldcylclc}). The constant coefficients $\kappa_{1(4)}$ and $\kappa_{m(6)}$ (with  $m=1 \cdots 29$ except for the value $m\neq 8$) are described in (\ref{kapQ234}) and (\ref{kapQ6}), separately. Recall that the distinct constant terms $\zeta_{n}$'s are given in (\ref{AppenA}).

\subsection{Coefficient Functions of $\mathscr{O}^{(7)}_{\mu\nu}$-Part:}\label{AppenH}
Thereafter, the expressions for coefficient functionals $\Theta_{k}^{[7]j} A_l $ (where $j=1\cdots 41$) pertaining to the compact term $\mathscr{O}^{(7)}_{\mu\nu}$ in (\ref{o7partexplicit}) are sequentially as follows
\begingroup
\allowdisplaybreaks
\begin{align}
\Theta^{[7]1}_\mu A_\nu&=-\Xi^{(1)}_{\mu\nu}, \nonumber\\
\Theta^{[7]2}_{\mu\alpha\nu\beta} A_\lambda&=-(\kappa_{7(6)}/2) A_\lambda R_{\mu\alpha\nu\beta}, \nonumber\\
\Theta^{[7]3\alpha}_\mu A^\sigma&=[\Theta^{[6]29}_\mu A^\alpha+(\zeta_6/2)R_\mu{^\alpha}]A^\sigma, \nonumber\\
\Theta^{[7]4\alpha}_\mu A^\sigma&=[\Theta^{[3]4}_\mu A^\alpha+(\zeta_9/2)R_\mu{^\alpha}]A^\sigma, \nonumber\\
\Theta_{[7]5}^{\alpha\beta} A^\sigma&=\kappa_{1(7)} \nabla^\alpha \nabla^\beta A^\sigma+(\zeta_5/2)\widehat{\boxdot}^{\beta\sigma} A^\alpha+[\kappa_{2(7)}\nabla^\alpha A^\beta+(\kappa_{3(7)}+\zeta_5) \nabla^\beta A^\alpha+(\zeta_{21}/2) A^\alpha A^\beta]A^\sigma, \nonumber\\
\Theta_{[7]6}^{\alpha\sigma} A^\beta&=(\zeta_6/4) [\nabla^\alpha F^{\sigma\beta}-2 A^\alpha F^{\sigma\beta}+A_\lambda (2R^{\sigma\alpha\beta\lambda}+R^{\sigma\beta\alpha\lambda})]-[\zeta_{17}\nabla^\alpha A^\beta-\kappa_{4({Q_7})}\nabla^\alpha A^\beta -\nonumber\\
&(\zeta_9/2) R^{\alpha\beta}] A^\sigma,\nonumber\\
\Theta^{[7]7\alpha\sigma\beta}_\mu A^\lambda&=-36\alpha_2 A^\lambda R_\mu{^{\alpha\sigma\beta}}, \nonumber\\
\Theta_{[7]8}^\alpha A^\beta&=(1/2)[\zeta_5 \nabla^\alpha A^\beta-\kappa_{7(6)} \triangle^\beta A^\alpha],\nonumber\\
\Theta_{[7]9} A&=\kappa_{7(7)} \nabla \cdot A+\kappa_{8(7)} A^2,\nonumber\\
\Theta_{[7]10}A^\alpha&=-(\kappa_{7(6)}/2)\widehat{\boxdot} A^\alpha+\kappa_{11(7)} \nabla^\alpha \nabla \cdot A+\kappa_{12(7)} A^\sigma \nabla^\alpha A_\sigma,\nonumber\\
\Theta^{[7]11}_{\mu\alpha} A_\beta&=\kappa_{13(7)} \nabla_\mu \nabla_\alpha A_\beta+\kappa_{5(6)}[\nabla_\mu \nabla_\beta A_\alpha-2A_\mu \nabla_\alpha A_\beta]+(\zeta_{22}/4)\nabla_\alpha \nabla_\beta A_\mu+A_\beta [\kappa_{15(7)}\nabla_\alpha A_\mu \nonumber\\
&+\kappa_{16(7)}\nabla_\mu A_\alpha+(\zeta_6/2)R_{\mu\alpha}]-48\alpha_8 A_\mu \nabla_\alpha A_\beta-2\kappa_{7(6)} [(\nabla_\beta A_{(\mu})A_{\alpha)}-(1/2)A^\sigma R_{\beta(\mu\alpha)\sigma}]\nonumber\\
&+A_\alpha [\kappa_{4(5)} \nabla_\mu A_\beta+(\zeta_{15}/2)A_\mu A_\beta ]+\zeta_5 A^\sigma R_{\alpha\mu\beta\sigma},\nonumber\\
\Theta^{[7]12}_{\mu\alpha} A_\beta&=\kappa_{17(7)}\nabla_\mu \nabla_\alpha A_\beta+\kappa_{5(6)}[\nabla_\alpha \nabla_\beta A_\mu+A^\sigma R_{\alpha\mu\beta\sigma}]+\kappa_{6(6)}A_\mu \nabla_\alpha A_\beta+A_\alpha [\kappa_{18(7)}\nabla_\mu A_\beta \nonumber\\
&+\kappa_{19(7)}\nabla_\beta A_\mu+(1/2)(\zeta_{21}A_\mu A_\beta-\zeta_9 R_{\mu\beta})],\nonumber\\
\Theta^{[7]13} A^\alpha&=\kappa_{5(6)}[\square A^\alpha+A_\beta R^{\beta\alpha}]+\kappa_{20(7)}\nabla^\alpha\nabla\cdot A-A_\beta[\kappa_{12(7)}\nabla^\alpha A^\beta-\kappa_{6(6)}\nabla^\beta A^\alpha] \nonumber\\
&+(1/2)A^\alpha \Big[\zeta_{23}\, \triangle_{(2)} \cdot A+\zeta_{10} R \Big], \nonumber\\
\Theta_{[7]14\mu}^\alpha A_\nu &=-\kappa_{23(7)} \nabla^\alpha \nabla_\mu A_\nu+(\kappa_{12(5)}/2) \widehat{\boxdot}_{\mu\nu}A^\alpha+\kappa_{3(5)}A_\mu \nabla^\alpha A_\nu+A^\alpha [(\kappa_{12(5)}+\kappa_{24(7)}) \nabla_\mu A_\nu- \nonumber\\
&(1/4)(\zeta_{15}A_\mu A_\nu-\zeta_6 R_{\mu\nu})], \nonumber\\
 \Theta_{[7]15} A_\mu &=(\kappa_{3(5)}/4) \widehat{\boxdot} A_\mu+\kappa_{25(7)} \nabla_\mu \nabla \cdot A+2 \kappa_{8(5)}A^\alpha \nabla_\mu A_\alpha, \nonumber\\
\Theta_{[7]16}^\alpha A^\beta &=-(\kappa_{7(6)}/2)\,\square \nabla^\alpha A^\beta+\kappa_{26(7)} \nabla^\alpha \nabla^\beta \nabla \cdot A+A^\alpha[\kappa_{4(7)}\square A^\beta+\kappa_{27(7)}\nabla^\beta \nabla \cdot A] \nonumber\\
&+A_\sigma [8\kappa_{1(4)}\nabla^\alpha \nabla^\beta A^\sigma-\zeta_5 \nabla^\sigma \nabla^\alpha A^\beta+\kappa_{3(7)}A^\beta R^{\alpha\sigma}]+[\kappa_{28(7)}\nabla^\alpha A^\sigma+\kappa_{29(7)}\nabla^\sigma A^\alpha \nonumber\\
&+\kappa_{31(7)}A^\alpha A^\sigma+\kappa_{34(7)}R^{\alpha\sigma}] \nabla^\beta A_\sigma+[\kappa_{30(7)} \nabla^\sigma A^\alpha+\zeta_{21}A^\alpha A^\sigma-(\zeta_9/2) R^{\alpha\sigma}]\nabla_\sigma A^\beta \nonumber\\
&-[\zeta_5 \nabla^\alpha A^\beta-(\zeta_{21}/2)A^\alpha A^\beta-\kappa_{32(7)} R^{\alpha\beta}] \nabla \cdot A+\Xi^{(5)}_{\sigma\lambda}R^{\alpha\sigma\beta\lambda}+\kappa_{33(7)} A^2 R^{\alpha\beta}, \nonumber\\\
\Theta^{[7]17\alpha\beta}_\mu A^\sigma &=(\kappa_{7(6)}/2)\nabla_\mu \nabla^\alpha \nabla^\beta A^\sigma+A_\mu [\zeta_5 \nabla^\alpha \nabla^\beta A^\sigma+(\zeta_{15}/2) A^\sigma \nabla^\alpha A^\beta-2\zeta_5 A_\lambda R^{\alpha\beta\sigma\lambda}+\kappa_{61(7)}A^\sigma R^{\alpha\beta}]\nonumber\\
&+A^\alpha [\kappa_{35(7)}\nabla_\mu \nabla^\beta A^\sigma-\hat{\Sigma}_C A_\lambda R_\mu{^{\lambda\beta\sigma}}+\zeta_6 A^\beta R^\sigma{_\mu}]+A^\sigma[\kappa_{36(7)}\nabla_\mu \nabla^\beta A^\alpha+\kappa_{37(7)}\nabla^\alpha \nabla^\beta A_\mu \nonumber\\
&-\kappa_{2(1)} \nabla_\mu \nabla^\alpha A^\beta+A_\lambda  (\kappa_{2(1)} R_\mu{^{\alpha\beta\lambda}}+\kappa_{51(7)} R_\mu{^{\beta\alpha\lambda}}+\hat{O}_A R_\mu{^{\lambda\alpha\beta}})]+\nabla^\beta A^\sigma [\kappa_{38(7)} \nabla_\mu A^\alpha +\nonumber\\
& \kappa_{39(7)} \nabla^\alpha A_\mu-(\zeta_{15}/2) A_\mu A^\alpha ]+\nabla_\mu A^\sigma [ \kappa_{40(7)} \nabla^\alpha A^\beta+ \kappa_{41(7)} \nabla^\beta A^\alpha+ \kappa_{44(7)} A^\alpha A^\beta+ \kappa_{59(7)} R^{\alpha\beta}] \nonumber\\
&+\nabla^\sigma A_\mu [ \kappa_{42(7)} \nabla^\alpha A^\beta+\kappa_{43(7)} \nabla^\beta A^\alpha+(3\zeta_{21}/2)A^\alpha A^\beta+\kappa_{60(7)} R^{\alpha\beta}]-A_\lambda [(\zeta_6/2)(\nabla_\mu R^{\alpha\beta\sigma\lambda}+ \nonumber\\
&\nabla^\sigma R_\mu{^{\beta\alpha\lambda}})-(\zeta_5/2)\nabla^\alpha R_\mu{^{\beta\sigma\lambda}}+\zeta_4 \nabla^\sigma R_\mu{^{\lambda\alpha\beta}}]-2(\Theta^{[7]1} A)R_\mu{^{\alpha\beta\sigma}}+\nabla_\lambda A^\sigma [\kappa_{45(7)}R_\mu{^{\alpha\beta\lambda}} +\nonumber \\
&\kappa_{50(7)}R_\mu{^{\beta\alpha\lambda}}+\kappa_{52(7)}R_\mu{^{\lambda\alpha\beta}}]+\nabla^\sigma A_\lambda [\kappa_{46(7)}R_\mu{^{\alpha\beta\lambda}}+\kappa_{49(7)}R_\mu{^{\beta\alpha\lambda}}+\kappa_{53(7)}R_\mu{^{\lambda\alpha\beta}}]\nonumber\\
&+\nabla^\alpha A_\lambda [\kappa_{47(7)}R_\mu{^{\beta\sigma\lambda}}+\kappa_{54(7)}R_\mu{^{\lambda\beta\sigma}}]+\nabla_\lambda A^\alpha  [\kappa_{48(7)}R_\mu{^{\beta\sigma\lambda}}+\kappa_{55(7)}R_\mu{^{\lambda\beta\sigma}}]+72 \alpha_2 A^2 R_\mu{^{\beta\alpha\sigma}}\nonumber\\
&+\nabla_\mu A_\lambda [\kappa_{56(7)}R^{\alpha\beta\sigma\lambda}+\kappa_{57(7)}R^{\alpha\lambda\beta\sigma}]+\nabla_\lambda A_\mu  [\zeta_3 R^{\alpha\beta\sigma\lambda}+\kappa_{58(7)}R^{\alpha\lambda\beta\sigma}]+\nonumber\\
&[(\kappa_{4(5)}/2)\nabla^\alpha A^\beta+\kappa_{62(7)}\nabla^\beta A^\alpha]R^\sigma{_\mu}, \nonumber\\
\Theta^{[7]18}_{\mu\nu\alpha} A_\beta &=\kappa_{12(5)}\nabla_\mu\nabla_\nu\nabla_\alpha A_\beta-(\kappa_{4(5)}/2)\nabla_\alpha\nabla_\beta\nabla_\mu A_\nu+\nabla_\mu A_\alpha [\kappa_{65(7)} \nabla_\beta A_\nu+\kappa_{68(7)} \nabla_\nu A_\beta+\nonumber\\
&(\zeta_{15}/2)A_\nu A_\beta+(\kappa_{7(6)}/2) R_{\nu\beta}]+\nabla_\alpha A_\mu[\kappa_{63(7)} \nabla_\beta A_\nu+\kappa_{69(7)} A_\beta A_\nu+\kappa_{72(7)}R_{\nu\beta}]+\nonumber\\
&A_\alpha [\kappa_{64(7)}\nabla_\mu \nabla_\nu A_\beta+\kappa_{67(7)}\nabla_\beta \nabla_\mu A_\nu]+\nabla_\alpha A_\beta [\kappa_{66(7)}\nabla_\mu A_\nu+\kappa_{70(7)}A_\mu A_\nu-(\kappa_{3(5)}/2) R_{\mu\nu}]\nonumber\\
&+A_\alpha A_\beta [\kappa_{71(7)}\nabla_\mu A_\nu-\zeta_{24}A_\mu A_\nu+\frac{\kappa_{3(5)}}{2} R_{\mu\nu}]+A_\mu [\zeta_6 \nabla_\alpha \nabla_\beta A_\nu+\zeta_9 \nabla_\nu \nabla_\alpha A_\beta+\kappa_{73(7)}A_\alpha R_{\beta\nu}], \nonumber\\
 \Theta_{[7]19}^\beta A_\mu&=(\kappa_{4(5)}/2)\square \nabla^\beta A_\mu+\kappa_{74(7)} \nabla_\mu \nabla^\beta \nabla \cdot A+A^\beta [\kappa_{75(7)}\square A_\mu+\kappa_{76(7)}\nabla_\mu \nabla \cdot A]-A_\mu [\zeta_9 \square A^\beta+\nonumber\\
&\zeta_6 \nabla^\beta \nabla \cdot A]+A^\alpha [\nabla_\alpha \Theta^{[6]22}_\mu A^\beta+4 \zeta_8 \nabla^\beta \nabla_\mu A_\alpha]+\nabla_\mu A_\alpha [\kappa_{77(7)}\nabla^\alpha A^\beta+\kappa_{78(7)}\nabla^\beta A^\alpha+\nonumber\\
&\kappa_{31(7)}A^\alpha A^\beta]+\nabla_\alpha A_\mu [\kappa_{79(7)}\nabla^\alpha A^\beta+\kappa_{80(7)}\nabla^\beta A^\alpha+\kappa_{82(7)}A^\alpha A^\beta]+\nabla \cdot A [\zeta_9 \nabla_\mu A^\beta+\nonumber\\
&\kappa_{81(7)}\nabla^\beta A_\mu+\kappa_{83(7)}A_\mu A^\beta+4\zeta_1 R_\mu{^\beta}]-A^2[\Theta_{[1]8}^\beta A_\mu+4\kappa_{4(1)}R_\mu{^\beta}]+A_\mu A_\alpha [\kappa_{82(7)}\nabla^\alpha A^\beta+\nonumber\\
&\kappa_{4(1)}\nabla^\beta A^\alpha]+\Big[\zeta_9 \nabla^\alpha A^\beta+\frac{\kappa_{17(6)}}{2} \nabla^\beta A^\alpha+\hat{\Sigma}_D A^\alpha A^\beta\Big] R_{\mu\alpha},\nonumber\\
\Theta_{[7]20} A &=\kappa_{84(7)}\square \nabla \cdot A-2\zeta_8 [A^\alpha \square A_\alpha+(\nabla_\alpha A_\beta)^2]+2(\Theta_{[2]1} A)R+(\zeta_6/2)[(\nabla \cdot A)^2+2A^\alpha \nabla_\alpha \nabla \cdot A\nonumber\\
&+(\nabla_\alpha A_\beta)(\nabla^\beta A^\alpha)], \nonumber\\
\Theta^{[7]21}_\mu A_\nu&= \kappa_{85(7)}\nabla_\mu \nabla_\nu \nabla \cdot A+\nabla_\nu A^\alpha [\kappa_{89(7)}\nabla_\mu A_\alpha-\zeta_{27}A_\mu A^\alpha]+\nabla^\alpha A_\nu [\kappa_{86(7)}\nabla_\mu A_\alpha+\kappa_{88(7)}\nabla_\alpha A_\mu\nonumber\\
&+\zeta_{27}A_\mu A^\alpha]+\zeta_{10}A_\mu [\square A_\nu-\nabla_\nu \nabla \cdot A]+\nabla \cdot A [\kappa_{87(7)}\nabla_\mu A_\nu+\zeta_{29}A_\mu A_\nu]+A^\alpha[4\zeta_{11}\nabla_\mu \nabla_\nu A_\alpha\nonumber\\
&-\zeta_{10}\nabla_\alpha \nabla_\mu A_\nu]+A^2 [\zeta_{29}\nabla_\mu A_\nu+2\zeta_{30}A_\mu A_\nu],\nonumber \\
\Theta^{[7]22}_\mu A^\alpha&=\kappa_{90(7)}\nabla_\mu A^\alpha+\kappa_{91(7)}\nabla^\alpha A_\mu+\kappa_{92(7)}A_\mu A^\alpha, \nonumber\\
\Theta^{[7]23} A_\mu&=\hat{\Sigma}_C \square A_\mu +2 \kappa_{21(6)}\nabla_\mu \nabla \cdot A+A^\alpha [\kappa_{31(7)} \nabla_\mu A_\alpha+\kappa_{82(7)} \nabla_\alpha A_\mu]+A_\mu [\kappa_{83(7)} \nabla \cdot A-\zeta_{24}A^2], \nonumber\\ 
\Theta^{[7]24}_{\mu\alpha} A_\beta&=\kappa_{93(7)}\nabla_\mu \nabla_\alpha A_\beta+\kappa_{94(7)}\nabla_\alpha \nabla_\beta A_\mu+A_\alpha [2\kappa_{82(7)}\nabla_\beta A_\mu+\kappa_{95(7)}\nabla_\mu A_\beta]-\kappa_{95(7)}A_\mu \nabla_\alpha A_\beta, \nonumber\\ 
 \Theta_{[7]25}^\alpha A^\beta&=\kappa_{96(7)}\nabla^\alpha A^\beta-\kappa_{82(7)}A^\alpha A^\beta, \nonumber\\
 \Theta_{[7]26}^\alpha A^\beta&=\kappa_{97(7)}\nabla^\alpha A^\beta+\kappa_{98(7)}\nabla^\beta A^\alpha +\kappa_{99(7)}A^\alpha A^\beta, \nonumber \\
\Theta^{[7]27}_{\mu\nu} A_\alpha&=\kappa_{19(5)}\nabla_\mu \nabla_\nu A_\alpha+\kappa_{100(7)}\nabla_\alpha \nabla_\mu A_\nu+\kappa_{101(7)}A_\alpha \nabla_\mu A_\nu+A_\mu [\kappa_{92(7)}\nabla_\nu A_\alpha+\kappa_{102(7)}\nabla_\alpha A_\nu], \nonumber \\
\Theta^{[7]28}_\mu A^\alpha&=\kappa_{103(7)}\nabla_\mu A^\alpha+\kappa_{104(7)}\nabla^\alpha A_\mu +\kappa_{82(7)}A_\mu A^\alpha, \nonumber\\
\Theta^{[7]29}_{\mu\nu} A_\alpha&=-\kappa_{2(1)}\nabla_\mu \nabla_\nu A_\alpha+\kappa_{105(7)}\nabla_\alpha \nabla_\mu A_\nu+\kappa_{106(7)}A_\mu\nabla_\alpha  A_\nu-A_\alpha [\kappa_{31(7)}\nabla_\mu  A_\nu+\zeta_{24}A_\mu A_\nu], \nonumber\\
\Theta^{[7]30}_\mu A_\nu&=\kappa_{107(7)}\nabla_\mu A_\nu+\kappa_{108(7)}A_\mu A_\nu, \nonumber\\
\Theta^{[7]31}_{\mu\alpha} A_\beta &=\kappa_{109(7)}\nabla_\mu \nabla_\alpha A_\beta+\kappa_{110(7)}\nabla_\mu \nabla_\beta A_\alpha+A_\alpha [\kappa_{112(7)} \nabla_\mu A_\beta+\kappa_{95(7)} \nabla_\beta A_\mu]+A_\beta [\kappa_{111(7)} \nabla_\mu A_\alpha \nonumber\\
&+\kappa_{31(7)} \nabla_\alpha A_\mu]+A_\mu [\kappa_{95(7)} \nabla_\beta A_\alpha+2\zeta_{24}A_\alpha A_\beta], \nonumber\\
\Theta^{[7]32} A_\mu &=\kappa_{113(7)}\nabla_\mu \nabla \cdot A+A^\alpha [\kappa_{114(7)} \nabla_\mu A_\alpha+\kappa_{92(7)} \nabla_\alpha A_\mu]-A_\mu [\kappa_{102(7)}\nabla \cdot A-\zeta_{24} A^2], \nonumber\\
\Theta^{[7]33} A &=\kappa_{115(7)}\nabla \cdot A+\kappa_{116(7)} A^2, \nonumber\\
\Theta^{[7]34}_\mu A_\nu &=\kappa_{117(7)}\nabla_\mu A_\nu+\kappa_{118(7)}A_\mu A_\nu, \nonumber \\
\Theta^{[7]35} A&=[\kappa_{119(7)}\nabla^\beta A^\alpha+\kappa_{120(7)}A^\alpha A^\beta]\nabla_\alpha A_\beta + [\kappa_{121(7)}\nabla \cdot A-\zeta_{42}A^2]A^2, \nonumber\\
\Theta_{[7]36} A^\beta &=-A_\alpha [2\kappa_{82(7)}\nabla^\alpha A^\beta+\kappa_{31(7)}\nabla^\beta A^\alpha]+A^\beta [\kappa_{101(7)}\nabla \cdot A+\zeta_{24}A^2], \nonumber\\
\Theta_{[7]37} A^\beta&=A_\alpha [\kappa_{99(7)}\nabla^\alpha A^\beta+\kappa_{122(7)}\nabla^\beta A^\alpha]+A^\beta [\kappa_{123(7)}\nabla \cdot A-4\zeta_{30}A^2], \nonumber\\
\Theta^{[7]38}_\mu A^\alpha &=[\kappa_{125(7)}\nabla_\mu A_\beta+\kappa_{124(7)}\nabla_\beta A_\mu+\kappa_{127(7)}A_\mu A_\beta]\nabla^\alpha A^\beta +[\kappa_{126(7)}\nabla_\beta A_\mu+\kappa_{128(7)}A_\beta A_\mu]\nabla^\beta A^\alpha \nonumber\\
&+ [\kappa_{130(7)}\nabla_\mu A^\alpha+\kappa_{129(7)}\nabla^\alpha A_\mu+2\zeta_{24} A_\mu A^\alpha]\nabla \cdot A+[\kappa_{132(7)}\nabla_\mu A_\beta+\kappa_{131(7)}\nabla_\beta A_\mu]A^\alpha A^\beta \nonumber\\
&+[\kappa_{134(7)}\nabla_\mu A^\alpha+\kappa_{133(7)}\nabla^\alpha A_\mu+2\zeta_{24}A_\mu A^\alpha]A^2, \nonumber \\
 \Theta^{[7]39}_\mu A^\alpha&= [\kappa_{135(7)}\nabla^\alpha A^\beta+\zeta_{34} A^\alpha A^\beta]\nabla_\beta A_\mu+[\kappa_{137(7)}\nabla^\alpha A^\beta+\kappa_{136(7)}\nabla^\beta A^\alpha]A_\mu A_\beta \nonumber\\
&+[\kappa_{138(7)}\nabla^\alpha A_\mu+2 \zeta_{24}A_\mu A^\alpha]\nabla \cdot A+[\kappa_{118(7)}\nabla^\alpha A_\mu+2 \zeta_{24}A_\mu A^\alpha] A^2, \nonumber\\
\Theta^{[7]40} A &= [\kappa_{139(7)}\nabla^\beta A^\alpha+2\zeta_{4} A^\alpha A^\beta]\nabla_\alpha A_\beta-2 \zeta_{42} A^2 (\nabla \cdot A), \nonumber\\
\Theta^{[7]41}_\mu A_\nu &=\kappa_{140(7)}\nabla_\mu A_\nu+\kappa_{141(7)}A_\mu A_\nu.
\label{appb7}
\end{align}
\endgroup
Here, the explicit form of $\Xi^{(1)}_{\mu\nu}$ is given in (\ref{xiterms}). Additionally, the distinct values of couplings $\kappa_{m(1)}$ (with $m=2, 4$), $\kappa_{n(5)}$ (with $n=3, 4, 8, 12, 19$), $\kappa_{k(6)}$ (with $k=5, 6, 7, 21$) and $\kappa_{l(7)}$ (with $l=1 \cdots 141$) can be arrived in (\ref{kapQ1}), (\ref{kapQ5}), (\ref{kapQ6}) and (\ref{kapQ7}), respectively. Further, the constant terms $\hat{O}_A, \hat{\Sigma}_C$ and $\hat{\Sigma}_D$ are
\begin{equation}
\hat{O}_A=-72 \alpha_2-2\zeta_{12}+\zeta_{13}, \quad \hat{\Sigma}_C=2(15\alpha_2+8\alpha_5+6\alpha_6), \quad \hat{\Sigma}_D=\hat{\Sigma}_C+6 \alpha_2.
\label{O7coefzt2}
\end{equation}
As before, the discrete constant terms $\zeta_{n}$'s in (\ref{appb7}) and also (\ref{O7coefzt2}) are provided in (\ref{AppenA}). Observe that just for the sake of compactness, the following derivatives in (\ref{appb7}) are also defined
\begin{equation}
\begin{aligned}
& \triangle^\alpha_{(1)} A^\beta= \nabla^\alpha A^\beta-2A^\alpha A^\beta, \,\,\,\,\,\,\,\triangle^\alpha_{(2)} A^\beta= \nabla^\alpha A^\beta+A^\alpha A^\beta\\
&\widehat{\boxdot}^{\mu\nu}A^\alpha=\nabla^\mu \nabla^\nu A^\alpha- 2A^\alpha (\nabla^\mu A^\nu)-2A^\nu (\nabla^\mu A^\alpha)+A_\lambda R^{\mu\alpha\nu\lambda}, 
\end{aligned}  
\end{equation}
where $\widehat{\boxdot} A_\mu =\widehat{\boxdot}_\alpha{^\alpha} A_\mu$.

\subsection{Coefficient Functions of $\mathscr{O}_{(8)}$-Part:} \label{AppenI}
Finally, the explicit expressions of functions $\Theta_{k}^{[8]j} A_l $ (with $j=1\cdots 23$) associated with the term $\mathscr{O}_{(8)}$ in (\ref{o8partexplicit}) successively become as
\begingroup
\allowdisplaybreaks
\begin{align}
\Theta^{[8]1} A&=\square \Phi^2, \nonumber\\
\Theta^{[8]2}_\alpha A_\sigma&=\hat{\Omega}^{(1)}_{\alpha\sigma}+(1/2)[\Xi^{\beta\lambda}_{(5)} \, R_{\alpha\lambda\sigma\beta}+\Xi^{(8)}_{\alpha\sigma} \cdot R+\Xi^{(11)}_{\alpha\sigma}]+\Xi^{(6)}\, R_{\alpha\sigma}+[\Xi^{(7)}_{\alpha\beta}-(\zeta_9/2) F_{\alpha\beta}]R^\beta{_\sigma},\nonumber\\
\Theta^{[8]3} A&=-\widehat{\Omega}_{(2)}+\Xi_{(8)}^{\alpha\beta} R_{\alpha\beta}+2 \Xi_{(9)} R+\Xi_{(12)}, \nonumber\\
\Theta_{[8]4}^{\sigma\lambda} A^\beta&=-(\zeta_6/4) \nabla^\sigma F^{\lambda\beta}+\kappa_{1(8)} A^\sigma F^{\lambda\beta}+[\kappa_{2(8)} \nabla^\lambda A^\sigma-\zeta_{16} \nabla^\sigma A^\lambda] A^\beta+(\zeta_6/4) A_\gamma [R^{\gamma\sigma\lambda\beta}+2 R^{\gamma\lambda\sigma\beta} ]\nonumber\\
&+(\zeta_9/2)A^\lambda R^{\sigma\beta}, \nonumber\\
\Theta^{[8]5}_\sigma A_\beta&=\kappa_{17(8)}\nabla_\sigma A_\beta+\kappa_{18(8)}\nabla_\beta A_\sigma-(\zeta_{21}/2) A_\beta A_\sigma, \nonumber\\
\Theta^{[8]6}_\beta A_\sigma&=\kappa_{19(8)}\nabla_\beta A_\sigma+\kappa_{20(8)}\nabla_\sigma A_\beta-\kappa_{16(5)} A_\beta A_\sigma, \nonumber\\
\Theta^{[8]7}_{\alpha\beta} A_\lambda&=[\Theta^{[3]4}_\alpha A_\beta+(1/2)(\zeta_{21} A_\alpha A_\beta+\zeta_9 R_{\alpha\beta})] A_\lambda, \nonumber\\
\Theta_{[8]8}^{\alpha\beta\sigma} A^\lambda&=\kappa_{25(8)} A^\alpha \nabla^\beta F^{\sigma\lambda}+(\zeta_6/8) F^{\alpha\beta}F^{\sigma\lambda}+(1/2) [3 \kappa_{2(1)}\nabla^\beta A^\lambda+(\kappa_{1(1)}-\zeta_6) \nabla^\lambda A^\beta]\nabla^\alpha A^\sigma-\nonumber\\
&(\kappa_{3(1)}/2) \nabla^\sigma A^\alpha\, \nabla^\lambda A^\beta-2 [ \kappa_{99(7)}\nabla^\beta A^\lambda+(\zeta_{21}/4) \nabla^\lambda A^\beta ] A^\alpha A^\sigma-(\zeta_6/2) A_\gamma \nabla^\sigma R^{\alpha\lambda\beta\gamma}\nonumber\\
&+(\kappa_{2(1)}/2) A^\sigma A_\gamma R^{\alpha\beta\lambda\gamma}-[(\zeta_6/2) \nabla^\sigma A_\gamma+\kappa_{28(8)}A^\sigma A_\gamma]R^{\alpha\lambda\beta\gamma}+(\zeta_9/2) F^{\alpha\sigma} R^{\beta\lambda}, \nonumber\\
\Theta_{[8]9}^\alpha A^\beta&=-(\kappa_{29(8)}/\zeta_6)\, \Xi^{\alpha\beta}_{(5)}, \nonumber\\
\Theta_{[8]10}^{\alpha\beta} A^\sigma&=(\kappa_{4(5)}/2) \nabla^\alpha \nabla^\beta A^\sigma+\kappa_{31(8)} \nabla^\sigma\nabla^\alpha A^\beta-[\kappa_{4(5)}\nabla^\alpha A^\beta+(\zeta_{21}/2) A^\alpha A^\beta ] A^\sigma+[\kappa_{32(8)}\nabla^\beta A^\sigma-\nonumber\\
&\kappa_{18(7)}\nabla^\sigma A^\beta+(\zeta_9/2) R^{\beta\sigma} ]A^\alpha+(\kappa_{4(5)}/2) A_\lambda R^{\alpha\sigma\beta\lambda}, \nonumber\\
\Theta_{[8]11} A^\alpha&=-(\kappa_{12(5)}/2) \Theta_{[5]22} A^\alpha-\kappa_{7(5)} \nabla^\alpha \nabla \cdot A-2\kappa_{8(5)} A_\beta \nabla^\alpha A^\beta+[\kappa_{37(8)}\nabla \cdot A-(1/4) (\zeta_{23}A^2+\nonumber\\
&\zeta_{10}R)]A^\alpha, \nonumber\\
\Theta_{[8]12}^\alpha A^\beta&=-\kappa_{12(5)} \square \nabla^\alpha A^\beta+\kappa_{38(8)} \nabla^\alpha\nabla^\beta \nabla \cdot A+[-\kappa_{64(7)} \square A^\beta+\kappa_{40(8)}\nabla^\beta \nabla \cdot A] A^\alpha-2[\zeta_8 \nabla^\alpha \nabla^\beta A^\sigma+\nonumber\\
&(\zeta_9/2) \nabla^\sigma \nabla^\alpha A^\beta]A_\sigma-[((4\zeta_8+\kappa_{78(7)})/4) \nabla^\beta A_\sigma+\kappa_{11(5)}\nabla_\sigma A^\beta] \nabla^\alpha A^\sigma+(\Theta_{[1]3}^\alpha A^\beta) R\nonumber\\
&+((\kappa_{3(1)}-2\kappa_{64(7)})/2) \nabla^\sigma A^\alpha\,\nabla_\sigma A^\beta-[\kappa_{6(5)} \nabla \cdot A+(\zeta_{23}/2) A^2] \nabla^\alpha A^\beta-[\zeta_{23}\nabla^\beta A^\sigma+\nonumber\\
&(3\zeta_{21}/2) \nabla^\sigma A^\beta] A^\alpha A_\sigma-[\kappa_{44(8)} \nabla \cdot A+\zeta_{24}A^2]A^\alpha A^\beta+[\zeta_9 \nabla^\alpha A_\sigma+((\kappa_{3(1)}-2\zeta_9)/2) A^\alpha A_\sigma]R^{\sigma\beta}, \nonumber\\
\Theta_{[8]13} A&=((-\zeta_{10}+32\alpha_7)/2) \square \nabla \cdot A-4\zeta_{11} [A^\alpha \square A_\alpha+(\nabla_\alpha A_\beta)^2]+\zeta_{10}[A^\alpha \nabla_\alpha \nabla \cdot A+\nonumber\\
&(1/2) ( \nabla^\alpha A^\beta \, \nabla_\beta A_\alpha+ (\nabla \cdot A)^2)],\nonumber\\
\Theta_{[8]14}^\alpha A^\beta&=(\kappa_{2(1)}+2\kappa_{1(2)}) \nabla^\alpha A^\beta-\kappa_{98(7)}  \nabla^\beta A^\alpha -\kappa_{99(7)}A^\alpha A^\beta, \nonumber\\
\Theta_{[8]15}^\alpha A^\beta &=((\zeta_{20}-2\kappa_{17(5)})/2) \nabla^\alpha A^\beta-\kappa_{18(5)} \nabla^\beta A^\alpha-\kappa_{44(8)} A^\alpha A^\beta, \nonumber\\
\Theta_{[8]16} A^\alpha&=(\kappa_{2(1)}/2) \square A^\alpha+(-\kappa_{19(5)}+\zeta_{17}) \nabla^\alpha \nabla \cdot A-[\kappa_{122(7)} \nabla^\alpha A^\beta+\kappa_{99(7)} \nabla^\beta A^\alpha] A_\beta+\nonumber\\
&[((4\zeta_{29}+\kappa_{4(1)})/2) \nabla \cdot A+4\zeta_{30}A^2 ] A^\alpha, \nonumber\\
\Theta_{[8]17} A^\alpha&=((\kappa_{3(1)}-2\kappa_{115(7)})/2) \nabla^\alpha \nabla \cdot A-[\kappa_{114(7)} \nabla^\alpha A^\beta+2 \kappa_{44(8)} \nabla^\beta A^\alpha] A_\beta-[\kappa_{56(8)} \nabla \cdot A+\zeta_{24} A^2] A^\alpha, \nonumber\\
\Theta_{[8]18} A&=\kappa_{115(7)} \nabla \cdot A+\kappa_{116(7)} A^2, \nonumber\\
\Theta_{[8]19}^{\alpha\beta} A^\sigma&=-(\kappa_{14(5)}/2) \nabla^\alpha \nabla^\beta A^\sigma+((\zeta_{19}-8\zeta_{25}-\kappa_{2(1)})/2)\,\nabla^\sigma \nabla^\alpha A^\beta-2[\kappa_{99(7)} \nabla^\beta A^\sigma+(3\zeta_{21}/4)\nabla^\sigma A^\beta]A^\alpha\nonumber\\
&-[\kappa_{111(7)} \nabla^\alpha A^\beta+\zeta_{24} A^\alpha A^\beta]A^\sigma, \nonumber\\
\Theta_{[8]20} A&=((\kappa_{4(1)}+4 \zeta_{29})/2) \nabla \cdot A+4 \zeta_{30}A^2, \nonumber\\
\Theta_{[8]21} A&=\kappa_{44(8)} \nabla \cdot A+(\zeta_{24}/2) A^2, \nonumber\\
\Theta_{[8]22}^\alpha A_\sigma&=\kappa_{111(7)} \nabla^\alpha A_\sigma+(\zeta_{21}/2) \nabla_\sigma A^\alpha+2 \zeta_{24}A^\alpha A_\sigma, \nonumber\\
\Theta^{[8]23}_\alpha A_\beta&=\zeta_{24} [\nabla_\alpha A_\beta\, \nabla \cdot A+(1/2) \nabla_\alpha A_\sigma\,\nabla_\beta A^\sigma]-4\zeta_{30} \nabla_\sigma A_\alpha\,\nabla^\sigma A_\beta,
\label{appb8}
\end{align}
\endgroup
wherein the {\it pure} curvature terms $\hat{\Omega}^{(1)}_{\alpha\sigma}$ and $\widehat{\Omega}_{(2)}$ are given in (\ref{firstexpression}) and (\ref{secondexpression}), successively. As in previous terms, the specific terms $\Xi_{(i)}$ are given in (\ref{xiterms}), whereas the distinct values of the coupling constants $\kappa_{m(1)}$, $\kappa_{1(2)}$, $\kappa_{n(5)}$, $\kappa_{k(7)}$ and $\kappa_{l(8)}$ can be reached in (\ref{kapQ1}), (\ref{kapQ234}), (\ref{kapQ5}), (\ref{kapQ7}), and (\ref{kapQ8}), respectively.

\section{Coefficients $\kappa_{j(i)}$ of functions $\Theta_{k}^{[i]j} A_l$} 
In this part, we collect all the constant coefficients $\kappa_{j(i)}$ emerging during the calculations of field equations and respectively associated with the functions $\Theta_{k}^{[i]j} A_l $ (with $i=1\cdots9$) which are explicitly given in the previous sections. 

\subsection{$\kappa_{j(1)}$ of $\Theta_{k}^{[1]j} A_l $} \label{kapQ1}
 First of all, the coefficients $\kappa_{j(1)}$ (with $j=1\cdots4$) corresponding to the functionals $\Theta_{k}^{[1]j} A_l $ in (\ref{appb1}) belonging to the compact term $\mathscr{O}_{(1)}$-part are 
\begingroup
\allowdisplaybreaks
\begin{align}
\kappa_{1(1)}=2\zeta_{16}-\zeta_{17}-\zeta_{18}+\zeta_{19}, \quad \kappa_{2(1)}=-\zeta_{16}+\zeta_{17}, \quad \kappa_{3(1)}=-\zeta_{16}+\zeta_{18}, \quad \kappa_{4(1)}=\zeta_{21}+\zeta_{22}.
\end{align}
\endgroup

\subsection{$\kappa_{j(i)}$ of $\Theta_{k}^{[2]j} A_l $, $\Theta_{k}^{[3]j} A_l $ and $\Theta_{k}^{[4]j} A_l $} \label{kapQ234}
Secondly, the constant coefficients $\kappa_{1(2)}, \kappa_{1(3)}$ and $\kappa_{j(4)}$ (for $j=1, 2$) respectively relevant to the functions $\Theta_{k}^{[2]j} A_l $, $\Theta_{k}^{[3]j} A_l $ and $\Theta_{k}^{[4]j} A_l $ in (\ref{appb2}), (\ref{appb3}) and (\ref{appb4}) read as 
\begingroup
\allowdisplaybreaks
\begin{align}
\kappa_{1(2)}=2\zeta_{25}+\zeta_{26}, \quad \kappa_{1(3)}=\zeta_{23}+\zeta_{24}, \quad \kappa_{1(4)}=\zeta_2+18 \alpha_2, \quad \kappa_{2(4)}=-\zeta_4+36 \alpha_2=\zeta_5/2.
\end{align}
\endgroup

\subsection{$\kappa_{j(5)}$ of $\Theta_{k}^{[5]j} A_l $}\label{kapQ5}
Thirdly, the coefficients $\kappa_{j(5)}$ (where $j=1\cdots 26$) pertaining to the functions $\Theta_{k}^{[5]j} A_l $ in (\ref{appb5}) associated with the term $\mathscr{O}^{(5)}_{\mu}$ expressly turn out to be as follows 
\begingroup
\allowdisplaybreaks
\begin{align}
&\kappa_{1(5)}=-\zeta_6+\zeta_{18}, \quad \kappa_{2(5)}=\zeta_6-\zeta_{16}+\frac{\zeta_{17}}{2}+\frac{\zeta_{18}}{2}+\frac{\zeta_{19}}{2}, \quad\kappa_{3(5)}=\zeta_6+2\zeta_{10}, \quad \kappa_{4(5)}=\zeta_6+\zeta_9,\nonumber\\
&\kappa_{5(5)}=2\zeta_7-\frac{\zeta_9}{2}, \quad \kappa_{6(5)}=\zeta_9-\frac{\zeta_{20}}{2}, \quad \kappa_{7(5)}=\zeta_7-32 \alpha_7, \quad \kappa_{8(5)}=\zeta_8+4\zeta_{11},\quad \kappa_{9(5)}=-\frac{\zeta_9}{2}+\nonumber\\
&2\zeta_{10},\quad \kappa_{10(5)}=-4\zeta_{10}-\frac{\kappa_{1(1)}}{2}, \quad \kappa_{11(5)}=\zeta_9-\kappa_{3(1)}, \quad \kappa_{12(5)}=-\frac{\zeta_9}{2}+\zeta_{10},\quad \kappa_{13(5)}=\frac{\zeta_{10}}{2}-\nonumber\\
&32 \alpha_7, \quad \kappa_{14(5)}=-\kappa_{2(1)}-8\zeta_{25}-4\zeta_{26},\quad \kappa_{15(5)}=-\frac{\kappa_{1(1)}}{2}+8\zeta_{25},\quad \kappa_{16(5)}=-\frac{\kappa_{4(1)}}{2}+2\zeta_{27},\nonumber\\
& \kappa_{17(5)}=-\zeta_{16}+\frac{\zeta_{17}}{2}-\frac{\zeta_{19}}{2},\quad \kappa_{18(5)}=\zeta_{16}-\frac{\zeta_{18}}{2},\quad \kappa_{19(5)}=-\frac{(\kappa_{1(1)}+\zeta_{20})}{2},\quad \kappa_{20(5)}=-\kappa_{3(1)}-\nonumber\\
&\frac{\zeta_{20}}{2},\quad \kappa_{21(5)}=\zeta_{20}+\frac{\zeta_{21}}{2},\quad \kappa_{22(5)}=\frac{\zeta_{20}}{2}+4\zeta_{28},\quad \kappa_{23(5)}=-\frac{\zeta_{23}}{2}+2\zeta_{29},\quad \kappa_{24(5)}=\zeta_{23}+2\zeta_{27},\nonumber\\
&\kappa_{25(5)}=\frac{\kappa_{4(1)}}{2}+4\zeta_{29},\quad \kappa_{26(5)}=-\frac{\kappa_{4(1)}}{2}+\zeta_{23}.
\end{align}
\endgroup

\subsection{$\kappa_{j(6)}$ of $\Theta_{k}^{[6]j} A_l $}\label{kapQ6}
As for those of the functionals $\Theta_{k}^{[6]j} A_l $ in (\ref{appb6}) belonging to the term $\mathscr{O}_{(6)\mu\nu}^\alpha$, the corresponding coefficients $\kappa_{j(6)}$ (with $j=1\cdots29$) successively become as follows 
\begingroup
\allowdisplaybreaks
\begin{align}
&\kappa_{1(6)}=\zeta_{16}-\kappa_{2(1)}, \,\,\,\, \kappa_{2(6)}=\zeta_{16}-\kappa_{3(1)}, \,\,\,\,\, \kappa_{3(6)}=\zeta_6+4\zeta_7, \,\,\,\, \,\kappa_{4(6)}=\zeta_{17}-\zeta_{19}, \,\,\,\, \kappa_{5(6)}=\zeta_4-\nonumber\\
&\frac{\zeta_9}{2}-36 \alpha_2, \quad \kappa_{6(6)}=\zeta_5+\zeta_9, \,\,\, \kappa_{7(6)}=2\zeta_4+\zeta_6-72 \alpha_2, \quad \kappa_{8(6)}=\zeta_5-\zeta_6, \quad \kappa_{9(6)}=8\zeta_1+\frac{\zeta_5}{2},\nonumber\\
& \kappa_{10(6)}=4\zeta_1+\zeta_7, \quad \kappa_{11(6)}=4\zeta_2-\zeta_8+72 \alpha_2, \quad \kappa_{12(6)}=\zeta_6-\zeta_{19}, \quad \kappa_{13(6)}=4 \zeta_6+\zeta_{18}-\zeta_{19},\nonumber\\
&\kappa_{14(6)}=2\zeta_5+\zeta_{18},\quad \kappa_{15(6)}=2\zeta_6+\zeta_9,\quad \kappa_{16(6)}=\zeta_9+\zeta_{14},\quad \kappa_{17(6)}=\zeta_6-\zeta_9,\quad \kappa_{18(6)}=\zeta_6+\nonumber\\
&2\zeta_9, \quad \kappa_{19(6)}=\zeta_9+2\zeta_{12}-\zeta_{13}-\zeta_{14}+72\alpha_2, \quad \kappa_{20(6)}=2(2\zeta_{12}-\zeta_{13}-\zeta_{14}+72\alpha_2)-\zeta_{20},\nonumber\\
&\quad \kappa_{21(6)}=-2(2\zeta_{12}-\zeta_{13}+72\alpha_2)-\zeta_{20}, \quad \kappa_{22(6)}=\zeta_{15}+2\zeta_{20}, \quad \kappa_{23(6)}=-2\zeta_{12}+\zeta_{13}-72\alpha_2\nonumber\\
&+\zeta_{19}, \quad \kappa_{24(6)}=2\zeta_{14}+\zeta_{18}-\zeta_{19}, \quad \kappa_{25(6)}=2\zeta_{14}-\zeta_{18}, \quad \kappa_{26(6)}=\zeta_{14}+\zeta_{20}, \quad \kappa_{27(6)}=2\zeta_{16}+\nonumber\\
&\zeta_{17}, \quad \kappa_{28(6)}=\kappa_{4(1)}+4\zeta_{23},\quad \kappa_{29(6)}=2\zeta_6+2\zeta_{12}-\zeta_{13}+72\alpha_2.
\end{align}
\endgroup

\subsection{$\kappa_{j(7)}$ of $\Theta_{k}^{[7]j} A_l $}\label{kapQ7}
Next, the constant coefficients $\kappa_{j(7)}$ (where $j=1\cdots 141$) associated to the coefficient functionals $\Theta_{k}^{[7]j} A_l $ in (\ref{appb7}) of the compact term $\mathscr{O}^{(7)}_{\mu\nu}$ are sequentially described as  
\begingroup
\allowdisplaybreaks
\begin{align}
&\kappa_{1(7)}=\frac{\zeta_5}{2}-\zeta_6, \quad\,\,\,\, \kappa_{2(7)}=2\zeta_6+\zeta_{16}-\frac{\kappa_{1(1)}}{2}, \quad\,\,\, \kappa_{3(7)}=-\kappa_{3(1)}-\zeta_5, \quad \,\,\,\kappa_{4(7)}=\zeta_6-\frac{\kappa_{1(1)}}{2}\nonumber\\
&\kappa_{5(7)}=-\frac{\zeta_5}{2} -\kappa_{4(5)}, \quad \kappa_{6(7)}=\zeta_6+\zeta_{10},\quad \kappa_{7(7)}=2\zeta_1-15\zeta_7,\quad \kappa_{8(7)}=-2\kappa_{1(4)}+\zeta_8+2\zeta_{11},\nonumber\\
&\kappa_{9(7)}=\frac{\kappa_{5(6)}}{2}+\frac{\kappa_{3(5)}}{4}), \quad\,\, \kappa_{10(7)}=-\frac{\kappa_{7(6)}}{2}-\kappa_{12(5)},\quad\,\, \kappa_{11(7)}=-\frac{\kappa_{7(6)}}{2}-2\kappa_{10(6)}, \quad\,\, \kappa_{12(7)}=\nonumber\\
&16(\kappa_{1(4)}+\frac{\zeta_8}{4}), \quad \kappa_{13(7)}= -\kappa_{5(6)}-\frac{\kappa_{7(6)}}{2}, \quad \kappa_{14(7)}=\kappa_{4(5)}-\frac{\kappa_{7(6)}}{2},\quad \kappa_{15(7)}=-2\kappa_{6(6)}-2\zeta_{12}+\nonumber\\
&\zeta_{13}-72\alpha_2,\,\,\,\, \kappa_{16(7)}=\kappa_{4(5)}+2 \zeta_{12}-\zeta_{13}-\zeta_{14}+72\alpha_2, \,\,\,\, \kappa_{17(7)}=\kappa_{3(5)}-\frac{\kappa_{6(6)}}{2},\quad \kappa_{18(7)}=-2\zeta_6\nonumber\\
&+\kappa_{10(5)},\quad \kappa_{19(7)}=\kappa_{6(6)}-\kappa_{3(1)}, \,\,\,\, \kappa_{20(7)}=-\kappa_{11(7)}-\frac{\kappa_{4(5)}}{2}, \,\,\,\, \kappa_{21(7)}=-\kappa_{12(7)}, \,\,\,\, \kappa_{22(7)}=\zeta_5+\nonumber\\
&\kappa_{6(5)}, \,\,\,\, \kappa_{23(7)}=\frac{\kappa_{15(6)}+2\zeta_{10}}{4}, \,\,\,\, \kappa_{24(7)}=-\kappa_{12(5)}+\frac{(\zeta_{13}+\zeta_{14})}{4}-(3\alpha_2+8\alpha_5+4\alpha_6), \,\,\,\, \kappa_{25(7)}=\nonumber\\
&\frac{\kappa_{3(5)}}{4}+\kappa_{7(5)}, \quad \kappa_{26(7)}=-4\zeta_1+\frac{\zeta_5}{2}, \quad \kappa_{27(7)}=\kappa_{20(5)}-\zeta_5, \quad \kappa_{28(7)}=8\kappa_{1(4)}+\zeta_{17}, \quad \kappa_{29(7)}=\nonumber\\
&-\zeta_5+2\zeta_{16}-\frac{1}{2} (\zeta_{17}+3\zeta_{18}-\zeta_{19}), \quad \kappa_{30(7)}=\zeta_6+\frac{1}{2} (\zeta_{17}+\zeta_{18}-\zeta_{19}), \quad \kappa_{31(7)}=\frac{\kappa_{4(1)}}{2}+\zeta_{23},\nonumber\\
& \kappa_{32(7)}=-2(2\zeta_1+\zeta_7), \quad \kappa_{33(7)}=\kappa_{11(6)}-\zeta_8, \quad \kappa_{34(7)}=-\kappa_{5(6)}+ \frac{\zeta_5}{2}, \quad \kappa_{35(7)}=\kappa_{3(1)}-\zeta_6+\nonumber\\
&2\zeta_{12}-\zeta_{13}+72\alpha_2, \quad \kappa_{36(7)}=-\kappa_{4(7)}-8(9\alpha_2+4\alpha_5+3\alpha_6), \quad \kappa_{37(7)}=-\kappa_{1(1)} +\zeta_5+2\zeta_{12}-\nonumber\\
&\zeta_{13}+72\alpha_2, \quad \kappa_{38(7)}=\kappa_{3(1)}-\zeta_6+2(57\alpha_2+8\alpha_5+4\alpha_6), \quad \kappa_{39(7)}=\kappa_{38(7)}+216\alpha_2+\zeta_5+\zeta_6,\nonumber\\
&\kappa_{40(7)}=-3\zeta_{12}+2\zeta_{13}-\zeta_{16}+3\zeta_{17}+4(9\alpha_2+4\alpha_5+3\alpha_6), \quad \kappa_{41(7)}=4(15\alpha_2+9\alpha_3-8\alpha_5+ \nonumber\\
&15 \alpha_6+24\alpha_8),\quad \kappa_{42(7)}=2(21\alpha_2-6\alpha_3+32\alpha_5+14\alpha_6+48\alpha_8),\quad \kappa_{43(7)}=2(-45\alpha_2-24\alpha_3\nonumber\\
&+16\alpha_5-38\alpha_6+24\alpha_8), \quad \kappa_{44(7)}=-\frac{(\zeta_{15}+\zeta_{21})}{2}, \quad \kappa_{45(7)}=-\zeta_3+\frac{\zeta_5}{2}, \quad \kappa_{46(7)}=\zeta_3-36\alpha_2, \nonumber\\
& \kappa_{47(7)}=-2\zeta_3+\frac{1}{2}(\zeta_5+\zeta_6),\quad \kappa_{48(7)}=-2(\zeta_3+\zeta_4)+\frac{\zeta_6}{2}, \quad \kappa_{49(7)}=\zeta_3-\zeta_6, \quad \kappa_{50(7)}=-\zeta_3+\nonumber\\
&\zeta_4-\frac{\zeta_6}{2}+72\alpha_2, \quad \kappa_{51(7)}=-\zeta_5+2\zeta_6+4(9\alpha_2+4\alpha_5+3\alpha_6), \quad \kappa_{52(7)}=-2(\zeta_4+18\alpha_2), \nonumber\\
& \kappa_{53(7)}=-2(\zeta_4-18\alpha_2), \,\,\,\, \kappa_{54(7)}=-\frac{\zeta_3}{2}+\zeta_4-18\alpha_2, \,\,\,\, \kappa_{55(7)}=\frac{\zeta_3-\zeta_4+\zeta_6}{2}, \,\,\,\,\kappa_{56(7)}=-\zeta_3-\nonumber\\
&\frac{\zeta_6}{2}+108\alpha_2, \,\,\,\, \kappa_{57(7)}=-\frac{(\zeta_3-\zeta_6)}{2}+\zeta_4+18\alpha_2,\,\,\,\, \kappa_{58(7)}=\frac{(\zeta_3+3\zeta_4)}{2}-36\alpha_2,\,\,\,\, \kappa_{59(7)}=-\frac{\zeta_5}{2}\nonumber\\
&+\frac{3\zeta_9}{2}-\zeta_6,\quad \kappa_{60(7)}=\kappa_{59(7)}-\zeta_9,\,\,\,\, \kappa_{61(7)}=\zeta_5+ 2\kappa_{17(6)},\quad \kappa_{62(7)}=-\frac{(3\zeta_6+\zeta_9)}{2},\quad  \kappa_{63(7)}=\nonumber\\
&-2(33\alpha_2+32\alpha_5+10\alpha_6+24\alpha_8), \quad \kappa_{64(7)}=-\frac{\kappa_{1(1)}}{2}-2\zeta_{10}, \quad \kappa_{65(7)}=\kappa_{1(1)}-\kappa_{3(1)}+\zeta_9,\nonumber \\
& \kappa_{66(7)}=\zeta_9+\zeta_{20}+\frac{(\zeta_{13}+\zeta_{14})}{2}-2(3\alpha_2+8\alpha_5+4\alpha_6),  \quad \kappa_{67(7)}=\zeta_9+\zeta_{14}-\kappa_{3(1)}, \quad \kappa_{68(7)} =\nonumber\\
&-\kappa_{29(7)}+\zeta_5-2\zeta_{10}+4(9\alpha_2+4\alpha_5+3\alpha_6),\quad \kappa_{69(7)}=-\zeta_{15}+\kappa_{4(1)},\quad \kappa_{70(7)}=-\frac{\zeta_{15}}{2}-\zeta_{23},\nonumber\\
& \kappa_{71(7)}=-\frac{\kappa_{22(6)}}{2}+\frac{\zeta_{21}}{2},\quad \kappa_{72(7)}=-\frac{(\zeta_5-\zeta_9)}{2},\quad  \kappa_{73(7)}=\zeta_5-\kappa_{4(5)}, \quad \kappa_{74(7)}=\frac{\zeta_6}{2}+\kappa_{5(5)},\nonumber\\
& \kappa_{75(7)}=-\zeta_9-2\zeta_{12}+\zeta_{13}-72\alpha_2, \quad \kappa_{76(7)}=\kappa_{6(5)}+8(9\alpha_2+4\alpha_5+3\alpha_6),\quad \kappa_{77(7)}=\frac{\kappa_{1(1)}}{2}+\nonumber\\
&\kappa_{19(6)},\,\,\, \kappa_{78(7)}=2(\kappa_{2(1)}+2\zeta_8), \,\,\, \kappa_{79(7)}=\kappa_{3(1)}-2\zeta_9-2\zeta_{12}+\zeta_{13}-72\alpha_2,\,\,\, \kappa_{80(7)}=-\zeta_6+\kappa_{1(1)},\nonumber\\
& \kappa_{81(7)}=-\zeta_6+\zeta_{20},\quad \kappa_{82(7)}=\frac{\zeta_{15}-\zeta_{21}}{2},\,\,\,\, \kappa_{83(7)}=\frac{\zeta_{15}}{2}-\zeta_{20},\,\,\,\, \kappa_{84(7)}=-\frac{\zeta_6}{2}-\zeta_7,\quad \kappa_{85(7)}=\nonumber\\
&\frac{\zeta_{10}}{2}-16\alpha_7,\quad \,\,\,\kappa_{86(7)}=-\zeta_{10}-4 \zeta_{25},\quad \,\,\,\kappa_{87(7)}=-\zeta_{10}+2 \zeta_{28},\quad \,\,\, \kappa_{88(7)}=\zeta_{10}+2 \zeta_{25}+\zeta_{26},\nonumber\\
&\kappa_{89(7)}=4\zeta_{11}+2 \zeta_{25}+\zeta_{26},\,\,\, \kappa_{90(7)}=-\frac{(\kappa_{1(1)}+\zeta_{20})}{2}+2\zeta_{12}-\zeta_{13}-\zeta_{14}+72\alpha_2,\,\,\, \kappa_{91(7)}=-\kappa_{3(1)}\nonumber\\
&-\frac{\zeta_{20}}{2}-2\zeta_{12}+\zeta_{13}-72\alpha_2,\,\, \kappa_{92(7)}=\frac{(\zeta_{15}+\zeta_{21})}{2}+\zeta_{20},\,\, \kappa_{93(7)}=\kappa_{29(7)}+\zeta_5-2\zeta_{12}+\zeta_{13}-72\alpha_2,\nonumber\\
& \kappa_{94(7)}=\kappa_{3(1)}+2(15\alpha_2+8\alpha_5+6\alpha_6),\quad \,\,\kappa_{95(7)}=\frac{\zeta_{15}}{2}+\zeta_{21},\quad \,\,\kappa_{96(7)}=\kappa_{18(5)}+\frac{\kappa_{4(6)}}{2}+\zeta_{14},\nonumber\\
& \quad \kappa_{97(7)}=-\zeta_{17}-4 \zeta_{25}-2\zeta_{26},\quad  \kappa_{98(7)}=-\frac{\kappa_{1(1)}}{2}+ 4\zeta_{25},\quad \kappa_{99(7)}=-\frac{\kappa_{4(1)}}{2}+\zeta_{27},\quad \kappa_{100(7)}=\nonumber\\
&-\kappa_{3(1)}+\zeta_{14}+\frac{\zeta_{20}}{2},\,\,\,\, \kappa_{101(7)}=\kappa_{92(7)}-\frac{3\zeta_{15}}{2},\,\,\,\, \kappa_{102(7)}=-\frac{\zeta_{15}}{2}-2\zeta_{20}, \,\,\, \kappa_{103(7)}=\frac{\kappa_{1(1)}}{2}-2\zeta_{12}+\nonumber\\
&\zeta_{13}-72\alpha_2, \,\,\,\, \kappa_{104(7)}=\kappa_{3(1)}+2\zeta_{12}-\zeta_{13}-\zeta_{14}+72\alpha_2,\,\,\,\, \kappa_{105(7)}=-\frac{\kappa_{1(1)}}{2}+\zeta_{14}, \,\,\,\, \kappa_{106(7)}=-\zeta_{15}\nonumber\\
&-\frac{\zeta_{21}}{2}, \,\,\,\, \kappa_{107(7)}=\frac{\zeta_{20}}{2}+\frac{(\zeta_{13}+\zeta_{14})}{2}-2(3\alpha_2+8\alpha_5+4\alpha_6), \,\,\,\, \kappa_{108(7)}=-\frac{(\zeta_{15}+\zeta_{20})}{2}, \,\,\,\, \kappa_{109(7)}=\nonumber\\
&-2 \kappa_{1(2)}, \,\,\,\, \kappa_{110(7)}= \kappa_{98(7)}+4(9\alpha_2+4\alpha_5+3\alpha_6), \,\,\, \kappa_{111(7)}= \kappa_{16(5)}+\zeta_{23}, \,\,\, \kappa_{112(7)}= -\kappa_{4(1)}+\zeta_{27}, \nonumber\\
& \kappa_{113(7)}=-\frac{\zeta_{20}}{2}-2\zeta_{28}+4(9\alpha_2+4\alpha_5+3\alpha_6),\quad \,\,\kappa_{114(7)}= -\kappa_{25(5)}+\zeta_{23},\quad\,\, \kappa_{115(7)}=\kappa_{113(7)}-\nonumber\\
&4(9\alpha_2+4\alpha_5+3\alpha_6), \quad \kappa_{116(7)}=\frac{\zeta_{23}}{2}-\zeta_{29}, \quad \kappa_{117(7)}= -\kappa_{31(7)}+2\zeta_{31},\quad \kappa_{118(7)}= -\zeta_{24}+2\zeta_{32}, \nonumber\\
&\kappa_{119(7)}= -\kappa_{82(7)}-2\zeta_{31}+\zeta_{38},\,\,\,\, \kappa_{120(7)}=2(\zeta_4+\zeta_{24}), \,\,\,\, \kappa_{121(7)}=\zeta_{24}-2\zeta_{41},\,\,\,\, \kappa_{122(7)}= \zeta_{23}+\zeta_{27},\nonumber\\
& \kappa_{123(7)}= -\frac{\kappa_{4(1)}}{2}-2\zeta_{39},\quad \kappa_{124(7)}=\frac{\zeta_{15}}{2}+\zeta_{21}+3\zeta_{33}, \quad \kappa_{125(7)}=\kappa_{111(7)}-\zeta_{33}+\zeta_{36},\quad \kappa_{126(7)}=\nonumber\\
&\kappa_{31(7)}-2\zeta_{33}+\zeta_{36},\,\, \kappa_{127(7)}=\kappa_{131(7)}-\zeta_{37},\,\,\kappa_{128(7)}=2(\zeta_{24}+\zeta_{34}), \,\, \kappa_{129(7)}=\kappa_{92(7)}-4\zeta_{31}+2\zeta_{38},\nonumber\\
& \kappa_{130(7)}=2\kappa_{116(7)}+\kappa_{117(7)},\,\,\,\kappa_{131(7)}=2(\zeta_{24}-\zeta_{34})+\zeta_{35}, \,\,\, \kappa_{132(7)}=-8\zeta_{30}+\zeta_{34}-\zeta_{35},\,\,\, \kappa_{133(7)}=\nonumber\\
&\zeta_{24}-4\zeta_{32}+2\zeta_{39},\,\,\,\, \kappa_{134(7)}=-2(2\zeta_{30}-\zeta_{32}), \,\,\,\, \kappa_{135(7)}=-\kappa_{95(7)}-\zeta_{33}+\zeta_{36}, \,\,\,\, \kappa_{136(7)}=-2\zeta_{34}+\nonumber\\
&\zeta_{35}-\zeta_{37},\quad \kappa_{137(7)}=-2(\kappa_{131(7)}+\zeta_{34}), \quad \kappa_{138(7)}=\kappa_{83(7)}+2\zeta_{31}, \quad \kappa_{139(7)}=-2\zeta_{32}+\zeta_{39},\nonumber\\
& \kappa_{140(7)}=-\kappa_{83(7)}-3\zeta_{40},\quad  \kappa_{141(7)}=-\zeta_{41}-3\zeta_{43}.
\end{align}
\endgroup

\subsection{$\kappa_{j(8)}$ of $\Theta_{k}^{[8]j} A_l $}\label{kapQ8}
Lastly, the coefficients $\kappa_{j(8)}$ (with $j=1\cdots61$) corresponding to the functionals $\Theta_{k}^{[8]j} A_l $ in (\ref{appb8}) belonging to the term $\mathscr{O}_{(8)}$ respectively become as below
\begingroup
\allowdisplaybreaks
\begin{align}
 &\kappa_{1(8)}=\frac{2\zeta_6-\zeta_{18}}{4}, \quad \,\, \kappa_{2(8)}=\kappa_{30(7)}+\zeta_{16}, \quad \,\, \kappa_{3(8)}=\kappa_{3(5)}, \quad \,\, \kappa_{4(8)}=\frac{\kappa_{4(5)}}{2}, \quad \,\,\kappa_{5(8)}=-\kappa_{7(5)}, \nonumber\\
&\kappa_{6(8)}=-\kappa_{12(5)}, \quad \,\, \kappa_{7(8)}=-\kappa_{5(5)},\quad\,\, \kappa_{8(8)}=-\kappa_{6(5)},\quad\,\, \kappa_{9(8)}=-\kappa_{9(5)},\quad \,\, \kappa_{10(8)}=-\kappa_{10(5)}, \nonumber\\
& \kappa_{11(8)}=\kappa_{11(5)}, \quad\,\, \kappa_{12(8)}=-\kappa_{13(5)},\quad \,\,\kappa_{13(8)}=-\kappa_{19(5)},\quad \,\,\kappa_{14(8)}=-\kappa_{20(5)},\quad\,\, \kappa_{15(8)}=-\kappa_{21(5)},\nonumber\\
&\kappa_{16(8)}=\kappa_{22(5)},\quad \,\, \kappa_{17(8)}=-\frac{\kappa_{4(6)}}{2}-\zeta_{16},\quad\,\, \kappa_{18(8)}=\zeta_{16}+\frac{\zeta_{18}}{2},\quad \,\,\kappa_{19(8)}=4\kappa_{1(2)}+\zeta_{16}+\zeta_{17},\nonumber\\
& \kappa_{20(8)}=-\kappa_{2(8)}+\zeta_{6}-8\zeta_{25},\,\, \kappa_{21(8)}=-\kappa_{16(5)},\,\,\kappa_{22(8)}=\kappa_{25(5)},\,\, \kappa_{23(8)}=-\kappa_{24(5)},\,\, \kappa_{24(8)}=-\kappa_{26(5)},\nonumber\\
&\kappa_{25(8)}=-\frac{\zeta_6}{2}+2\zeta_{25},\quad\,\,\,\, \kappa_{26(8)}=\frac{\kappa_{80(7)}}{2},\quad \,\, \,\,\kappa_{27(8)}=-2\kappa_{99(7)},\quad \,\,\,\, \kappa_{28(8)}=\frac{\zeta_6+\zeta_{19}}{2}-4\zeta_{25},\nonumber\\
&\kappa_{29(8)}=\frac{\kappa_{18(6)}-2\zeta_{10}}{4},\quad \,\,\kappa_{30(8)}=\kappa_{4(5)},\quad\,\, \kappa_{31(8)}=-\frac{\kappa_{17(6)}}{2}-2\zeta_{10},\quad \,\,\kappa_{32(8)}=-\kappa_{4(5)}+\kappa_{3(1)}, \nonumber\\
& \kappa_{33(8)}=-\kappa_{18(7)},\,\,\,\kappa_{34(8)}=-\kappa_{7(5)}-\frac{\kappa_{12(5)}}{2},\,\,\, \kappa_{35(8)}=2\kappa_{8(5)},\,\,\, \kappa_{36(8)}=\kappa_{23(5)},\,\,\,\kappa_{37(8)}=\kappa_{12(5)}+\nonumber\\
&\frac{\zeta_{20}}{4},\quad \kappa_{38(8)}=-\zeta_7+\frac{\zeta_9}{2},\,\,\,\, \kappa_{39(8)}=-\kappa_{64(7)},\,\,\,\,\kappa_{40(8)}=-\kappa_{6(5)}+\kappa_{3(1)},\quad \kappa_{41(8)}=-\frac{\kappa_{78(7)}}{4}-\zeta_{8},\nonumber\\
& \kappa_{42(8)}=-\kappa_{64(7)}+\frac{\kappa_{3(1)}}{2},\,\,\, \kappa_{43(8)}=-\kappa_{6(5)},\,\,\, \kappa_{44(8)}=\frac{\zeta_{20}+\zeta_{21}}{2},\,\,\,\, \kappa_{45(8)}=\frac{\kappa_{3(1)}}{2}-\zeta_{9},\,\,\,\, \kappa_{46(8)}=\nonumber\\
&-\frac{\zeta_{10}}{2}+16 \alpha_7,\quad\,\,\,\, \kappa_{47(8)}=\kappa_{2(1)}+2\kappa_{1(2)},\quad \,\,\,\, \kappa_{48(8)}=-\kappa_{98(7)},\quad\,\,\,\, \kappa_{49(8)}=-\kappa_{17(5)}+\frac{\zeta_{20}}{2},\nonumber\\
&\kappa_{50(8)}=-\kappa_{18(5)},\quad \,\,\,\,\kappa_{51(8)}=-\kappa_{19(5)}+\zeta_{17},\quad \,\,\,\,\,\, \kappa_{52(8)}=\kappa_{122(7)},\quad \,\,\,\, \,\,\kappa_{53(8)}=\frac{\kappa_{4(1)}}{2}+2\zeta_{29},\nonumber\\
&\kappa_{54(8)}=\frac{\kappa_{3(1)}}{2}-\kappa_{115(7)},\quad \,\, \,\, \kappa_{55(8)}=-\kappa_{114(7)},\quad \,\, \,\, \kappa_{56(8)}=2\zeta_{20}+\frac{\zeta_{21}}{2},\quad \,\, \,\, \kappa_{57(8)}=-\kappa_{115(7)},\nonumber\\
&\kappa_{58(8)}=-\kappa_{116(7)},\quad \kappa_{59(8)}=-\frac{\kappa_{14(5)}}{2},\quad  \kappa_{60(8)}=-\frac{(\kappa_{2(1)}-\zeta_{19})}{2}-4\zeta_{25},\quad \kappa_{61(8)}=-\kappa_{111(7)}.
\end{align}
\endgroup

\section{Coefficient Functionals $\mathscr{X}_{[A]i}$'s of Gauge Field Equation}\label{functionalsofvectorfields}
The explicit forms of coefficient functionals $\mathscr{X}_{[A]i}$ (with $i=1\cdots 11$) belonging to the gauge field equation in (\ref{vectfeqcmpc}) are described successively as follows
\begingroup
\allowdisplaybreaks
\begin{align}
\mathscr{X}_{[A]1}&=\zeta_1 R^2_{\alpha\beta\nu\lambda}+\zeta_7 R^2_{\alpha\beta}-8 \alpha_7 R^2 -(\Theta_{[1]6}^\nu A^\alpha)R_{\nu\alpha}-[2\zeta_{28} \nabla \cdot A +\zeta_{29} A^2] R,\nonumber\\
\mathscr{X}_{[A]2}&=-4 \zeta_{31} (\nabla_{(\alpha} A_{\beta)}) \nabla^\alpha A^\beta-[\zeta_{38} \nabla^\nu A^\alpha+2 \zeta_{24}A^\nu A^\alpha] \nabla_\alpha A_\nu+3 \zeta_{40} (\nabla \cdot A)^2+[2\zeta_{41} \nabla \cdot A \nonumber\\
&+\zeta_{42} A^2] A^2, \nonumber \\
\mathscr{X}^{[A]3}_{\mu\nu}&=-[\zeta_4 R_{\alpha\beta\lambda\nu}+72 \alpha_2 R_{\lambda\alpha\beta\nu}] R_\mu{^{\lambda\alpha\beta}}+\widehat{L}^{\alpha\beta} R_{\mu\alpha\nu\beta}+72 \alpha_2 F^{\alpha\beta} R_{\mu\nu\alpha\beta}+2[\Theta^\alpha_{[5]10}A_\mu -\frac{\zeta_9}{2} R_\mu{^\alpha}] R_{\alpha\nu}\nonumber\\
&-(\Theta^{[6]2} A) R_{\mu\nu}-\widehat{{K}}^\alpha{_\nu} R_{\mu\alpha}+[4\zeta_{25} \nabla_\mu A_\nu+\kappa_{A7} \nabla_\nu A_\mu+\zeta_{27} A_\mu A_\nu]R, \nonumber\\
\mathscr{X}^{[A]4}_{\mu\nu}&=[\kappa_{A8} \nabla \cdot A+\kappa_{A9} A^2] \nabla_\mu A_\nu-4 \Xi_{(13)} \nabla_\nu A_\mu-[3\zeta_{33} \nabla_\mu A_\alpha-\kappa_{A10} \nabla_\alpha A_\mu -\kappa_{A11}A_\mu A_\alpha] \nabla^\alpha A_\nu \nonumber\\
&+[2\kappa_{A10} \nabla_{(\mu}A_{\alpha)}-\kappa_{A12}A_\mu A_\alpha]\nabla_\nu A^\alpha+[\kappa_{A11} \nabla_\mu A^\alpha-2\zeta_{34} \nabla^\alpha A_\mu] A_\nu A_\alpha-\Xi^{(18)} A_\mu A_\nu \nonumber\\
\mathscr{X}^{[A]5}_\mu&=[2\zeta_1 \nabla_\mu R^{\alpha\beta\nu\lambda}-\zeta_4 \nabla^\lambda R_\mu{^{\nu\alpha\beta}}-72\alpha_2 \nabla^\lambda R_\mu{^{\alpha\beta\nu}}+2A_\mu (\zeta_2 R^{\alpha\beta\nu\lambda}+36 \alpha_2 R^{\alpha\nu\beta\lambda})\nonumber\\
&-2\zeta_5 A^\lambda R_\mu{^{\alpha\beta\nu}}]R_{\alpha\beta\nu\lambda}-[\zeta_4 R_\mu{^{\nu\alpha\beta}}+72\alpha_2 R_\mu{^{\alpha\beta\nu}}] \nabla^\lambda R_{\alpha\beta\nu\lambda}+144\alpha_2 (\nabla^\nu A^\alpha) \nabla^\beta R_{\mu\beta\nu\alpha}\nonumber\\
&+[\kappa_{A13} \nabla^\nu \nabla^\alpha A^\beta+\kappa_{A14} A^\alpha \nabla^\nu A^\beta-\zeta_{15} A^\nu (\nabla^\alpha A^\beta+2\nabla^\beta A^\alpha+\frac{\kappa_{A15}}{\zeta_{15}}R^{\alpha\beta})+2A_\lambda (72 \alpha_2 R^{\alpha\nu\beta\lambda}\nonumber\\
&+\kappa_{A2} R^{\nu\beta\alpha\lambda})+\zeta_6 \nabla^\nu R^{\alpha\beta}]R_{\mu\alpha\nu\beta}-\widehat{L}^{\alpha\nu} \nabla^\beta R_{\mu\alpha\nu\beta}, \nonumber\\
\mathscr{X}^{[A]6}_\mu&=-[\Theta_{[1]6}^\nu A^\alpha-2\zeta_7 R^{\nu\alpha}] \nabla_\mu R_{\nu\alpha}+[\kappa_{A16}\nabla_\mu \nabla_\nu A_\alpha-\kappa_{A4} \nabla_\nu \nabla_\alpha A_\mu+A_\nu(\kappa_{A17}\nabla_\mu A_\alpha+\kappa_{A18}\nabla_\alpha A_\mu \nonumber\\
&+\kappa_{A19}A_\mu A_\alpha+\kappa_{A20} R_{\mu\alpha})-A_\mu (\kappa_{A21} \nabla_\nu A_\alpha+2 \zeta_8 R_{\nu\alpha})]R^{\nu\alpha}-[\widehat{K}^{\nu\alpha}+\zeta_9 R^{\nu\alpha}] \nabla_\alpha R_{\mu\nu} \nonumber\\
&+[-\kappa_{A6} \square A^\nu+\kappa_{A16} \nabla^\nu \nabla \cdot A+A^\nu (\kappa_{A22}\nabla \cdot A+\kappa_{A23}A^2+\kappa_{A24}R)+\kappa_{A25}A_\alpha \nabla^\nu A^\alpha \nonumber\\
&+\kappa_{A26}\nabla^\nu R]R_{\mu\nu}, \nonumber\\
\mathscr{X}^{[A]7}_\mu&=[\Theta_{[7]33}A+\kappa_{A29}R] \nabla_\mu R+[\kappa_{A7} \square A_\mu+\kappa_{A30} \nabla_\mu \nabla \cdot A+\kappa_{A31} A^\nu (\nabla_\mu A_\nu)-A_\mu (\kappa_{A31}\nabla \cdot A-\nonumber\\
&4\zeta_{30}A^2+2\zeta_{11}R)]R+[\kappa_{A32}\nabla_\mu A_\nu+\kappa_{A33}\nabla_\nu A_\mu+\kappa_{A34}A_\mu A_\nu]\nabla^\nu R, \nonumber\\
\mathscr{X}^{[A]8}_\mu&=-[\kappa_{A10} R_{\mu\nu\alpha\beta}-3\zeta_{33}R_{\mu\alpha\nu\beta}] A^\beta \nabla^\nu A^\alpha, \nonumber\\ 
\mathscr{X}^{[A]9}_\mu &=[\kappa_{A35}\nabla^\nu A^\alpha+\kappa_{A19}A^\nu A^\alpha]\nabla_\mu \nabla_\nu A_\alpha+\kappa_{A36} (\nabla^\nu A^\alpha) \nabla_\mu \nabla_\alpha A_\nu +2 [\kappa_{A10} \nabla^\nu A^\alpha-\nonumber\\
&\zeta_{34}A^\nu A^\alpha] \nabla_\nu \nabla_\alpha A_\mu+[\kappa_{A36} \nabla_\mu A_\nu+\kappa_{A35} \nabla_\nu A_\mu+\kappa_{A19} A_\mu A_\nu] \nabla^\nu \nabla \cdot A+[\kappa_{A37} \nabla \cdot A+\nonumber\\
&\kappa_{A38}A^2] \nabla_\mu \nabla \cdot A, \nonumber\\
\mathscr{X}^{[A]10}_\mu &=-4\Xi^{(13)}\square A_\mu+ [2 \kappa_{A10} \nabla_{(\mu}A_{\alpha)}+\kappa_{A12} A_\mu A_\alpha]\square A^\alpha, \nonumber\\
\mathscr{X}^{[A]11}_\mu &=[\kappa_{A39}(\nabla_\nu A_\alpha)^2+\kappa_{A40} (\nabla \cdot A)^2]A_\mu+[A_\alpha (\kappa_{A41}\nabla^\nu A^\alpha+\kappa_{A42} \nabla^\alpha A^\nu)+A^\nu (\kappa_{A43}\nabla \cdot A+\nonumber\\
&\kappa_{A44} A^2)] \nabla_\mu A_\nu-2[A_\alpha (4\zeta_{32}\nabla^\nu A^\alpha+\zeta_{34} \nabla^\alpha A^\nu )+A^\nu (\kappa_{A45}\nabla \cdot A+\zeta_{24}A^2 )]\nabla_\nu A_\mu \nonumber\\
&+[\kappa_{A46} (\nabla_\nu A_\alpha)(\nabla^\alpha A^\nu)-2A^2 (\kappa_{A47}\nabla \cdot A+3\zeta_{43}A^2)]A_\mu,
\end{align}
\endgroup
where the particular coefficient functionals $\widehat{{K}}^{\alpha\beta}$ and $\widehat{L}^{\alpha\beta}$ are described as follows 
\begingroup
\allowdisplaybreaks
\begin{align}
\widehat{{K}}^{\alpha\beta}&=2 \Theta_{[3]3}^\alpha A^\beta-\zeta_{21} A^\alpha A^\beta, \nonumber\\
\widehat{L}^{\alpha\beta}&=\kappa_{A1} \nabla^\alpha A^\beta+\kappa_{A2} \nabla^\beta A^\alpha+\zeta_{15} A^\alpha A^\beta+\zeta_6 R^{\alpha\beta}.
\end{align}
\endgroup
Here, the explicit form of the distinct values of terms $\Xi^{(m)}$'s and $\zeta_n$'s can be found in (\ref{xiterms}) and (\ref{AppenA}), respectively. Further, the coefficient functionals $\Theta_{[1]6}^\nu A^\alpha$, $\Theta^\alpha_{[5]10}A_\mu$, $\Theta_{[6]2} A$ and $\Theta_{[7]33} A$ are exhibited in (\ref{appb1}), (\ref{appb5}), (\ref{appb6}), and (\ref{appb7}), successively. Notice also that the constant coefficients $\kappa_{Aj}$ (where $j=1\cdots47$) belonging to the gauge field are presented in (\ref{constansofgaugefldseq}).

\section{Constant Coefficients $\kappa_{Aj}$'s of Gauge Field Equations}\label{constansofgaugefldseq}

Finally, the particular constant coefficients $\kappa_{Aj}$ (with $j=1\cdots47$) associated with the gauge field equation in (\ref{vectfeqcmpc}) expressly are described as below
\begingroup
\allowdisplaybreaks
\begin{align}
& \kappa_{A1}=2(2\zeta_{12}-\zeta_{13}-\zeta_{14}),\,\,\,\, \kappa_{A2}=2(-2\zeta_{12}+\zeta_{13}),\,\,\,\,\kappa_{A3}=\kappa_{1(1)}, \,\,\,\, \kappa_{A4}=2\kappa_{2(1)},\,\,\,\, \kappa_{A5}=\kappa_{4(1)}, \nonumber\\
&\kappa_{A6}=2 \kappa_{3(1)}, \quad \kappa_{A7}=-2 \kappa_{1(2)}, \quad \kappa_{A8}=2(2\zeta_{31}-\zeta_{38}), \quad\kappa_{A9}=-2 \kappa_{139(7)}, \quad \kappa_{A10}=\zeta_{33}-\zeta_{36}, \nonumber\\
&\kappa_{A11}=-\kappa_{136(7)}, \,\,\kappa_{A12}=2(\zeta_{34}-\zeta_{35}), \,\kappa_{A13}=2 (2\zeta_{12}-\zeta_{13}-\zeta_{14}+72\alpha_2), \,\, \kappa_{A14}=-\kappa_{A10}+\zeta_{15}, \nonumber\\
&\kappa_{A15}=-2\kappa_{4(7)},\,\,\,\, \kappa_{A16}=2\kappa_{19(5)}, \,\,\,\, \kappa_{A17}=2\zeta_{20}-\zeta_{21}-3\zeta_{33},\,\,\,\, \kappa_{A18}=\kappa_{A10}, \,\,\,\, \kappa_{A19}=-\kappa_{127(7)},\nonumber\\
&  \kappa_{A20}=-\kappa_{1(1)}-2\zeta_9, \,\, \kappa_{A21}=2\kappa_{31(7)}, \,\,\,\kappa_{A22}=-2 \zeta_{20}+\zeta_{21}+\kappa_{A8}, \,\,\, \kappa_{A23}=-16 \{ 14(3\alpha_2+\alpha_6)+\nonumber\\
&3 \alpha_3+4[3(\alpha_4-\alpha_7)-4\alpha_5]\},\,\,\kappa_{A24}=64 [\alpha_4-\frac{1}{2} \alpha_5+\frac{1}{2} \alpha_6-\frac{5}{4} \alpha_7], \,\,\kappa_{A25}=2 \kappa_{31(7)}, \,\, \kappa_{A26}=\kappa_{12(5)}, \nonumber\\
& \kappa_{A27}=\kappa_{115(7)}, \quad \kappa_{A28}=\kappa_{116(7)}, \quad \kappa_{A29}=\kappa_{85(7)}, \quad \kappa_{A30}=4\zeta_{25}-2\zeta_{28}, \quad  \kappa_{A31}=-\zeta_{27}-2\zeta_{29},\nonumber\\
& \kappa_{A32}= \kappa_{98(7)}, \quad \kappa_{A33}=-2 \kappa_{1(2)}-\kappa_{2(1)},\quad \kappa_{A34}=\kappa_{99(7)}, \quad \kappa_{A35}=\kappa_{A10}-4 \zeta_{32}, \quad \kappa_{A36}=\kappa_{A8}-\nonumber\\
&3 \zeta_{33}, \quad \kappa_{A37}=\kappa_{A8}+6 \zeta_{40}, \quad \kappa_{A38}=\kappa_{A9}+2 \zeta_{41},\quad \kappa_{A39}=\kappa_{A12}+4 \zeta_{32},\quad \kappa_{A40}=-2(\zeta_{24}+\zeta_{41}),\nonumber\\
&\kappa_{A41}=-\kappa_{A11}+2 \kappa_{A9}-2\zeta_{24},\quad \kappa_{A42}=\kappa_{A19}+2\zeta_{34}, \quad \kappa_{A43}=\kappa_{A19}-2 \kappa_{A40}+2\zeta_{24}, \quad \kappa_{A44}=\nonumber\\
&4(\zeta_{24}+\zeta_{42}), \quad \kappa_{A45}=\frac{1}{2} \kappa_{128(7)},\quad  \kappa_{A46}=\kappa_{A19}-\kappa_{A9}+2\zeta_{24},\quad \kappa_{A47}=\frac{1}{4}\kappa_{A44}+\zeta_{42},
\end{align}
\endgroup
where the functional $\Theta_{[3]3}^\alpha A^\beta$ and the particular values of terms $\zeta_n$'s are respectively given in (\ref{appb3}) and (\ref{AppenA}). As in the terms so far, the explicit expressions for the different values of constants $\kappa_{m(1)}$, $\kappa_{1(2)}$, $\kappa_{n(5)}$, and $\kappa_{k(7)}$ can be arrived in (\ref{kapQ1}), (\ref{kapQ234}), (\ref{kapQ5}) and (\ref{kapQ7}), independently. Note also that all the terms $\alpha$'s which have occurred in almost all expressions through the paper are the relative coupling constants defined in the main action (\ref{wecgaction1}).

\end{document}